\RequirePackage{fix-cm} 
\RequirePackage[reqno]{amsmath}
\RequirePackage[l2tabu, orthodox]{nag}
\documentclass[a4paper, twoside, dvips, 12pt]{amsart}
\usepackage{etex}
\usepackage{fixltx2e}   

\usepackage[T1]{fontenc}
\usepackage[latin1]{inputenc}
\usepackage{lmodern}

\usepackage{esint}
\usepackage{amsbsy}
\usepackage{dsfont}
\usepackage{empheq}
\usepackage{xspace}
\usepackage{amsgen}
\usepackage{amsthm}
\usepackage{manfnt}
\usepackage{pifont}
\usepackage{amssymb}
\usepackage{upgreek}
\usepackage{amsfonts}
\usepackage{extarrows}
\usepackage{mathtools}
\usepackage{textgreek}
\usepackage[nice]{nicefrac}

\usepackage{mathabx}
\usepackage{relsize}
\usepackage{textcomp}
\usepackage[mathcal]{euscript}

\usepackage{rsfso}
\usepackage{mathrsfs}
\DeclareMathAlphabet{\mathscrbf}{OMS}{mdugm}{b}{n}

\DeclareFontFamily{U}{dutchcal}{\skewchar\font=45 }
\DeclareFontShape{U}{dutchcal}{m}{n}{<-> s*[1.0] dutchcal-r}{}
\DeclareFontShape{U}{dutchcal}{b}{n}{<-> s*[1.0] dutchcal-b}{}
\DeclareMathAlphabet{\mathlcal}{U}{dutchcal}{m}{n}
\SetMathAlphabet{\mathlcal}{bold}{U}{dutchcal}{b}{n}

\usepackage{a4wide}

\usepackage[dvipsnames, table]{xcolor}
\usepackage[most]{tcolorbox}

\definecolor{bckg}{RGB}{20.8, 20.8, 20.8}
\definecolor{oneblue}{rgb}{0.0, 0.0, 0.85}
\definecolor{Lightblue}{RGB}{214, 214, 214}
\definecolor{bluepigment}{rgb}{0.2, 0.2, 0.6}
\definecolor{charcoal}{rgb}{0.21, 0.27, 0.31}
\definecolor{denimblue}{rgb}{0.08, 0.38, 0.74}
\definecolor{Lightgray}{rgb}{0.89, 0.89, 0.89}
\definecolor{darkgrey}{rgb}{0.273, 0.281, 0.30}
\definecolor{darkelectricblue}{rgb}{0.33, 0.41, 0.47}

\usepackage[sort&compress, comma, square, numbers]{natbib}

\usepackage{tikz}
\usepackage{psfrag}
\usepackage{graphicx}
\usepackage{subfigure}
\usepackage{morefloats}
\usepackage{indentfirst}

\usepackage{acronym}
\usepackage{microtype}
\usepackage[labelsep=period,%
            labelfont={bf,sf,color=bluepigment},%
            justification=raggedright]{caption}

\usepackage[perpage, symbol]{footmisc}

\usepackage[usenames, dvipsnames, pdf]{pstricks}
\usepackage{pst-grad} 
\usepackage{pst-plot} 

\usepackage[colorlinks,
           urlcolor=oneblue,
           linkcolor=denimblue,
           citecolor=NavyBlue,
           bookmarksopen=false,
           pdfpagemode=UseNone,
           plainpages=false,
           pagebackref]{hyperref}

\usepackage[explicit]{titlesec}

\titleformat{\section}
  {\color{NavyBlue}\Large\sffamily\bfseries}
  {}
  {0em}
  {\colorbox{bckg!5}{\parbox{\dimexpr\linewidth-2\fboxsep\relax}{\centering\thesection. #1}}}
  [\vspace*{0.33em}]

\titleformat{name=\section, numberless}
  {\color{NavyBlue}\Large\sffamily\bfseries}
  {}
  {0.0em}
  {\colorbox{bckg!5}{\parbox{\dimexpr\linewidth-2\fboxsep\relax}{\centering#1}}}
  [\vspace*{0.33em}]

\titleformat{\subsection}
  {\color{NavyBlue}\large\sffamily\bfseries}
  {}
  {0.0em}
  {\colorbox{bckg!5}{\parbox{\dimexpr\linewidth-5\fboxsep\relax}{\centering\thesubsection. #1}}}
  [\vspace*{0.33em}]

\titleformat{name=\subsection, numberless}
  {\color{NavyBlue}\Large\sffamily\bfseries}
  {}
  {0em}
  {\colorbox{bckg!5}{\parbox{\dimexpr\linewidth-5\fboxsep\relax}{\centering#1}}}
  [\vspace*{0.33em}]

\titleformat{\subsubsection}
  {\color{bluepigment}\sffamily\normalsize\bfseries}
  {}
  {0.0em}
  {\colorbox{bckg!5}{\parbox{\dimexpr\linewidth-5\fboxsep\relax}{\centering\thesubsubsection. #1}}}
  [\vspace*{0.33em}]

\titleformat{name=\subsubsection, numberless}
  {\color{bluepigment}\sffamily\normalsize\bfseries}
  {}
  {0.0em}
  {\colorbox{bckg!5}{\parbox{\dimexpr\linewidth-5\fboxsep\relax}{\centering#1}}}
  [\vspace*{0.33em}]

\titleformat{\paragraph}[runin]
  {\color{bluepigment}\sffamily\small\bfseries}
  {}
  {0em}
  {#1}

\titlespacing{\section}{1.0em}{1.5em plus 2pt minus 2pt}%
{1.0em plus 2pt minus 2pt}[0em]
\titlespacing{\subsection}{1.0em}{1.5em plus 2pt minus 2pt}%
{1.0em}[0em]
\titlespacing{\subsubsection}{1.0em}{1.5em plus 2pt minus 2pt}%
{1.0em plus 2pt minus 2pt}[0em]

\usepackage{titletoc}

\contentsmargin{0.5em}
\newlength{\tocsep} 
\titlecontents{section}[\tocsep]
  {\addvspace{10pt}\bfseries\sffamily}
  {\contentslabel[\thecontentslabel]{\tocsep}}
  {}
  {\ \titlerule*[0.75pc]{.}\ \thecontentspage}
  []

\titlecontents{subsection}[\tocsep]
  {\addvspace{8pt}\sffamily}
  {\contentslabel[\thecontentslabel]{\tocsep}}
  {}
  {\ \titlerule*[0.5pc]{.}\ \thecontentspage}
  []

\titlecontents{subsubsection}[\tocsep]
  {\addvspace{4pt}\footnotesize\sffamily}
  {\contentslabel[\thecontentslabel]{\tocsep}\hspace*{0.5em}}
  {}
  {\ \titlerule*[0.35pc]{.}\ \thecontentspage}
  []

\makeatletter
\def\@setauthors{%
  \begingroup
  \def\thanks{\protect\thanks@warning}%
  \trivlist
  \centering\footnotesize \@topsep30\p@\relax
  \advance\@topsep by -\baselineskip
  \item\relax
  \author@andify\authors
  \def\\{\protect\linebreak}%
  \textsc{\normalsize\textcolor{darkelectricblue}{\authors}}%
  \ifx\@empty\contribs
  \else
    ,\penalty-3 \space \@setcontribs
    \@closetoccontribs
  \fi
  \endtrivlist
  \endgroup
}
\def\@settitle{\begin{center}%
  \baselineskip14\p@\relax
    \bfseries
    \textsc{\Large\textcolor{charcoal}{\@title}}
  \end{center}%
}
\makeatother

\usepackage{enumitem}
\setlist[description]{%
  topsep = 30pt,               
  itemsep = 8pt,               
  labelsep = 10pt,
  font={\bfseries\color{NavyBlue}}, 
}

\usepackage{fancyhdr}
\usepackage{lastpage}

\newcommand*\Title{\textcolor{bluepigment}{Wave/partially immersed body interaction. Part I}}
\newcommand*\Authors{\textcolor{bluepigment}{G.~Khakimzyanov \& D.~Dutykh}}
\newcommand*{\plogo}{\textcolor{gray}{{\texttt{arXiv.org} / \textsc{hal}}}} 

\numberwithin{equation}{section}

\theoremstyle{plain}
\newtheorem{theorem}{Theorem}

\theoremstyle{remark}
\newtheorem{remark}{Remark}

\newcommand{\up}[1]{$^{\mathrm{\small\textsf{#1}}}$} 

\newtcbox{\mymath}[1][]{%
    nobeforeafter, math upper, tcbox raise base,
    enhanced, colframe = black!35,
    colback = black!5, boxrule = 1pt, arc = 0mm,
    #1}

\newcommand{\N}{\mathds{N}}
\newcommand{\R}{\mathds{R}}

\newcommand{\Ee}{\mathds{E}}
\newcommand{\Id}{\mathds{I}}
\newcommand{\ud}{\mathrm{d}}

\renewcommand{\O}{\mathcal{O}}

\renewcommand{\S}{\mathcal{S}}
\newcommand{\const}{\mathrm{const}}
\newcommand{\No}{$\mathrm{N}^\circ$}

\newcommand{\pt}{\hat{p}}
\newcommand{\pc}{\check{p}}
\newcommand{\qc}{\check{q}}
\newcommand{\A}{\mathds{A}}
\newcommand{\vphi}{\varphi}
\renewcommand{\psi}{\uppsi}
\newcommand{\E}{\mathscr{E}}
\newcommand{\Q}{\mathscr{Q}}
\newcommand{\s}{\mathlcal{s}}
\newcommand{\D}{\mathfrak{D}}
\newcommand{\Pp}{\mathscr{P}}
\renewcommand{\Pr}{\powerset}
\newcommand{\qs}{\mathfrak{q}}
\newcommand{\fs}{\mathfrak{f}}
\newcommand{\rs}{\mathfrak{s}}
\renewcommand{\H}{\mathcal{H}}
\newcommand{\Kin}{\mathcal{K}}
\newcommand{\Pot}{\mathcal{T}}
\newcommand{\Qq}{\mathfrak{Q}}
\newcommand{\g}{\boldsymbol{g}}
\newcommand{\n}{\boldsymbol{n}}
\newcommand{\x}{\boldsymbol{x}}
\renewcommand{\delta}{\updelta}
\renewcommand{\omega}{\upvarpi}
\renewcommand{\a}{\boldsymbol{a}}
\renewcommand{\b}{\boldsymbol{b}}
\renewcommand{\r}{\boldsymbol{r}}
\renewcommand{\u}{\boldsymbol{u}}
\renewcommand{\v}{\boldsymbol{v}}
\newcommand{\cl}{\mathfrak{cl}\,}
\renewcommand{\mapsto}{\longmapsto}
\newcommand{\taub}{\boldsymbol{\tau}}
\newcommand{\Om}{\boldsymbol{\Omega}}
\newcommand{\ub}{\bar{\boldsymbol{u}}}
\newcommand{\Ddt}{\under{\mathcal{D}}}
\newcommand{\Rrt}{\under{\mathscr{R}}}
\newcommand{\bigma}{\boldsymbol{\sigma}}
\newcommand{\ut}{\under{\boldsymbol{u}}}
\renewcommand{\kappa}{\mbox{\textkappa}}
\newcommand{\Et}{\widetilde{\mathscr{E}}}
\newcommand{\B}{\boldsymbol{\mathfrak{B}}}
\newcommand{\Rrb}{\bar{\strut\mathscr{R}}}
\newcommand{\Ddb}{\bar{\strut\mathcal{D}}}
\newcommand{\bepsilon}{\boldsymbol{\epsilon}}
\newcommand{\rbv}{\breve{\strut\boldsymbol{r}}}
\newcommand{\zbv}{\breve{\strut\boldsymbol{0}}}
\renewcommand{\ni}{\!\stackrel{\rightarrow}{\n}}
\renewcommand{\eta}{\hspace{0.05em}\mbox{\texteta}}
\newcommand{\rb}{\breve{\raisebox{0pt}[1.2\height]{r}}}
\newcommand{\zb}{\breve{\raisebox{0pt}[1.2\height]{0}}}
\newcommand{\kappar}{\ring{\raisebox{0pt}[1.2\height]{\kappa}}}
\newcommand{\powerset}{\raisebox{.15\baselineskip}{\Large\ensuremath{\wp}}}
\newcommand{\under}[1]{\mkern 1.5mu\underline{\mkern-1.5mu#1\mkern-1.5mu}\mkern 1.5mu}

\newcommand{\St}{\mathrm{St}}

\newcommand{\NS}{\boldsymbol{\mathsf{NS}}}
\newcommand{\SW}{\boldsymbol{\mathsf{SV}}}
\newcommand{\Eu}{\boldsymbol{\mathsf{Eul}}}
\newcommand{\Po}{\boldsymbol{\mathsf{Pot}}}
\newcommand{\SGN}{\boldsymbol{\mathsf{SGN}}}
\newcommand{\Bouss}{\boldsymbol{\mathsf{Bouss}}}

\newcommand{\eps}{\varepsilon}
\renewcommand{\leq}{\leqslant}
\renewcommand{\geq}{\geqslant}

\newcommand{\ie}{\emph{i.e.}\xspace}
\newcommand{\eg}{\emph{e.g.}\xspace}
\newcommand{\etc}{\emph{etc.}\xspace}

\newcommand{\eqass}{\models}

\newcommand{\rot}{\grad\bwedge}
\renewcommand{\div}{\grad\scal}
\newcommand{\Mat}{\mathrm{Mat}}

\newcommand{\bydef}{:\Rightarrow}
\newcommand{\dprime}{\prime\prime}
\newcommand{\bwedge}{\pmb{\wedge}}
\newcommand{\grad}{\boldsymbol{\nabla}}

\newcommand{\asseq}{\reflectbox{$\models$}}

\newcommand{\Fin}[1]{\boldsymbol{#1}^{\,\leqslant}}
\newcommand{\abs}[1]{\left\lvert\, #1\, \right\rvert}

\newcommand{\eqdef}{\mathop{\stackrel{\,\mathrm{def}}{:=}\,}}
\newcommand{\defeq}{\mathop{\stackrel{\,\mathrm{def}}{=:}\,}}

\newcommand{\pd}[2]{\frac{\partial\hspace{0.0556em} #1}{\partial\/ #2}}
\newcommand{\pdd}[2]{\dfrac{\partial\hspace{0.0556em} #1}{\partial\/ #2}}
\newcommand{\od}[2]{\frac{\mathrm{d}\hspace{0.0556em} #1}{\mathrm{d}\/#2}}

\newcommand{\scal}{\;\raisebox{0.25ex}{\tikz\filldraw[black, x=1pt, y=1pt](0,0) circle (1);}\;}

\newcommand{\half}{{\textstyle{1\over2}}}

\makeatletter
\renewcommand*\env@matrix[1][\arraystretch]{%
  \edef\arraystretch{#1}%
  \hskip -\arraycolsep
  \let\@ifnextchar\new@ifnextchar
  \array{*\c@MaxMatrixCols c}}
\makeatother

\usepackage{acronym}
\acrodef{BVP}{Boundary Value Problem}
\acrodef{NSWE}{Nonlinear Shallow Water Equations}

\begin{document}

\title[\Title]{Long wave interaction with a partially immersed body. Part I: Mathematical models}

\author[G.~Khakimzyanov]{Gayaz Khakimzyanov}
\address{\textbf{G.~Khakimzyanov:} Institute of Computational Technologies of Siberian Branch of the Russian Academy of Sciences, Novosibirsk 630090, Russia}
\email{Khak@ict.nsc.ru}
\urladdr{https://www.researchgate.net/profile/Gayaz\_Khakimzyanov3/\vspace*{0.5em}}

\author[D.~Dutykh]{Denys Dutykh$^*$}
\address{\textbf{D.~Dutykh:} Univ. Grenoble Alpes, Univ. Savoie Mont Blanc, CNRS, LAMA, 73000 Chamb\'ery, France and LAMA, UMR 5127 CNRS, Universit\'e Savoie Mont Blanc, Campus Scientifique, 73376 Le Bourget-du-Lac Cedex, France}
\email{Denys.Dutykh@univ-smb.fr}
\urladdr{http://www.denys-dutykh.com/}
\thanks{$^*$ Corresponding author}



\begin{titlepage}
\clearpage
\pagenumbering{arabic}
\thispagestyle{empty} 
\noindent
{\Large Gayaz \textsc{Khakimzyanov}}\\
{\textit{\textcolor{gray}{Institute of Computational Technologies of SB RAS, Russia}}}
\\[0.02\textheight]
{\Large Denys \textsc{Dutykh}}\\
{\textit{\textcolor{gray}{CNRS, Universit\'e Savoie Mont Blanc, France}}}
\\[0.16\textheight]

\vspace*{0.99cm}

\colorbox{Lightblue}{
  \parbox[t]{1.0\textwidth}{
    \centering\huge
    \vspace*{0.75cm}
    
    \textsc{\textcolor{bluepigment}{Long wave interaction with a partially immersed body. Part I: Mathematical models}}
    
    \vspace*{0.75cm}
  }
}

\vfill 

\raggedleft     
{\large \plogo} 
\end{titlepage}


\cleardoublepage
\thispagestyle{empty} 
\par\vspace*{\fill}   
\begin{flushright} 
{\textcolor{denimblue}{\textsc{Last modified:}} \today}
\end{flushright}


\clearpage
\maketitle
\thispagestyle{empty}


\begin{abstract}

In the present article we consider the problem of wave interaction with a partially immersed, but floating body. We assume that the motion of the body is prescribed. The general mathematical formulation for this problem is presented in the framework of a hierarchy of mathematical models. Namely, in this first part we formulate the problem at every hierarchical level. The special attention is payed to fully nonlinear and weakly dispersive models since they are most likely to be used in practice. For this model we have to consider separately the inner (under the body) and outer domains. Various approached to the gluing of solutions at the boundary is discussed as well. We propose several strategies which ensure the global conservation or continuity of some important physical quantities.

\bigskip\bigskip
\noindent \textbf{\keywordsname:} floating body; wave/body interaction; free surface flows; nonlinear dispersive waves; long waves \\


\end{abstract}


\newpage
\pagestyle{empty}
\tableofcontents
\clearpage
\pagestyle{fancy}


\clearpage
\bigskip\bigskip
\section{Introduction}

In the design process of various floating devices, one of the main parameters to take into account is the expected wave run-up magnitude during unavoidable over-topping events in rough seas. Clearly, there is a need to analyze large parameter spaces, \ie wave/wind states, relative dimensions, orientations, \etc in order to obtain a nearly-optimal design. In this way we arrive naturally to the need of development of fast and accurate numerical algorithms to simulate wave fields interaction with floating structures. The methods developed in coastal/naval engineering communities are based on an important number of simplifying assumptions (see \eg \cite{Newman1977, Goda2010}) to obtain quick estimations of required parameters. There are also well-developed analytical methods restricted essentially to linear problems \cite{Linton2001}. The goal of our study is to introduce into this topic a more nonlinear description along with efficient methods to solve equations numerically. As a result, we would like to be able to estimate even local characteristics of the flow. We shall make at some point two main simplifying assumptions:
\begin{enumerate}
  \item The waves are long, \ie \emph{weakly} dispersive,
  \item The object is floating, but fixed in horizontal directions\footnote{In our modelling we allow the object to move freely in the vertical direction according to a prescribed law.} (in contrast to freely floating objects).
\end{enumerate}
Because of the second assumption, we shall speak below about a \emph{partially immersed body}. Nevertheless, the model predictions will be checked against spare experimental data \cite{Kamynin2010}. Despite these simplifying assumptions, there are practically important situations, where they hold true. To give an example, the resonant wave pumping device analyzed in \cite{Carmigniani2017a} falls perfectly in the framework presented in this study.

The well-studied topic is the wave generation by moving structures such as ships. Practical interest of such works is quite obvious. The first analytical steps in this direction were done by Lord \textsc{Kelvin}. We can refer also, for example, to early numerical (finite difference) attempts to compute the wave field behind a ship \cite{Miyata1985}. In the problem considered in our study the structure is fixed and we are interested in wave interaction with it. Moreover, we are looking at generated wave fields behind \emph{and} in front of the floating object.

The topic of numerical modelling of the wave/(floating or immersed) body interaction attracted much attention in the recent years. In water wave theory the most studied situation both by analytical and numerical techniques is the wave interaction with a single \cite{Isaacson1982, Zhou2016} or with an array \cite{Zhao2007} of floating/fixed circular cylinders. In \cite{Isaacson1982} the cylinder was allowed to move in vertical direction. Most of the numerical studies are based on the boundary integral equations method, while analytical investigations focus mainly on linear or, exceptionally, weakly nonlinear formulations. However, there are a few exceptions. A mixed \textsc{Eulerian}--\textsc{Lagrangian} method was applied to describe wave-induced motions of a floating body in \cite{Kashiwagi2000}. On the other hand, an \textsc{Eulerian} spectral element method was recently applied to wave/body interaction problems in \cite{Engsig-Karup2017}. The ultimate goal of such investigations is to propose a robust and efficient methodology for the simulation of floating real world objects such as boats \cite{Parolini2005} and/or wave energy converters \cite{Rafiee2016, Folley2012, Carmigniani2017}. In such complex applications sometimes even the full \textsc{Navier}--\textsc{Stokes} equations are solved using state-of-the-art computational techniques \cite{Zhuang1997, Lin2006a}. For instance, in \cite{Lin2006a} the \textsc{Navier}--\textsc{Stokes} ($\NS$) equations were solved using multiple-layer $\sigma-$coordinate method in the presence of submerged obstacles in the flow.

The problem of wave interaction with a partially immersed body is presented below in the framework of a hierarchy of mathematical models:
\begin{itemize}
  \item Rotational incompressible ideal fluid flow model \cite{Lamb1932} ($\Eu$),
  \item Potential flow model \cite{Stoker1957} ($\Po$),
  \item Fully nonlinear weakly dispersive wave model \cite{Serre1953, Serre1953a, Green1976, Green1974} ($\SGN$),
  \item \textsc{Boussinesq}-type weakly nonlinear and weakly dispersive model \cite{Boussinesq1877, DMII, Brocchini2013}
  \item Nonlinear shallow water (nonlinear non-hydrostatic or \textsc{Saint-Venant}) equations \cite{SV1871} ($\SW$).
\end{itemize}
Every subsequent model on this list can be obtained from the previous one by applying one simplifying assumption. The need to consider a sequence of models can be attributed to the celebrated \textsc{Ockham} razor principle (\emph{lex parsimoniae}), which states that among competing hypothetical answers to a problem, one should select the answer that makes the fewest assumptions \cite{Ariew1976}.

In our work we shall consider in more or less detail all models listed above along with corresponding discretization algorithms (to be described in the second part \cite{Khakimzyanov2018b}). In practice, this hierarchical approach allows us to assess the accuracy of various models together with corresponding simplifying assumptions. The numerical algorithm is based on the moving grid technique \cite{Khakimzyanov2015a} to have a balanced and adaptive grid distribution. For shallow water equations the computational domain is split in two parts: under the partially immersed body and the rest. The beauty of our approach consists in the fact that at this modelling level, governing equations can be solved \emph{almost} analytically under the object. Thus, the numerical treatment is needed only in the outer domain. The local surgery of solutions at various sub-domains boundaries is done using compatibility conditions.

Using developed algorithms we can study the influence of incident wave amplitude, body elongation and immersion on the induced wave fields in the vicinity of the body and the generated flow under the body. A particular attention is payed to the situation where the partially immersed body is located near a vertical wall. The reason is that transmitted waves may enter into a resonance with waves reflected from the wall. This resonance would lead to anomalously high amplitudes in the interval between the immersed body and the wall \cite{Khakimzyanov2018}. Moreover, we provide the comparisons of numerical solutions to various models. Based on these comparisons we can draw some practical conclusions on the validity range of approximate models for the wave/body interaction problems, which are more peculiar than free wave propagation. These results will be described in the second part of this study \cite{Khakimzyanov2018b}.

The present manuscript is organized as follows. The wave/obstacle interaction problem is described in Section~\ref{sec:prob} as a hierarchy of mathematical models in the inner (\ie under the obstacle) and outer (\ie outside) domains. Namely, in Section~\ref{sec:eul} we describe the $\Eu$ model, the potential flow model $\Po$ in Section~\ref{sec:pot}. The fully nonlinear weakly dispersive model $\SGN$ is discussed in Section~\ref{sec:sgn}, while its non-dispersive hydrostatic version $\SW$ in Section~\ref{sec:sv}. The coupling (or transmission) conditions for $\SGN$ and $\SW$ models are derived and discussed in Section~\ref{sec:compa}. Finally, the article is completed by outlining the main conclusions and perspectives of the present study in Section~\ref{sec:concl}. We would like to mention that all notations used in our study are summarized in Appendix ``\textrm{Nomenclature}'' on page~\pageref{sec:N}. This manuscript contains also one Appendix, which describes an alternative derivation of compatibility conditions.


\section{Hierarchy of mathematical models}
\label{sec:prob}

The problem of wave/floating body interaction should be considered as a hierarchy of descriptions of various complexity levels. Schematically, this situation can be depicted as:
\begin{align}\label{eq:hie}
  & \xLeftarrow{\text{Simplified}} \nonumber \\
  \SW\ \Longrightarrow\ \Bouss\ \Longrightarrow\ \SGN\ &\Longrightarrow\ \ldots\ \Longrightarrow\ \Po\ \Longrightarrow\ \Eu\ \Longrightarrow\ \NS \\
  & \xRightarrow{\text{More complete}} \nonumber
\end{align}
The arrows show the direction of increasing complexity\footnote{In the present study we do not consider the last class of models $\NS$, which is rather reserved for CFD-type applications.}. Moreover, the last sequence \eqref{eq:hie} of models is \emph{exact} in the sense that to move in the opposite direction, at each level only one simplifying assumption is needed. This hierarchical approach will allow us to assess the `price to pay' for each simplification. In the present work we shall pay more attention to models from the hierarchy \eqref{eq:hie} since they are mostly used in applications due to their reduced complexity. In all this study we shall neglect the influence of compressibility, friction, wind stress, currents, \etc In the upcoming works we shall focus on the hard part ($\Po\ \Longleftarrow\ \Eu$) of the hierarchy \eqref{eq:hie}.


\subsection{Euler equations}
\label{sec:eul}

\begin{figure}
  \centering
  \includegraphics[width=0.99\textwidth]{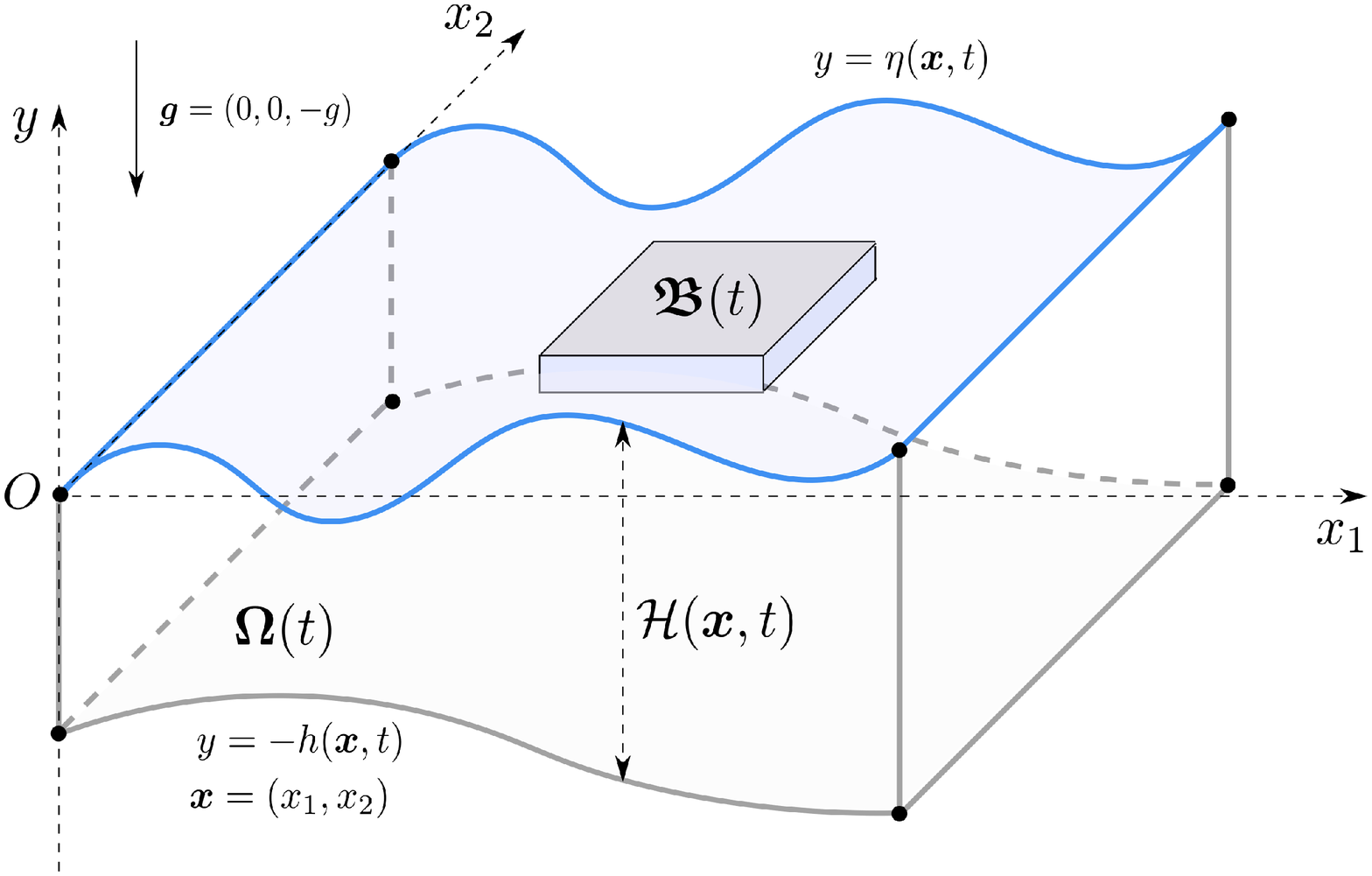}
  \caption{\small\em Sketch of the fluid domain with a fixed partially immersed body. A side view of the fluid/solid domain is shown in Figure~\ref{fig:side}.}
  \label{fig:sketch}
\end{figure}

Consider a flow of an ideal and incompressible fluid, which occupies some three-dimensional simply connected domain $\Om\ \subseteq\ \R^{\,3}\,$. The domain $\Om$ may be unbounded on the sides in theoretical investigations. However, in practical simulations we assume that $\Om$ is bounded by vertical lateral walls on the sides\footnote{This geometric configuration mimics the configuration of a traditional laboratory wave tank.}. The domain $\Om$ is bounded from below by a prescribed solid bottom $y\ =\ -\,h\,(\x,\,t)$ and from above by the free surface $y\ =\ \eta\,(\x,\,t)\,$. In order to describe analytically the boundaries, we introduced implicitly a \textsc{Cartesian} system of coordinates $O\,x_{\,1}\,x_{\,2}\,y\,$. The vector $\x$ is the canonical projection of the $3$D vector $(\x,\,y)$ on the horizontal coordinate plane, \ie $\x\ \cong\ \bigl(x_{\,1},\,x_{\,2}\bigr)\ \in\ \R^{\,2}\,$. The vertical axis $O\,y$ points in the direction opposite to the gravity vector $\g\ \cong\ (0,\,0,\,-g)\ \in\ \R^{\,3}\,$. Traditionally we assume also that the coordinate plane $y\ =\ 0$ coincides with the still water level. See Figure~\ref{fig:sketch} for an illustration of the fluid domain definition. To underline the fluid dynamic character of the problem, we shall denote the fluid domain as $\Om\,(t)\,$, where $t\ \geq\ 0$ is the time coordinate.

The projection of a $3$D domain $\Om\,(t)$ onto the horizontal plane will be denoted by $\under{\Om}\ \eqdef\ \pi_{\,\x;\,t}\,\bigl(\Om\,(t)\bigr)\,$, \ie
\begin{equation*}
  \pi_{\,\x;\,t}:\ \Om\,(t)\ \subseteq\ \R^{\,2}\,\times\,\R\ \longrightarrow\ \R^{\,2}\,, \qquad (\x,\,y)\ \stackrel{\pi_{\,\x;\,t}}{\mapsto}\ \x\,.
\end{equation*}
Thus, $\pi_{\,\x;\,t}$ at any fixed time $t$ is the canonical projection of product space. The fluid domain $\Om\,(t)$ has to be non-degenerate, \ie 
\begin{equation*}
  \H\,(\x,\,t)\ \eqdef\ (h\ +\ \eta\,)\,(\x,\,t)\ \geq\ h_{\,0}\ >\ 0\,, \qquad
  \forall\,t\ \geq\ 0\,,\quad \forall\,\x\ \in\ \under{\Om}\,,
\end{equation*}
where we introduced also the \emph{total water depth} variable $\H:\ \under{\Om}\,\times\,\R^{\,+}_{\,0}\ \longrightarrow\ \R^{\,+}\,$. Moreover, we assume that the fluid density $\rho$ and the gravity acceleration $g$ are constant throughout our experiments. All other forces such as friction, wind forcing, \textsc{Coriolis}, \etc are neglected in the present study.

In the general $3$D \textsc{Eulerian} description, one has to determine the field of fluid particle velocities $\bigl(\u,\,v)\,:\ \Om\,(t)\,\times\,\R^{\,+}_{\,0}\ \longrightarrow\ \R^{\,3}\,$, with $\u$ being the horizontal velocity vector $\u\ \cong\ (u_{\,1},\,u_{\,2})$ and $v$ the vertical component, the pressure distribution $p\,:\ \Om\,(t)\,\times\,\R^{\,+}_{\,0}\ \longrightarrow\ \R$ and the free surface excursion $\eta\,:\ \under{\Om}\,\times\,\R^{\,+}_{\,0}\ \longrightarrow\ \R\,$, which satisfy in domain $\Om\,(t)$ the system $\Eu$ of incompressible \textsc{Euler} equations:
\begin{equation}\label{eq:cont}
  \div\u\ +\ v_{\,y}\ =\ \zb\,,
\end{equation}
\begin{equation}\label{eq:hormom}
  \rho\,\u_{\,t}\ +\ \rho\,(\u\scal\grad)\,\u\ +\ \rho\,v\,\u_{\,y}\ +\ \grad\,p\ =\ \zbv\,,
\end{equation}
\begin{equation}\label{eq:vertmom}
  \rho\,v_{\,t}\ +\ \rho\,\u\scal\grad\,v\ +\ \rho\,v\,v_{\,y}\ +\ p_{\,y}\ =\ -\rho\,g\,,
\end{equation}
where subscripts denote partial derivatives, \ie $(\cdot)_{\,t}\ \eqdef\ \pd{(\cdot)}{t}$ or $(\cdot)_{\,y}\ \eqdef\ \pd{(\cdot)}{y}\,$. On the right hand side of the first two equations, $\zb$ ($\zbv$) denotes a (vector$-$) function, which takes a constant zero value (zero vector), \ie 
\begin{equation*}
  \zb\,:\ \Om\,(t)\,\times\,\R^{\,+}_{\,0}\ \longrightarrow\ \R\,, \qquad (\x,\,y,\,t)\ \stackrel{\zb}{\mapsto}\ 0\,.
\end{equation*}
Notice, that the domain of $\zb$ ($\zbv$) might be restricted depending on the equations where this map appears. This happens, for example, when we specify the boundary conditions. The restricted domain can be inferred, in general, from the context. Thus, we shall not make any special comments on this point in each particular situation.

We introduced also the horizontal gradient symbol $\grad\ \cong\ \bigl(\pd{}{x_{\,1}},\,\pd{}{x_{\,2}}\bigr)\,$. It is to be manipulated as usual:
\begin{equation*}
  \div\u\ \equiv\ \pd{u_{\,1}}{x_{\,1}}\ +\ \pd{u_{\,2}}{x_{\,2}}\,, \qquad
  \u\scal\grad\ \equiv\ u_{\,1}\,\pd{}{x_{\,1}}\ +\ u_{\,2}\,\pd{}{x_{\,2}}\,.
\end{equation*}
On the free surface we have to satisfy two boundary conditions:
\begin{equation}\label{eq:fskc}
  \eta_{\,t}\ +\ \u\scal\grad\,\eta\ -\ v\ =\ \zb\,, \qquad y\ =\ \eta\,(\x,\,t)\,, \qquad \x\ \in\ \under{\Om}\ \setminus\ \under{\B}\,, \qquad t\ \geq\ 0\,,
\end{equation}
\begin{equation}\label{eq:fsdc}
  p\ =\ \zb\,, \qquad y\ =\ \eta\,(\x,\,t)\,, \qquad \x\ \in\ \under{\Om}\ \setminus\ \under{\B}\,, \qquad t\ \geq\ 0\,,
\end{equation}
where as before $\under{\B}\ \eqdef\ \pi_{\,\x;\,t}\,\bigl(\B\,(t)\bigr)$ (the definition of domain $\B\,(t)$ will be properly discussed in Section~\ref{sec:bobo}). On the moving bottom we have a kinematic impermeability condition:
\begin{equation}\label{eq:bkc}
  h_{\,t}\ +\ \u\scal\grad\,h\ +\ v\ =\ \zb\,, \qquad y\ =\ -\,h\,(\x,\,t)\,, \qquad \x\ \in\ \under{\Om}\,, \qquad t\ \geq\ 0\,.
\end{equation}
Of course, we have to admit that all functions are sufficiently smooth to have at least first order derivatives.


\subsubsection{Energy conservation}

In order to derive the total energy conservation equation for the full \textsc{Euler} equations given above, we obtain the evolutions of the kinetic $\Kin\,:\ \Om\,(t)\,\times\,\R^{\,+}_{\,0}\ \longrightarrow\ \R^{\,+}_{\,0}\,$, $\Kin\ \eqdef\ \dfrac{\rho}{2}\;\bigl(\,\abs{\u}^{\,2}\ +\ v^{\,2}\,\bigr)$ and potential $\Pot\,:\ \Om\,(t)\,\times\,\R^{\,+}_{\,0}\ \longrightarrow\ \R\,$, $\Pot\,(\scal,\,y,\,\cdot)\ \eqdef\ \rho\,g\,y$ energies separately for the sake of simplicity. The kinetic energy equation is obtained by multiplying\footnote{The multiplication here is understood in the sense of the standard scalar product in $\Ee^{\,2}\,$.} Equation~\eqref{eq:hormom} by $\u$ and Equation~\eqref{eq:vertmom} by $v$ and summing them up we obtain:
\begin{multline}\label{eq:kin}
  \qquad\u\scal\text{\eqref{eq:hormom}}\ +\ v\cdot\text{\eqref{eq:vertmom}}\ \Longrightarrow\\
  \Kin_{\,t}\ +\ \u\scal\grad\,(\Kin\ +\ p)\ +\ v\,(\Kin\ +\ p)_{\,y}\ +\ \rho\,g\,v\ =\ \zb\,.\qquad
\end{multline}
The potential energy equation reads:
\begin{equation}\label{eq:pot}
  \Pot_{\,t}\ +\ \u\scal\grad\,\Pot\ +\ v\,\Pot_{\,y}\ -\ \rho\,g\,v\ =\ \zb\,.
\end{equation}
By summing up Equations~\eqref{eq:kin} and \eqref{eq:pot} we obtain the total energy $\E\,:\ \Om\,(t)\,\times\,\R^{\,+}_{\,0}\ \longrightarrow\ \R\,$, $\E\ \eqdef\ \Kin\ +\ \Pot$ conservation equation:
\begin{multline}\label{eq:ene}
  \qquad\qquad\text{\eqref{eq:kin}}\ +\ \text{\eqref{eq:pot}}\ \Longrightarrow\\
  \E_{\,t}\ +\ \u\scal\grad\,(\E\ +\ p)\ +\ v\cdot(\E\ +\ p)_{\,y}\ =\ \zb\,.\qquad\qquad
\end{multline}
Finally, by taking into account the incompressibility Condition~\eqref{eq:cont} we obtain the last Equation~\eqref{eq:ene} in the conservative form:
\begin{multline}\label{eq:toteu}
  \qquad\text{\eqref{eq:ene}}\ +\ (\E\ +\ p)\cdot\text{\eqref{eq:cont}}\ \Longrightarrow\\
  \E_{\,t}\ +\ \div\bigl((\E\ +\ p)\,\u\bigr)\ +\ \bigl((\E\ +\ p)\,v\bigr)_{\,y}\ =\ \zb\,.\qquad
\end{multline}
The last equation holds everywhere in the fluid domain $\Om\,(t)\,$.


\subsubsection{Partially immersed body}
\label{sec:bobo}

We assume that the partially immersed body occupies the domain $\B\ \subseteq\ \R^{\,3}$ has vertical side walls, which are impermeable for fluid particles, as it is common in many practical problems. Moreover, we assume that the body cannot move in horizontal directions. In this sense it is fixed. However, we allow another degree of freedom: our body may perform vertical displacements according to a prescribed law. According to the last comment, it is judicious to denote the domain as $\B\,(t)$ to underline its dynamic character. The function $y\ =\ d\,(\x,\,t)$ prescribes the instantaneous position of the object bottom. Since the body is partially immersed into water, we have
\begin{equation*}
  d\,(\x,\,t)\ <\ 0\,, \qquad \forall\,t\ \geq\ 0\,, \qquad \forall\,\x\ \in\ \under{\B}\,.
\end{equation*}
The function $d:\ \under{\B}\,\times\,\R^{\,+}_{\,0}\ \longrightarrow\ \R^{\,-}$ is assumed to be known. The object bottom impermeability condition can be readily written:
\begin{equation}\label{eq:obbc}
  d_{\,t}\ +\ \u\scal\grad\,d\ -\ v\ =\ \zb\,,\qquad y\ =\ d\,(\x,\,t)\,, \qquad \x\ \in\ \under{\B}\,, \qquad t\ \geq\ 0\,.
\end{equation}
Similar formulations exist for all problems where the body bottom moves according to a prescribed trajectory. To give an example of practical importance, the resonance wave pumping set-up studied analytically and experimentally in \cite{Carmigniani2017, Carmigniani2017a} suits perfectly our framework. Additionally, our description admits objects $\B\,(t)$ with impermeable, but \emph{deformable} bottoms provided that lateral walls of $\B\,(t)$ remain always vertical and the deformation is known through the function $d\,$.

The impermeability condition of \emph{wet} lateral walls (of the immersed body $\B\,(t)$ \emph{and} of the wave tank $\Om\,(t)$) can be written as:
\begin{equation}\label{eq:sides}
  \u\scal\n\ =\ \zb\,, \qquad (\x,\,y)\ \in\ \cl\,\bigl(\Om\,(t)\bigr)\ \bigcap\ \cl\,\bigl(\B\,(t)\bigr)\,,
\end{equation}
where $\n\ \in\ \R^{\,2}$ is the vector of the unitary exterior normal to the wall containing only horizontal components\footnote{This boundary condition can be `derived' from $3$D by noticing that from our assumptions $\n_{\,3}\ =\ (\n,\,0)\ \in\ \R^{\,3}$ and
\begin{equation*}
  (\u,\,v)\scal\n_{\,3}\ =\ (\u,\,v)\scal(\n,\,0)\ =\ \u\scal\n\ +\ v\cdot 0\ \equiv\ \u\scal\n\,.
\end{equation*}
Thus, the vertical velocity component $v$ does not play any r\^ole in this boundary condition.}. Finally, to obtain a well-posed problem, one has to prescribe the compatible initial conditions: the initial position of the free surface excursion $\eta\,(\x,\,0)\,$, $\x\ \in\ \under{\Om}\ \setminus\ \under{\B}$ along with initial velocities of fluid particles $(\u,\,v)\,(\x,\,y,\,0)\,$, $(\x,\,y)\ \in\ \Om\,(0)\,$. We remind that the initial positions of the fluid and object bottoms are given through functions $-\,h\,(\x,\,0)\,$, $\x\ \in\ \under{\Om}$ and $d\,(\x,\,0)\,$, $\x\ \in\ \under{\B}$ correspondingly.


\subsubsection{Domain dissection}

From now on we introduce some shorthand notations for various useful domains in the plane $O\,x_{\,1}\,x_{\,2}$ to simplify the problem description. Two projections $\under{\Om}$ and $\under{\B}$ have already been introduced. In this Section we rename them to have uniform notation throughout the rest of our manuscript:
\begin{equation}\label{eq:dissect}
  \D^{\:\boxdot}\ \eqdef\ \under{\Om}\,, \qquad
  \D^{\,\blackdiamond}\ \eqdef\ \under{\B}\,, \qquad
  \D^{\,\diamond}\ \eqdef\ \D^{\:\boxdot}\ \setminus\ \D^{\,\blackdiamond}\,.
\end{equation}
By our assumptions made hereinabove, the domains $\D^{\:\boxdot}\,$, $\D^{\,\blackdiamond}$ and $\D^{\,\diamond}$ are fixed, \ie their locations do not change with time. Moreover, obviously we have $\D^{\:\boxdot}\ =\ \D^{\,\blackdiamond}\ \bigcup\ \D^{\,\diamond}\,$. We introduce also the two boundaries:
\begin{equation*}
  \Gamma^{\,\blackdiamond}\ \eqdef\ \partial\,\D^{\,\blackdiamond}\,, \qquad
  \Gamma^{\:\boxdot}\ \eqdef\ \partial\,\D^{\:\boxdot}\,.
\end{equation*}
We shall also assume that the interior domain $\D^{\,\blackdiamond}$ does not touch (\ie intersect with) the boundary of the fluid domain $\D^{\:\boxdot}\,$, \ie $\Gamma^{\,\blackdiamond}\ \bigcap\ \Gamma^{\:\boxdot}\ =\ \varnothing\,$.

\begin{figure}
  \centering
  \includegraphics[width=0.69\textwidth]{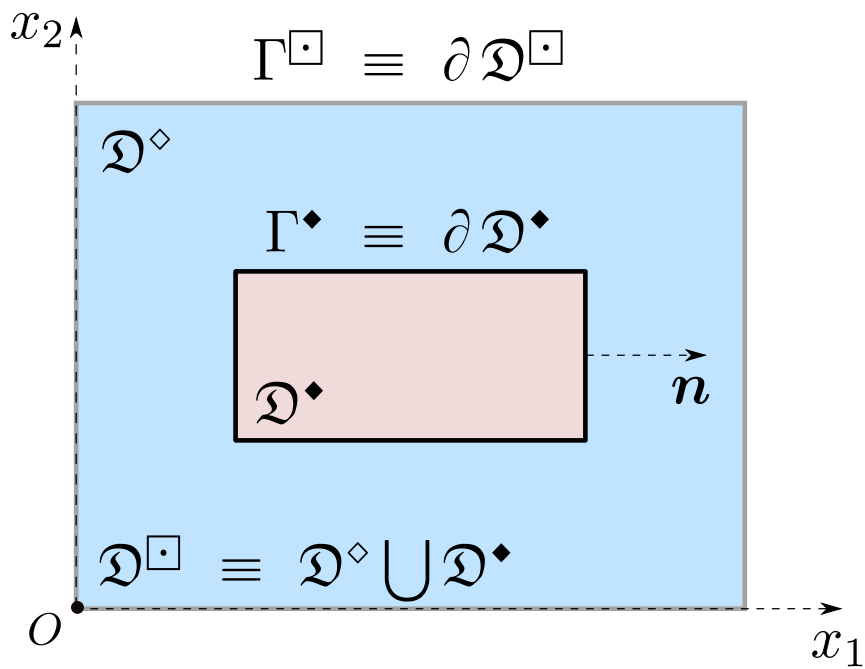}
  \caption{\small\em Disposition of various domains in the $O\,x_{\,1}\,x_{\,2}$ plane.}
  \label{fig:sketch2}
\end{figure}

In new notations, the boundary condition \eqref{eq:bkc} has to hold for $\forall\,\x\ \in\ \D^{\:\boxdot}\,$, Conditions \eqref{eq:fskc}, \eqref{eq:fsdc} hold for $\forall\,\x\ \in\ \D^{\,\diamond}$ and condition \eqref{eq:obbc} for $\forall\,\x\ \in\ \D^{\,\blackdiamond}\,$. Finally, the boundary condition \eqref{eq:sides} has to be satisfied for $\forall\,\x\ \in\ \Gamma^{\,\blackdiamond}\ \bigcup\ \Gamma^{\:\boxdot}\,$. Various domains are illustrated in Figure~\ref{fig:sketch2}. A side view of of the fluid domain is shown in Figure~\ref{fig:side}.

\bigskip
\paragraph*{The first lyrical digression.}
\addcontentsline{toc}{subsubsection}{The first lyrical digression}

The dissection of the computational domain presented above may appear to be artificial for 3D models ($\NS$, $\Eu$, $\Po$), however, this step is absolutely necessary for depth-integrated models ($\SGN$, $\SW$) for the reasons which will become clear below. The design of numerical algorithms follows closely this observation \cite{Khakimzyanov2018b}. In this digression, we would like to discuss the historical origins of the idea to dissect the fluid domain according to the partition \eqref{eq:dissect}.

The first author of the present manuscript has been working in the field of gas dynamics until approximatively 1984. In particular, he has been investigating the problem of gas flow in channels of varying cross-section \cite{Yaushev1980, Yaushev1980a}. These works were based on the pioneering paper by I.~K.~\textsc{Yaushev} (1967) \cite{Yaushev1967}. The main point is that there is a certain analogy between our problem and these gas flows. In particular, in points where there is a jump in channel cross-section, the conservation of momentum does not hold anymore and some extra conditions have to be used at the interface between two sub-domains.

At the end of the (19)80's the problem of wave/floating body interaction has been discussed in the Computational Mathematics seminar\footnote{We would like to add a little precision: the notion of a `\emph{seminar}' in the \textsc{Soviet Union} better corresponds to the modern notion of a `\emph{working group}'.} in \textsc{Krasnoyarsk}, where the first author was present at that time. This topic was stimulated by an experimental campaign conducted in \textsc{Leningrad}\footnote{Currently \textsc{Saint Petersburg}.} on wave/anchored pontoon interaction. The first mathematical models were proposed in the framework of the nonlinear shallow water equations ($\SW$). However, the question of numerical simulation was still open. It is at this moment that the analogy with the gas dynamics has been noticed, probably, for the first time and the previous knowledge \cite{Yaushev1967, Yaushev1980, Yaushev1980a} was implemented in shallow waters. So, definitely, the inspiration for this study comes from the gas dynamics. The first results of this work started to appear in the early (19)90's in the following papers \cite{Urusov1991, Nudner1991, Nudner1992}. To our knowledge, this approach was later rediscovered in \cite{Jiang2001, Henn2001}, where the modern era of wave-body interaction begins.


\subsubsection{Pressure boundary conditions}

We present also another reformulation of boundary condition \eqref{eq:sides} imposed on immersed body impermeable vertical sides $\Gamma^{\,\blackdiamond}\,$. We believe that this consequence might be helpful in designing, for example, numerical algorithms for this problem. Let us consider a regular point $\x\ \in\ \Gamma^{\,\blackdiamond}\,$. Let $\n$ be the vector of unit exterior normal to the object $\D^{\,\blackdiamond}$ in the same point $\x\,$. It is important to notice that vector $\n$ is constant for each side of $\D^{\,\blackdiamond}$ (provided that the immersed body $\B$ has a polygonal shape). By taking a $2$D scalar product of Equation~\eqref{eq:hormom} with normal vector $\n$ we obtain:
\begin{equation*}
  \frac{1}{\rho}\;\grad\,p\scal\n\ =\ -\,(\u\scal\n)_{\,t}\ -\ \u\scal\grad\,(\u\scal\n)\ -\ v\,(\u\scal\n)_{\,y}\,.
\end{equation*}
Taking into account the Condition \eqref{eq:sides}, we obtain the following consequence:
\begin{equation}\label{eq:dpdn}
  \pd{p}{\n}\ \eqdef\ \grad\,p\scal\n\ =\ \zb\,, \qquad \forall\,\x\ \in\ \Gamma^{\,\blackdiamond}\,, \qquad d\,(\x,\,t)\ \leq\ y\ \leq\ \eta\,(\x,\,t)\,, \qquad \forall\, t\ \geq\ 0\,.
\end{equation}
However, we do not stop here, since from the last identity, we can derive another useful consequence for the free surface behaviour on along the solid boundary. Indeed, the dynamic boundary condition \eqref{eq:fsdc} can be rewritten as
\begin{equation}\label{eq:fsdc2}
  p\,\bigl(\x,\,\eta\,(\x,\,t),\,t\bigr)\ =\ 0\,, \qquad \forall\,\x\ \in\ \D^{\,\diamond}\,, \qquad \forall\, t\ \geq\ 0\,.
\end{equation}
By taking the horizontal gradient operator $\grad$ of \eqref{eq:fsdc2}, we obtain:
\begin{equation*}
  \grad\,p\ +\ p_{\,y}\grad\,\eta\ =\ \zbv\,, \qquad \forall\,\x\ \in\ \D^{\,\diamond}\,, \qquad \forall\, t\ \geq\ 0\,.
\end{equation*}
By restricting the last differential identity to the points on boundary $\Gamma^{\,\blackdiamond}$ and taking the scalar product with $\n\,$, in accordance with consequence \eqref{eq:dpdn} we obtain:
\begin{equation*}
  \pd{\eta}{\n}\ =\ \zb\,, \qquad \forall\,\x\ \in\ \Gamma^{\,\blackdiamond}\,, \qquad \forall\, t\ \geq\ 0\,.
\end{equation*}
Thus, the free surface always meets the solid boundary at the right angle. It is not difficult to generalize the proof to more general situations. The case of curvilinear object boundary is treated in the following Section.


\subsubsection{Free surface behaviour in the vicinity of the partially immersed body}
\label{app:a}

As we supposed in the beginning of our study, we consider the case when the partially immersed body $\B\,(t)$ with vertical lateral boundaries $\forall\, t\ \geq\ 0\,$. More precisely, we can describe this boundary as
\begin{equation}\label{eq:sbody}
  \partial_{\,\parallel}\,\B\,(t)\ \bydef\ \Gamma^{\,\blackdiamond}\,\times\,\bigl[\,d\,\vert_{\,\Gamma^{\,\blackdiamond}}\,(\x,\,t),\,+\,\infty\,\bigr]\ \subseteq\ \R^{\,3}\,, \qquad \Gamma^{\,\blackdiamond}\ \equiv\ \pi_{\,\x;\,t}\,\bigl(\partial_{\,\parallel}\,\B\,(t)\bigr)\,.
\end{equation}
The function $d\,(\scal,\,t)\,:\ \D^{\,\blackdiamond}\ \longrightarrow\ \R^{\,-}$ gives us the instantaneous position of the object $\B\,(t)$ bottom. The second Identity in \eqref{eq:sbody} expresses mathematically the fact that the object remains vertical during its motion. The upper limit in $\partial_{\,\parallel}\,\B\,(t)$ is taken to be $+\,\infty$ (here $\abs{\infty}\ =\ \aleph_{\,1}$) for the sake of convenience. We just assume that the body is high enough to avoid the wave overtopping on it. More precisely, in this Section we shall work on the `wet' part of the boundary, where the traces of various fluidic fields can be computed:
\begin{equation*}
  \widetilde{\partial_{\,\parallel}\,\B\,(t)}\ \eqdef\ \Gamma^{\,\blackdiamond}\,\times\,\bigl[\,d\,\vert_{\,\Gamma^{\,\blackdiamond}}\,(\x,\,t),\,\eta\,\vert_{\,\Gamma^{\,\blackdiamond}}\,(\x,\,t)\,\bigr]\ \subseteq\ \partial_{\,\parallel}\,\B\,(t)\,.
\end{equation*}
Our strategy consists in studying first the normal derivative of the pressure $p$ described in $\Eu$ model along the `wet' points on $\widetilde{\partial_{\,\parallel}\,\B\,(t)}\,$. In this Section we treat in details the case when the boundary projection $\Gamma^{\,\blackdiamond}$ is a (piecewise) smooth oriented closed curve in $\R^{\,2}\,$. For the sake of simplicity, we shall choose a natural parametrization $\x\ =\ \r\,(\s)\,$, with $\s\ \in\ \bigl[\,0,\,\S\,\bigr]$ being the arc-length parameter increasing in the positive direction along the curve $\Gamma^{\,\blackdiamond}\,$, whose length is $\S\,$. The starting point is immaterial. The orientation is chosen\footnote{We mention that the change in orientation ($\circlearrowleft\ \leftsquigarrow\ \circlearrowright$) changes the direction of the tangent vector $\taub\,(\s)$ to the opposite $\taub\,(\s)\ \leftsquigarrow\ -\,\taub\,(\s)\,$. However, the principal normal $\n\,(\s)$ is invariant under this transformation since the direction of $\n\,(\s)$ depends on the local convexity properties of the curve $\Gamma^{\,\blackdiamond}\,$.} to be counter-clockwise $\circlearrowleft\,$. Then, vector $\taub\,(\s)\ \eqdef\ \dot{\r}\,(\s)$ is the unit tangent vector to the curve $\Gamma^{\,\blackdiamond}$ \cite{Pogorelov1959}. According to the first \textsc{Frenet--Serret} formula, vector $\ddot{\r}\,(\s)$ is directed along the principal normal $\n\,(\s)$ and
\begin{equation}\label{eq:fs}
  \taub\scal\n\ =\ \zb\,, \qquad
  \ddot{\r}\ =\ \kappa\,\n\,,
\end{equation}
where $\kappa\,:\ \bigl[\,0,\,\S\,\bigr]\ \longrightarrow\ \R^{\,+}$ is the \emph{curvature} defined as $\kappa\,(\s)\ \eqdef\ \abs{\ddot{\r}\,(\s)}\,$. The first relation in \eqref{eq:fs} implies that $\taub\,(\s)\perp\n\,(\s)\,$, $\forall\,\s\ \in\ \bigl[\,0,\,\S\,\bigr]\,$, while the second relation can be rewritten as
\begin{equation}\label{eq:curva}
  \dot{\taub}\ =\ \kappa\,\n\,.
\end{equation}
The curvature $\kappa\,(\s)$ is always non-negative and it vanishes only on rectilinear portions of the curve $\Gamma^{\,\blackdiamond}\,$. For the purposes of this Section it is more convenient to use another vector $\ni\,(\s)\,$, which is the unit exterior normal to the fluid domain (which projects on $\D^{\,\diamond}$) surrounding the object $\B\,(t)\,$. In other words, for $\forall\,\s\ \in\ \bigl[\,0,\,\S\,\bigr]$ the vector $\ni\,(\s)$ points inside $\B\,(t)\,$. This property explains the major advantage of the vector $\ni\,(\s)$ over the vector of the principal normal $\n\,(\s)\,$, which can point inside or outside $\B\,(t)$ depending on the local convexity properties\footnote{Indeed, everything depends on the rotation direction of the tangent vector $\taub\,(\s)\,$, when we move along the curve $\Gamma^{\,\blackdiamond}\,$. Namely, if along a local portion of the curve $\Gamma^{\,\blackdiamond}_{\,\mathrm{loc}}\ \subseteq\ \Gamma^{\,\blackdiamond}\ =\ \partial\,\D^{\,\blackdiamond}$ the domain $\D^{\,\blackdiamond}$ is locally convex (\ie for every pair of points on $\Gamma^{\,\blackdiamond}_{\,\mathrm{loc}}\,$, the straight line segment joining this pair of points entirely belongs to the closure $\cl(\,\D^{\,\blackdiamond})$), then $\taub\,(\s)$ rotates in the counter-clockwise direction and the principal normal $\n\,(\s)$ inside the body $\B\,(t)\,$. On the other hand, if the curve portion $\Gamma^{\,\blackdiamond}_{\,\mathrm{loc}}$ is concave, then $\taub\,(\s)$ rotates in the clockwise direction and $\n\,(\s)$ points to the exterior of $\B\,(t)\,$.} of the domain $\D^{\,\blackdiamond}\,$. Equation~\eqref{eq:curva} can be re-written in terms of the interior normal (to the body $\B\,(t)$) $\ni$ function, if one introduces the signed curvature $\kappar\,$, which is positive on convex and negative on concave portions of $\Gamma^{\,\blackdiamond}\,$:
\begin{equation}\label{eq:93}
  \dot{\taub}\ =\ \kappar\,\ni\,.
\end{equation}

According to the impermeability Condition~\eqref{eq:sides} and taking into account the fact that the curve $\Gamma^{\,\blackdiamond}$ is stationary (\ie does not evolve in time), we have the following representation of the horizontal velocity vector $\u\,(\x,\,y,\,t)$ in the points of $\widetilde{\partial_{\,\parallel}\,\B\,(t)}\,$:
\begin{equation*}
  \u\,(\x,\,y,\,t)\ =\ u_{\,\tau}\,(\x,\,y,\,t)\,\taub\,(\x)\,, \qquad (\x,\,y)\ \in\ \widetilde{\partial_{\,\parallel}\,\B\,(t)}\,,
\end{equation*}
where $u_{\,\tau}$ is the tangential component of the velocity vector $\u\,(\x,\,y,\,t)\,$:
\begin{equation*}
  u_{\,\tau}\ \eqdef\ \u\scal\taub\,.
\end{equation*}
We assume that Equation~\eqref{eq:hormom} is verified up to the boundary $\widetilde{\partial_{\,\parallel}\,\B\,(t)}$ and we multiply\footnote{The multiplication here is understood in the sense of the standard scalar product in $\Ee^{\,2}\,$.} it by $\ni$ on the right:
\begin{equation*}
  \rho\,\u_{\,t}\scal\ni\ +\ \rho\,\bigl(\,(\u\scal\grad)\,\u\,\bigr)\scal\ni\ +\ \rho\,v\,\u_{\,y}\scal\ni\ +\ \pd{p}{\ni}\ =\ \zb\,, \qquad (\x,\,y)\ \in\ \widetilde{\partial_{\,\parallel}\,\B\,(t)}\,.
\end{equation*}
Now, we take into account the fact that vector $\ni\,(\cdot)$ does not depend on (independent) variables $y$ and $t\,$:
\begin{equation*}
  \rho\,\underbrace{(\u\scal\ni)_{\,t}}_{\equiv\ \zb}\ +\ \rho\,\bigl(\,(\u\scal\grad)\,\u\,\bigr)\scal\ni\ +\ \rho\,v\,\underbrace{(\u\scal\ni)_{\,y}}_{\equiv\ \zb}\ +\ \pd{p}{\ni}\ =\ \zb\,, \qquad (\x,\,y)\ \in\ \widetilde{\partial_{\,\parallel}\,\B\,(t)}\,.
\end{equation*}
Thanks to the impermeability Condition~\eqref{eq:sides} two terms vanish. Thus, the last equation simplifies to
\begin{equation}\label{eq:94}
  \rho\,u_{\,\tau}\,\bigl((\taub\scal\grad)\cdot(u_{\,\tau}\,\taub)\bigr)\scal\ni\ +\ \pd{p}{\ni}\ =\ \zb\,, \qquad (\x,\,y)\ \in\ \widetilde{\partial_{\,\parallel}\,\B\,(t)}\,.
\end{equation}
After some elementary computations and using the identity\footnote{Let us provide some details about these computations. For any two vector functions $\a,\,\b\,:\ \R^{\,2}\ \longrightarrow\ \R^{\,2}$ and a scalar function $\alpha\,:\ \R^{\,2}\ \longrightarrow\ \R$ we have the following vectorial identity:
\begin{equation*}
  (\a\scal\grad)\cdot(\alpha\,\b)\ =\ (\a\scal\grad\,\alpha)\cdot\b\ +\ \alpha\,(\a\scal\grad)\,\b\,.
\end{equation*}
Thus, by taking $\a\ =\ \b\ \equiv\ \taub$ and $\alpha\ \equiv\ u_{\,\tau}$ in our case we readily obtain:
\begin{equation*}
  (\taub\scal\grad)\cdot(\u_{\,\tau}\,\taub)\ =\ (\taub\scal\grad\,\u_{\,\tau})\cdot\taub\ +\ \u_{\,\tau}\,(\taub\scal\grad)\,\taub\,.
\end{equation*}
Now we have to transform the last term on the right-hand side of the last equation. The curve $\Gamma^{\,\blackdiamond}$ is parametrized as $\x\ =\ \r\,(\s)\ \cong\ \bigl(\r_{\,1},\,\r_{\,2}\bigr)\,$. Let us introduce also the components of the tangent vector $\taub\ \cong\ (\taub_{\,1},\,\taub_{\,2})\,$. Then, in every regular point on $\Gamma^{\,\blackdiamond}$ we have:
\begin{equation*}
  \dot{\taub}\ \cong\ \begin{pmatrix}[1.2]
    \pd{\taub_{\,1}}{\s} \\
    \pd{\taub_{\,2}}{\s}
  \end{pmatrix}\ =\ 
  \begin{pmatrix}[1.2]
    \pd{\taub_{\,1}}{x_{\,1}}\cdot\od{\r_{\,1}}{\s}\ +\ \pd{\taub_{\,1}}{x_{\,2}}\cdot\od{\r_{\,2}}{\s} \\
    \pd{\taub_{\,2}}{x_{\,1}}\cdot\od{\r_{\,1}}{\s}\ +\ \pd{\taub_{\,2}}{x_{\,2}}\cdot\od{\r_{\,2}}{\s}
  \end{pmatrix}\ =\ 
  \begin{pmatrix}[1.2]
    \pd{\taub_{\,1}}{x_{\,1}}\cdot\taub_{\,1}\ +\ \pd{\taub_{\,1}}{x_{\,2}}\cdot\taub_{\,2} \\
    \pd{\taub_{\,2}}{x_{\,1}}\cdot\taub_{\,1}\ +\ \pd{\taub_{\,2}}{x_{\,2}}\cdot\taub_{\,2}
  \end{pmatrix}\ \cong\ (\taub\scal\grad)\,\taub\,.
\end{equation*}
Hence, we finally obtain that
\begin{equation*}
  (\taub\scal\grad)\cdot(\u_{\,\tau}\,\taub)\ =\ (\taub\scal\grad\,\u_{\,\tau})\cdot\taub\ +\ \u_{\,\tau}\,\dot{\taub}\,.
\end{equation*}
This completes our computational comment.} $(\taub\scal\grad)\,\taub\ =\ \dot{\taub}$ we obtain:
\begin{equation*}
  (\taub\scal\grad)\cdot(u_{\,\tau}\,\taub)\ =\ (\taub\scal\grad\,u_{\,\tau})\,\taub\ +\ u_{\,\tau}\,\dot{\taub}\,.
\end{equation*}
Henceforth, Equation~\eqref{eq:94} takes the form:
\begin{equation*}
  \rho\,u_{\,\tau}\,(\taub\scal\grad\,u_{\,\tau})\cdot(\taub\scal\ni)\ +\ \rho\,u_{\,\tau}^{\,2}\,(\dot{\taub}\scal\ni)\ +\ \pd{p}{\ni}\ =\ \zb\,, \qquad (\x,\,y)\ \in\ \widetilde{\partial_{\,\parallel}\,\B\,(t)}\,.
\end{equation*}
Since $\taub\,(\cdot)\perp\;\ni\,(\cdot)$ and thanks to the modified \textsc{Frenet--Serret} Equation~\eqref{eq:93} we obtain that $\dot{\taub}\scal\ni\ =\ \kappar\,$. Thus, we finally obtain:
\begin{equation}\label{eq:95}
  \pd{p}{\ni}\ =\ -\,\rho\,u_{\,\tau}^{\,2}\,\kappar\,, \qquad (\x,\,y)\ \in\ \widetilde{\partial_{\,\parallel}\,\B\,(t)}\,.
\end{equation}

In order to make a passage from the pressure normal derivative $\pd{p}{\ni}$ to that of the free surface excursion $\pd{\eta}{\ni}$ on the boundary $\Gamma^{\,\blackdiamond}\,$, we shall assume that the free surface dynamic boundary Condition~\eqref{eq:fsdc2} in the exterior domain $\D^{\,\diamond}$ holds up to the boundary $\Gamma^{\,\blackdiamond}$ as well:
\begin{equation*}
  p\,(\x,\,\eta\,(\x,\,\cdot\,),\,\cdot\,)\ =\ \zb\,, \qquad \forall\,\x\ \in\ \Gamma^{\,\blackdiamond}\,.
\end{equation*}
By taking the horizontal gradient $\grad\,\eqref{eq:fsdc2}$ and taking the trace at the boundary $\Gamma^{\,\blackdiamond}\,$, we readily obtain that
\begin{equation*}
  \grad\,p\ +\ p_{\,y}\,\grad\,\eta\ =\ \zbv\,, \qquad \forall\,\x\ \in\ \Gamma^{\,\blackdiamond}\,, \qquad y\ =\ \eta\,(\x,\,t)\,.
\end{equation*}
After taking the scalar product with vector $\ni$ on the right, we obtain:
\begin{equation*}
  \pd{p}{\ni}\ +\ p_{\,y}\;\pd{\eta}{\ni}\ =\ \zb\,, \qquad \forall\,\x\ \in\ \Gamma^{\,\blackdiamond}\,, \qquad y\ =\ \eta\,(\x,\,t)\,.
\end{equation*}
Finally, by using the previously derived Identity~\eqref{eq:95} at the free surface $y\ =\ \eta\,(\x,\,t)\,$, we arrive to the desired expression for the normal derivative of the free surface $\eta\,$:
\begin{equation}\label{eq:96}
  p_{\,y}\;\pd{\eta}{\ni}\ =\ \rho\,u_{\,\tau}^{\,2}\,\kappar\,, \qquad \forall\,\x\ \in\ \Gamma^{\,\blackdiamond}\,, \qquad y\ =\ \eta\,(\x,\,t)\,.
\end{equation}
We underline the fact that the last result does not depend on the chosen orientation of the curve $\Gamma^{\,\blackdiamond}\,$.

\begin{remark}
If $\pd{p}{y}\bigr\vert_{\,y\,=\,\eta\,(\x,\,t)}\,(\x,\,t)\ \neq\ 0\,$, then Equation~\eqref{eq:96} can be solved with respect to $\pd{\eta}{\ni}\,$. We have an intuition\footnote{Indeed, our intuition seems to be supported by some scientific studies. For an idea fluid in the absence of wind and surface tension, the vertical balances the local inertia of the wave field. Namely, the vertical momentum balance Equation~\eqref{eq:vertmom} can be rewritten as:
\begin{equation*}
  \pd{p}{y}\ =\ -\,\rho\,(\,a_{\,\updownarrows}\ +\ g\,)\,,
\end{equation*}
where $a_{\,\updownarrows}$ is the \textsc{Lagrangian} vertical particle acceleration. From the last equation one notices that $\pd{p}{y}$ vanishes if and only if $a_{\,\updownarrows}\ =\ -\,g\,$. Thus, the question on the sign of $\pd{p}{y}$ can be answered by comparing $a_{\,\updownarrows}$ with $g\,$. Historically, the equality $a_{\,\updownarrows}\ =\ -\,g$ was proposed first as a dynamical criterion for the onset of breaking by \textsc{Phillips} (1958) \cite{Phillips1958}. Later, \textsc{Longuet-Higgins} (1985) \cite{Longuet-Higgins1985} proved that $\abs{a_{\,\updownarrows}}\ <\ \frac{g}{2}$ for the limiting steady \textsc{Stokes} wave. Recent wave tank experiments performed and analysed by \textsc{Shemer} and \textsc{Noskowitz} (2013) \cite{Shemer2013a} confirmed \textsc{Longuet-Higgins}'s theoretical result by showing that $a_{\,\updownarrows}\ +\ g$ does not vanish even for unsteady waves and at the onset of breaking. This is more than enough to conclude that one can express $\pd{\eta}{\ni}$ from Equation~\eqref{eq:96} (we would like to thank Dr.~Francesco~\textsc{Fedele} (\textsc{Georgia} Institute of Technology, USA) for clarifying this situation for us).} that for non-breaking waves $\pd{p}{y}\bigr\vert_{\,y\,=\,\eta\,(\x,\,t)}\ <\ \zb\,$. Thus, Equation~\eqref{eq:96} tells us that the free surface $\eta$ meets the solid boundary at the right angle only if $\kappar\ =\ \zb$ or $u_{\,\tau}\ =\ \zb\,$. The former corresponds to (locally) flat boundaries and the former to the fluid at rest state.
\end{remark}


\addcontentsline{toc}{subsubsection}{A lyrical digression}
\subsubsection*{A lyrical digression}

\begin{figure}
  \centering
  \includegraphics[width=0.59\textwidth]{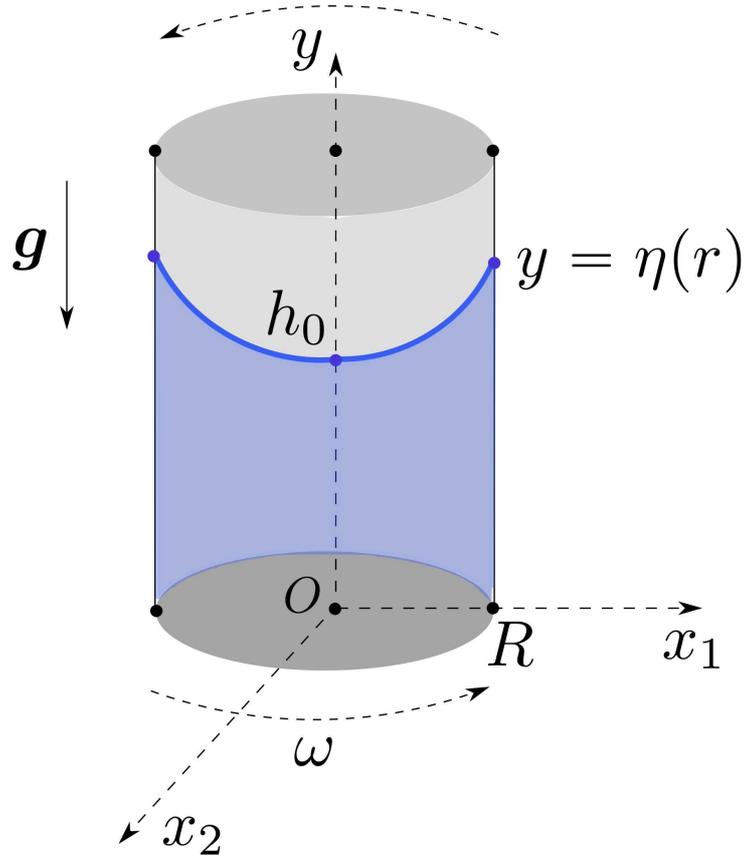}
  \caption{\small\em Sketch of the fluid domain in the rotating cylinder problem.}
  \label{fig:cyl}
\end{figure}

It is so widely believed that the free surface meets the solid boundary at the right angle, that we decided to provide here a simple analytical solution, which will allow us to check empirically the Relation~\eqref{eq:96}. Consider an ideal incompressible fluid (with constant density $\rho$) in a cylinder of radius $R$ rotating around its vertical axis $O\,y$ with constant angular velocity $\omega\,$. The cylinder is not entirely filled. So, the fluid is bounded above by the free surface and below by the solid cylinder base. The sketch of the fluid domain with rotating cylinder is shown in Figure~\ref{fig:cyl}. It is natural to introduce a cylindrical coordinate system $O\,r\,\vphi\,y$ to describe this situation. An axi-symmetric steady\footnote{The steady \textsc{Euler} equations in cylindrical coordinates read:
\begin{align*}
  \pd{u_{\,r}}{r}\ +\ \pd{u_{\,\vphi}}{\vphi}\ +\ \pd{v}{y}\ +\ \frac{u_{\,r}}{r}\ &=\ 0\,, \\
  u_{\,r}\;\pd{u_{\,r}}{r}\ +\ \frac{1}{r}\;u_{\,\vphi}\;\pd{u_{\,r}}{\vphi}\ +\ v\;\pd{u_{\,r}}{y}\ -\ \frac{u_{\,\vphi}^{\,2}}{r}\ +\ \frac{1}{\rho}\;\pd{p}{r}\ &=\ 0\,, \\
  u_{\,r}\;\pd{u_{\,\vphi}}{r}\ +\ \frac{1}{r}\;u_{\,\vphi}\;\pd{u_{\,\vphi}}{\vphi}\ +\ v\;\pd{u_{\,\vphi}}{y}\ +\ \frac{u_{\,r}\,u_{\,\vphi}}{r}\ +\ \frac{1}{\rho\,r}\;\pd{p}{\vphi}\ &=\ 0\,, \\
  u_{\,r}\;\pd{v}{r}\ +\ \frac{1}{r}\;u_{\,\vphi}\;\pd{v}{\vphi}\ +\ v\;\pd{v}{y}\ +\ \frac{1}{\rho}\;\pd{p}{y}\ &=\ -\,g\,,
\end{align*}
where $p$ is the pressure as above, $u_{\,r}$ is the radial component of the velocity, $u_{\,\vphi}$ is the rotational component and $v$ is the axial velocity.} analytical solution to the full \textsc{Euler} equations in this situation is well-known if one assumes the pressure to be hydrostatic \cite[Section~\textsection26]{Lamb1932}, \ie
\begin{equation*}
  p\,(r,\,y)\ =\ \rho\,g\,\bigl(\eta\,(r)\ -\ y\bigr)\,,
\end{equation*}
where $r\ \eqdef\ \abs{\x}\ \leq\ R$ is the radial coordinate. The free surface is then determined by the following formula:
\begin{equation}\label{eq:cex}
  \eta\,(r)\ =\ h_{\,0}\ +\ \frac{(\omega\,r)^{\,2}}{2\,g}\qquad \Longleftrightarrow\qquad \eta\,(\x)\ =\ h_{\,0}\ +\ \frac{\omega^{\,2}\,\abs{\x}^{\,2}}{2\,g}\,,
\end{equation}
where $h_{\,0}$ is the free surface height at $r\ =\ 0\,$. These two elements of the solution are already enough to discuss the contact of the free surface $\eta$ with cylinder boundary. However, to complete the solution we provide here also the velocity field in cylindrical
\begin{equation*}
  u_{\,r}\ =\ \zb\,, \qquad u_{\,\vphi}\,(r,\,\phi,\,y)\ =\ \omega\,r\,, \qquad v\ =\ \zb\,,
\end{equation*}
and \textsc{Cartesian} coordinates:
\begin{equation*}
  u_{\,1}\,(\x,\,y)\ =\ -\,\omega\,x_{\,2}\,, \qquad u_{\,2}\,(\x,\,y)\ =\ \omega\,x_{\,1}\,, \qquad v\ =\ \zb\,.
\end{equation*}

Since we have an explicit Expression~\eqref{eq:cex} for the free surface $\eta\,$, we can compute its normal derivative (in the outer direction $\n$) at the boundary:
\begin{equation*}
  \pd{\eta}{\n}\ \equiv\ \pd{\eta}{r}\ =\ \frac{\omega^{\,2}\,R}{g}\,, \qquad r\ =\ R\,.
\end{equation*}
We notice that the tangential component of the horizontal velocity field at the boundary $r\ =\ R$ is $u_{\,\tau}\ =\ \omega\,R\,$. Thus, the normal derivative of the free surface can be expressed in terms of $u_{\,\tau}$ as
\begin{equation}\label{eq:res}
  \pd{\eta}{\n}\ =\ \frac{u_{\,\tau}^{\,2}}{R\,g}\,.
\end{equation}
The last formula shows that the contact cannot happen at the right angle since in every point of the contact line $\pd{\eta}{\n}\ \neq\ \zb\,$. However, let us check it by comparing with the prediction of the previously derived general Formula~\eqref{eq:96}. Since the pressure is hydrostatic, we readily obtain $p_{\,y}\ =\ -\,\rho\,g\,$. The cylinder lateral boundary is the circle of radius $R\,$. Thus, its signed curvature is $\kappar\ =\ -\frac{\breve{\strut 1}}{R}\,$. By substituting all these elements into Equation~\eqref{eq:96} we obtain:
\begin{equation*}
  -\,\rho\,g\;\pd{\eta}{\n}\ =\ -\,\rho\;\frac{u_{\,\tau}^{\,2}}{R}\,.
\end{equation*}
It is not difficult to see after some simplifications that the last formula coincides with Equation~\eqref{eq:res}, which is an indirect confirmation of the boundary Condition~\eqref{eq:96}.

To conclude this digression we would like to stress one more time that the contact angle of the free surface with a solid boundary depends on the curvature of the vertical wall, the tangential fluid velocity in the point \emph{and} on the pressure gradient in the vertical direction.

\subsection{Potential flow model}
\label{sec:pot}

In this Section we perform the reduction from $\Eu$ to $\Po$. The main idea behind this reduction is to reduce the number of unknown functions by assuming that the velocity field derives from a \emph{potential} function $\phi:\ \Om\,(t)\,\times\,\R^{\,+}_{\,0}\ \longrightarrow\ \R$ in the following way:
\begin{equation}\label{eq:potdef}
  \u\ =\ \grad\,\phi\,, \qquad
  v\ =\ \phi_{\,y}\,, \qquad (\x,\,y)\ \in\ \Om\,(t)\,.
\end{equation}
In this way we obtain a model of incompressible and irrotational flows of an ideal fluid. Nevertheless, in water wave community the formulation presented in this Section is referred to as the \emph{full water wave problem} (see \eg \cite{Wu2011}). In mathematical terms one has to find a harmonic function $\phi$ in the fluid domain:
\begin{equation*}
  \grad^{\,2}\,\phi\ +\ \phi_{\,y\,y}\ =\ \zb\,, \qquad (\x,\,y)\ \in\ \Om\,(t)\,,
\end{equation*}
which satisfies certain \emph{nonlinear} boundary conditions, which can be readily obtained from those given in Section~\ref{sec:eul}. So, on the free surface we have to satisfy the kinematic and dynamic boundary conditions respectively:
\begin{equation*}
  \eta_{\,t}\ +\ \grad\,\phi\scal\grad\,\eta\ -\ \phi_{\,y}\ =\ \zb\,, \quad y\ =\ \eta\,(\x,\,t)\,, \quad \x\ \in\ \D^{\,\diamond}\,, \qquad \forall\, t\ \geq\ 0\,,
\end{equation*}
\begin{equation*}
  \phi_{\,t}\ +\ \half\;\abs{\grad\,\phi}^{\,2}\ +\ \half\;\phi_{\,y}^{\,2}\ +\ g\,\eta\ =\ \zb\,, \quad y\ =\ \eta\,(\x,\,t)\,, \quad \x\ \in\ \D^{\,\diamond}\,, \qquad \forall\, t\ \geq\ 0\,.
\end{equation*}
On the solid bottom of the wave tank we have the impermeability condition:
\begin{equation*}
  h_{\,t}\ +\ \grad\,\phi\scal\grad\,h\ +\ \phi_{\,y}\ =\ \zb\,, \quad y\ =\ -\,h\,(\x,\,t)\,, \quad \x\ \in\ \D^{\:\boxdot}\,, \qquad \forall\, t\ \geq\ 0\,.
\end{equation*}
A similar condition hold on the object bottom as well:
\begin{equation*}
  d_{\,t}\ +\ \grad\,\phi\scal\grad\,d\ -\ \phi_{\,y}\ =\ \zb\,, \quad y\ =\ d\,(\x,\,t)\,, \quad \x\ \in\ \D^{\,\blackdiamond}\,, \qquad \forall\, t\ \geq\ 0\,.
\end{equation*}
Finally, on lateral solid boundaries of the wave tank and of the immersed body, we have the following impermeability condition:
\begin{equation*}
  \pd{\phi}{\n}\ \eqdef\ \grad\,\phi\scal\n\ =\ \zb\,, \quad \x\ \in\ \Gamma^{\,\blackdiamond}\ \bigcup\ \Gamma^{\:\boxdot}\,, \qquad \forall\, t\ \geq\ 0\,,
\end{equation*}
\begin{align*}
  \x\ &\in\ \Gamma^{\,\blackdiamond}\,:\quad d\,(\x,\,t)\ \leq\ y\ \leq\ \eta\,(\x,\,t)\,, \\
  \x\ &\in\ \Gamma^{\:\boxdot}\,:\quad -\,h\,(\x,\,t)\ \leq\ y\ \leq\ \eta\,(\x,\,t)\,.
\end{align*}
To obtain a well-posed problem, we have to prescribe also the compatible initial conditions for the free surface elevation $\eta\,(\x,\,0)\,$, $\x\ \in\ \D^{\,\diamond}$ and the velocity potential\footnote{We notice that it is actually enough to specify the trace of the velocity potential on the free surface $\varphi\,(\x,\,0)\ \eqdef\ \phi\,\bigl(\x,\,\eta\,(\x,\,0),\,0\bigr)\,$, $\x\ \in\ \D^{\,\diamond}$ to obtain a well-posed problem, since the rest can be easily reconstructed by requiring that $\phi\,(\x,\,y,\,0)$ be a harmonic function.} $\phi\,(\x,\,y,\,0)\,$, $(\x,\,y)\ \in\ \Om\,(0)\,$.

The total energy density for the potential flow formulation $\Po$ is usually defined as:
\begin{equation*}
  \frac{\E}{\rho}\ =\ \frac{1}{2}\;\bigl(\,\abs{\grad\phi}^{\,2}\ +\ \phi_{\,y}^{\,2}\,\bigr)\ +\ g\,y\,.
\end{equation*}
The quantity $\iint_{\,\D}\,\int_{\,-\,h}^{\,\eta}\,\E\;\ud y\,\ud\x$ is conserved\footnote{Obviously, we can speak also about the energy conservation under the floating body $\B\,(t)\,$. However, to have the conservation of this quantity we have to require additionally the body to be fixed as well. Otherwise, it evolves in time according to the energy input/output due to the object $\B\,(t)$ motion.} (locally) in every sub-domain $\D\ \subseteq\ \D^{\,\diamond}$ if the fluid bottom is (globally) steady. The total energy conservation Equation~\eqref{eq:toteu} holds everywhere in the fluid domain $\Om\,(t)\,$. We only mention here that the depth-integrated total energy $\int_{\,-\,h}^{\,\eta}\,\E\;\ud y$ for potential flows $\Po$ plays a very important r\^ole of the \textsc{Hamiltonian} functional with the standard symplectic operator (here simply matrix) and corresponding canonical variables being $\varphi\,(\x,\,t)\ \eqdef\ \phi\,\bigl(\x,\,\eta\,(\x,\,t),\,t\bigr)$ and $\eta\,(\x,\,t)$ (see \eg \cite{Petrov1964, Zakharov1968, Broer1974}).

The fluid pressure in this formulation can be easily reconstructed in any point of the fluid by using the well-known \textsc{Cauchy--Lagrange} integral:
\begin{equation*}
  \frac{p}{\rho}\ =\ -\,\Bigl(\,\phi_{\,t}\ +\ \half\;\abs{\grad\,\phi}^{\,2}\ +\ \half\;\phi_{\,y}^{\,2}\ +\ g\,y\,\Bigr)\,, \qquad (\x,\,y)\ \in\ \Om\,(t)\,,\, \qquad t\ \geq\ 0\,.
\end{equation*}
Fluid particle velocities can be reconstructed by differentiating the velocity potential function according to \eqref{eq:potdef}. This completes the description of the potential model $\Po$.


\subsection{Fully nonlinear weakly dispersive models}
\label{sec:sgn}

The reduction $\Po\ \Longleftarrow\ \SGN$ for freely propagating waves over general bathymetries is well-understood nowadays. Consequently, we do not repeat the derivation here to focus on the specific issues related to the presence of a partially immersed body. For our approach to this derivation we refer to this recent review \cite{Khakimzyanov2016c} (and to \cite{Khakimzyanov2016a} for globally spherical geometries).

\begin{figure}
  \centering
  \includegraphics[width=0.99\textwidth]{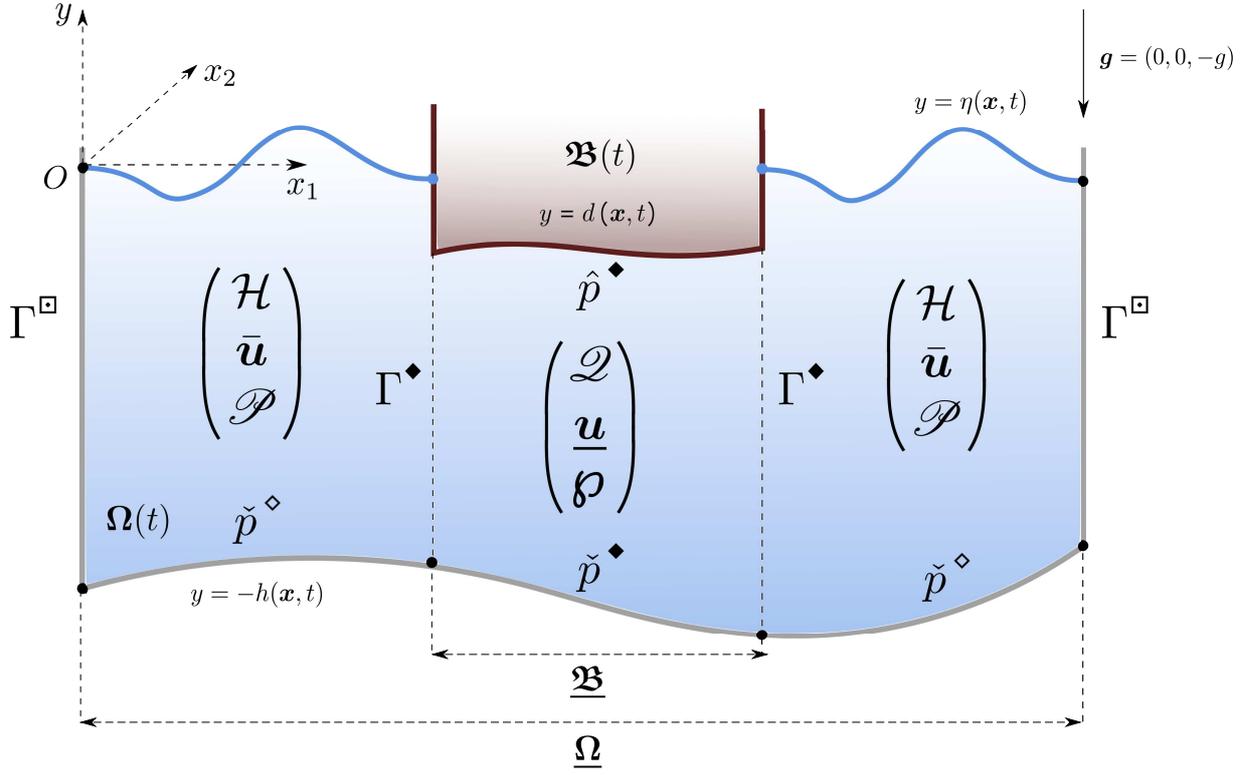}
  \caption{\small\em Side view of the computational domain along with the main variables used mainly in $\SGN$ and $\SW$ models.}
  \label{fig:side}
\end{figure}

Our task consists in specifying $\SGN$ equations in the outer domain $\D^{\,\diamond}\,$, under the object $\B\,(t)$ in the inner domain $\D^{\,\blackdiamond}$ and we have to specify how to glue these solutions along the boundary $\Gamma^{\,\blackdiamond}$ of two domains (see Figure~\ref{fig:sketch2} for an illustration). The peculiarity here is that under the object $\B\,(t)$ the fluid layer is bounded by two surfaces (impermeable boundaries) with prescribed motions. It turns out that this situation is actually simpler than the free surface regime in the outer domain $\D^{\,\diamond}\,$. Moreover, it is not difficult to see that solutions in two sub-domains influence each other and, thus, they are inter-dependent. It means that one has to construct both solutions simultaneously using gluing (or compatibility) conditions to be discussed below. A sketch of the side view of the computational domain is shown in Figure~\ref{fig:side}.

\subsubsection{Outer domain}

In our previous study \cite{Khakimzyanov2016c} we presented a unified derivation of $\SGN$ equations in the free surface regime. This model will be used in our study to model the fluid flow in the outer domain $\D^{\,\diamond}\,$. The $\SGN$ model equations are written in terms of the total water depth $\H$ and depth-averaged horizontal velocity vector:
\begin{equation}\label{eq:ave}
  \ub\,(\x,\,t)\ \eqdef\ \frac{1}{\H\,(\x,\,t)}\;\int_{\,-\,h\,(\x,\,t)}^{\,\eta\,(\x,\,t)}\,\u\,(\x,\,y,\,t)\;\ud y\,, \qquad \forall\,\x\ \in\ \D^{\,\diamond}\,, \qquad \forall\, t\ \geq\ 0\,.
\end{equation}
Thus, we have an induced map $\ub:\ \D^{\,\diamond}\,\times\,\R^{\,+}_{\,0}\ \longrightarrow\ \R^{\,2}\,$. The system of governing equations reads \cite{Khakimzyanov2016c}:
\begin{align}
  \H_{\,t}\ +\ \div(\H\,\ub)\ &=\ \zb\,,\label{eq:gn1} \\
  \ub_{\,t}\ +\ (\ub\cdot\grad)\,\ub\ +\ \frac{\grad\,\Pp}{\rho\,\H}\ &=\ \frac{\pc^{\,\diamond}\,\grad\,h}{\rho\,\H}\,.\label{eq:gn2}
\end{align}
The dynamics is completely defined by two functions $\H$ and $\ub\,$. To close the System \eqref{eq:gn1}, \eqref{eq:gn2}, we have to provide the expressions for the fluid pressure $\Pp$ and fluid pressure on the bottom $\pc^{\,\diamond}$ through variables $\H$ and $\ub\,$:
\begin{align}\label{eq:exprpp}
  \Pp\ &=\ \rho\,g\,\frac{\H^{\,2}}{2}\ -\ \rho\;\biggl(\frac{\H^{\,3}}{3}\;\Rrb_{\,1}\ +\ \frac{\H^{\,2}}{2}\;\Rrb_{\,2}\biggr)\,, \\
  \pc^{\,\diamond}\ &=\ \rho\,g\,\H\ -\ \rho\,\biggl(\frac{\H^{\,2}}{2}\;\Rrb_{\,1}\ +\ \H\,\Rrb_{\,2}\biggr)\,,\label{eq:exprpc}
\end{align}
where the terms $\Rrb_{\,1,\,2}\,:\ \D^{\,\diamond}\,\times\,\R^{\,+}_{\,0}\ \longrightarrow\ \R$ are defined as
\begin{equation}\label{eq:rr}
  \Rrb_{\,1}\ \eqdef\ \Ddb\,(\div\ub)\ -\ (\div\ub)^{\,2}\,, \qquad
  \Rrb_{\,2}\ \eqdef\ \Ddb^{\,2}\,h\,,
\end{equation}
$\Ddb$ being the total (or material) derivative operator:
\begin{equation*}
  \Ddb\,(\cdot)\ \eqdef\ (\cdot)_{\,t}\ +\ \ub\scal\grad\,(\cdot)\,, \qquad
  \Ddb^{\,n}\ =\ \Ddb^{\,n-1}\,\diamond\,\Ddb\,, \qquad n\ \geq\ 2\,.
\end{equation*}
System \eqref{eq:gn1}, \eqref{eq:gn2} was obtained in \cite{Khakimzyanov2016c} by assuming that the flow is weakly dispersive, \ie the dispersion parameter $\mu\ \ll\ 1\,$. We underline the fact that no assumption was made on the nonlinearity. Thus, the model is \emph{fully nonlinear}. We would like to mention also that the momentum balance Equation~\eqref{eq:gn2} can be rewritten in the following conservative form\footnote{Equation~\eqref{eq:mom} is a differential consequence of Equation~\eqref{eq:gn1}, \eqref{eq:gn2}. Namely, if one multiplies Equation~\eqref{eq:gn1} by $\ub$ and Equation~\eqref{eq:gn2} by $\H\,$, after some simple algebro-differential transformations, one arrives to \eqref{eq:mom}. This scheme can be summarized as:
\begin{equation*}
  \eqref{eq:gn1}\,\cdot\,\ub\ +\ \eqref{eq:gn2}\,\cdot\,\H\ \Longrightarrow\ \eqref{eq:mom}\,.
\end{equation*}}:
\begin{equation}\label{eq:mom}
  (\H\,\ub)_{\,t}\ +\ \div\Bigl(\,\H\,\ub\,\otimes\,\ub\ +\ \frac{\Pp}{\rho}\;\Id\,\Bigr)\ =\ \frac{\pc^{\,\diamond}\,\grad\,h}{\rho}\,,
\end{equation}
where $\Id\ \in\ \Mat_{\,2}\,(\R)$ is the identity matrix and operation $\otimes$ is the tensorial product, \ie if $\a,\,\b\ \in\ \R^{\,2}\,$, then
\begin{equation*}
  \a\,\otimes\,\b\ \defeq\ \A\,, \qquad \A_{\,i\,j}\ =\ \a_{\,i}\cdot\b_{\,j}\,, \qquad i\,,j\ \in\ \Fin{2}\,.
\end{equation*}
We underline the fact that Equation~\eqref{eq:mom} becomes a conservation law on even bottoms $\grad\,h\ \equiv\ \zbv\,$.

Even if the governing equations \eqref{eq:gn1}, \eqref{eq:gn2} are $(2+1)-$dimensional, its solutions $\H\,(\x,\,t)\,$, $\ub\,(\x,\,t)$ can be used to reconstruct approximatively the flow structure in $(3+1)-$dimensions. The pressure field in the fluid layer $-\,h\,(\x,\,t)\ \leq\ y\ \leq\ \eta\,(\x,\,t)\,$, $\x\ \in\ \D^{\,\diamond}$ can be reconstructed to the accuracy $\O\,(\mu^{\,4})$ using the following formula:
\begin{multline}\label{eq:press}
  p\,(\x,\,y,\,t)\ \eqass\ \rho\,\Bigl(\H\,(\x,\,t)\ -\ \bigl(y\ +\ h\,(\x,\,t)\bigr)\Bigr)\cdot\bigl(g\ -\ \Rrb_{\,2}\,(\x,\,t)\bigr)\\ 
  -\ \rho\;\biggl(\frac{\H^{\,2}\,(\x,\,t)}{2}\ -\ \frac{\bigl(y\ +\ h\,(\x,\,t)\bigr)^{\,2}}{2}\biggr)\;\Rrb_{\,1}\,(\x,\,t)\,, \quad \x\ \in\ \D^{\,\diamond}\,.
\end{multline}
We notice that expressions \eqref{eq:exprpp}, \eqref{eq:exprpc} can be easily derived from \eqref{eq:press} since
\begin{equation*}
  \Pp\,(\x,\,t)\ \equiv\ \int_{\,-\,h\,(\x,\,t)}^{\,\eta\,(\x,\,t)}\,p\,(\x,\,y,\,t)\;\ud y\,, \qquad
  \pc^{\,\diamond}\,(\x,\,t)\ \equiv\ p\,\bigl(\x,\,-\,h\,(\x,\,t),\,t\bigr)\,.
\end{equation*}
The horizontal components of the $3$D velocity field $\u\,(\x,\,y,\,t)$ are usually approximated to asymptotic order $\O\,(\mu^{\,2})\,$. However, for irrotational flows of the class $\Po$, one achieves the accuracy $\O\,(\mu^{\,4})$ \cite{Khakimzyanov2016c}:
\begin{multline}\label{eq:hor}
  \u\,(\x,\,y,\,t)\ \eqass\ \ub\,(\x,\,t)\ +\\
  \biggl(\frac{\H\,(\x,\,t)}{2}\ -\ \bigl(y\ +\ h\,(\x,\,t)\bigr)\biggr)\cdot\Bigl(\grad\,(\Ddb\,h\,(\x,\,t))\ +\ \grad\,h\,(\x,\,t)\,\bigl(\div\ub\,(\x,\,t)\bigr)\Bigr)\\
  +\ \biggl(\frac{\H^{\,2}\,(\x,\,t)}{6}\ -\ \frac{\bigl(y\ +\ h\,(\x,\,t)\bigr)^{\,2}}{2}\biggr)\;\grad\,(\div\ub\,(\x,\,t))\,, \qquad \x\ \in\ \D^{\,\diamond}\,.
\end{multline}
It is not difficult to check that the proposed asymptotic expression respects perfectly Definition \eqref{eq:ave}. The vertical velocity component $v\,(\x,\,y,\,t)$ can be reconstructed to accuracy $\O\,(\mu^{\,2})$ as follows:
\begin{equation}\label{eq:vert}
  v\,(\x,\,y,\,t)\ \eqass\ -\,\Ddb\,h\,(\x,\,t)\ -\ \bigl(y\ +\ h\,(\x,\,t)\bigr)\,(\div\ub)\,(\x,\,t)\,, \qquad \x\ \in\ \D^{\,\diamond}\,.
\end{equation}
Moreover, Equations \eqref{eq:gn1}, \eqref{eq:gn2} verify the following total energy (local) balance equation:
\begin{equation}\label{eq:en}
  (\H\,\E)_{\,t}\ +\ \div\Bigl(\bigl(\,\H\,\E\ +\ \Pp\,\bigr)\,\ub\Bigr)\ =\ -\,\pc^{\,\diamond}\,h_{\,t}\,, \qquad \x\ \in\ \D^{\,\diamond}\,.
\end{equation}
The derivation of the last equation can be found\footnote{We would like to underline the fact that Equation~\eqref{eq:en} can be derived at least by two different approaches. The first method consists in deriving the total energy conservation equations as a differential consequence of System~\eqref{eq:gn1}, \eqref{eq:gn2} similarly to the derivation given in Section~\ref{app:AE} for the inner domain $\D^{\,\blackdiamond}\,$. The other method consists in applying the depth-averaging operation to the total energy conservation Equation~\eqref{eq:toteu} in the full \textsc{Euler} model $\Eu$.}, for example, in \cite{Fedotova2014}. It is not difficult to see that on static bottoms, \ie $h\ =\ h\,(\x)\ \Longrightarrow\ h_{\,t}\ \equiv\ \zb\,$, Equation~\eqref{eq:en} becomes a conservation law. The total energy $\E$ is defined as \cite{Fedotova2014, Khakimzyanov2016c}
\begin{equation}\label{eq:defen}
  \frac{\E}{\rho}\ \eqdef\ \frac{\abs{\ub}^{\,2}}{2}\ +\ \frac{\H^{\,2}}{6}\;\bigl(\div\ub\bigr)^{\,2}\ +\ \frac{\H}{2}\;(\div\ub)\,\Ddb\,h\ +\ \frac{(\Ddb\,h)^{\,2}}{2}\ +\ g\;\frac{\H\ -\ 2\,h}{2}\,.
\end{equation}

Equations \eqref{eq:gn1}, \eqref{eq:gn2} have to be completed by appropriate initial and boundary conditions to obtain a well-posed problem. Since the wave tank is bounded by vertical impermeable walls, then for $\SGN$ model we impose
\begin{equation}\label{eq:bz}
  \ub\scal\n\ =\ \zb\,, \qquad \x\ \in\ \Gamma^{\:\boxdot}\,,
\end{equation}
where $\n$ is the unit exterior normal to $\Gamma^{\:\boxdot}\,$. However, some other types of boundary conditions might be imposed. Since the wave tank boundary $\Gamma^{\:\boxdot}$ has a polygonal shape, we can propose the following useful consequence of the Equation of motion \eqref{eq:gn2} and of non-permeability Condition~\eqref{eq:bz}:
\begin{equation*}
  \pd{\Pp}{\n}\ =\ \pc^{\,\diamond}\,\pd{h}{\n}\,, \qquad
  \x\ \in\ \Gamma^{\:\boxdot}\,,
\end{equation*}
where the pressure $\Pp$ is related to other dynamic variables $\H$ and $\ub$ through Equation~\eqref{eq:exprpp}. Moreover, on the exterior boundary we shall use the relation
\begin{equation}\label{eq:poly}
  \pd{\eta}{\n}\ =\ \zb\,, \qquad \x\ \in\ \Gamma^{\:\boxdot}\,,
\end{equation}
whose derivation is similar to the one given in Section~\ref{app:a}, taking into account the fact that the boundary $\Gamma^{\:\boxdot}$ consists of a polygon.

On the interior boundary $\Gamma^{\,\blackdiamond}$ we use an analogue of Condition~\eqref{eq:96} derived in Section~\ref{app:a} for the free surface excursion $\eta\,$, where we replace the pressure $p$ of the $3$D model by the reconstructed pressure \eqref{eq:press} and the horizontal velocity $\u$ is replaced by the depth-averaged one $\ub\,$:
\begin{equation*}
  p_{\,y}\;\pd{\eta}{\n}\ =\ \rho\,\bar{u}_{\,\tau}^{\,2}\,\kappar\,, \qquad \forall\,\x\ \in\ \Gamma^{\,\blackdiamond}\,,
\end{equation*}
where $\kappar$ is the signed curvature\footnote{For instance, if the boundary is flat as it is schematically depicted in Figures~\ref{fig:sketch} and \ref{fig:sketch2}, then $\kappa\ \equiv\ \zb\,$.} of the boundary $\Gamma^{\:\blackdiamond}\,$. The derivation of Equation~\eqref{eq:96} along with more precise definition of the signed curvature are given in Section~\ref{app:a}. In many practical problems the floating object can be approximated by a polygonal shape with straight lines as sides. In this case, the last condition simplifies to
\begin{equation*}
  \pd{\eta}{\n}\ =\ \zb\,, \qquad \x\ \in\ \Gamma^{\,\blackdiamond}\,.
\end{equation*}
However, there are also compatibility conditions to be satisfied on $\Gamma^{\,\blackdiamond}$ to glue solutions in inner $\D^{\,\blackdiamond}$ and outer $\D^{\,\diamond}$ domains. They will be discussed below in Section~\ref{sec:compa}.


\addcontentsline{toc}{subsubsection}{A lyrical digression}
\subsubsection*{A lyrical digression}

In this Section we would like to discuss the total energy Definition~\eqref{eq:defen} since it differs (in the potential part) from the classical definition, which can be found in \cite{Stoker1957}, to give an example. The same comments apply also to the total energy definition in the inner domain $\D^{\,\blackdiamond}$ that will be given below in the following Section. The energy density $\E$ is composed of the kinetic and potential energies densities of the flow:
\begin{equation*}
  \frac{\E}{\rho}\ \eqdef\ \underbrace{\frac{\abs{\ub}^{\,2}}{2}\ +\ \frac{\H^{\,2}}{6}\;\bigl(\div\ub\bigr)^{\,2}\ +\ \frac{\H}{2}\;(\div\ub)\,\Ddb\,h\ +\ \frac{(\Ddb\,h)^{\,2}}{2}}_{\displaystyle{\defeq\ \frac{\Kin}{\rho}}}\ +\ \underbrace{g\;\frac{\H\ -\ 2\,h}{2}}_{\displaystyle{\defeq\ \frac{\Pot}{\rho}}}\,.
\end{equation*}
It is desirable from physical and mathematical points of view to have a positive-definite energy definition. The positive-definiteness of the kinetic energy $\Kin$ can be demonstrated as follows. First, let us compute the contribution of the vertical velocity $v$ into the kinetic energy balance. To achieve this, we use the reconstruction Formula~\eqref{eq:vert} for $v\,$. A simple integration gives us:
\begin{multline*}
  \frac{1}{2\,\H}\;\int_{\,-h}^{\,\eta}\,v^{\,2}\,(\scal,\,y,\,\cdot)\,\ud y\ =\\ 
  \frac{1}{2\,\H}\;\int_{\,-h}^{\,\eta}\,\Bigl\{(\Ddb\,h)^{\,2}\ +\ 2\,(y\ +\ h)\,(\div\ub)\,\Ddb\,h\ +\ (y\ +\ h)^{\,2}\,(\div\ub)^{\,2}\Bigr\}\,\ud y\ =\\
  \frac{\H^{\,2}}{6}\;\bigl(\div\ub\bigr)^{\,2}\ +\ \frac{\H}{2}\;(\div\ub)\,\Ddb\,h\ +\ \frac{(\Ddb\,h)^{\,2}}{2}\,.
\end{multline*}
Hence, we can rewrite the kinetic energy density $\Kin$ in a way that its positive-definiteness becomes obvious:
\begin{equation*}
  \frac{\Kin}{\rho}\ =\ \frac{\abs{\ub}^{\,2}}{2}\ +\ \frac{1}{2\,\H}\;\int_{\,-h}^{\,\eta}\,v^{\,2}\,(\scal,\,y,\,\cdot)\,\ud y\ \geq\ 0\,.
\end{equation*}

Concerning the potential energy density $\Pot\ \bydef\ \rho\,g\;\frac{\H\ -\ 2\,h}{2}\,$, the situation is less unambiguous. In the current definition $\Pot$ is not positive definite since at the state of the rest $\Pot\ =\ \rho\,g\;\frac{\breve{0}\ +\ h\ -\ 2\,h}{2}\ =\ \rho\,g\;\frac{-\,h}{2}\ <\ \zb\,$. Everything depends on the choice of the zero potential energy level definition. To give an example, \textsc{Miles} \& \textsc{Salmon} (1985) define the potential energy with respect to the arbitrary level $y\ =\ y_{\,0}$ (see \cite[Equation~(3.4a), p.~523]{Miles1985}). A similar definition of the potential energy as in our work was adopted also in \cite{Bernetti2008}. However, this fact is not an obstacle to introduce a conserved and positive-definite energy together with the corresponding energy norm, at least for the case of the steady bottoms, \ie $h_{\,t}\ \equiv\ \zb\,$. For this, we consider the total energy:
\begin{equation*}
  \H\,\E\ =\ \rho\,\H\;\frac{\abs{\ub}^{\,2}}{2}\ +\ \frac{\rho}{2}\;\int_{\,-h}^{\,\eta}\,v^{\,2}\,(\scal,\,y,\,\cdot)\,\ud y\ +\ \rho\,g\,\frac{\eta^{\,2}}{2}\ -\ \rho\,g\,\frac{h^{\,2}}{2}\,.
\end{equation*}
The total energy satisfies the conservation Equation~\eqref{eq:en}. Let us integrate \eqref{eq:en} over any $2$D domain $\D\ \subseteq\ \R^{\,2}$ bounded by vertical walls:
\begin{equation}\label{eq:enint}
  \od{}{t}\;\iint_{\,\D}\,\H\,\E\;\ud\x\ +\ \oint_{\partial\,\D}\,\bigl(\,\H\,\E\ +\ \Pp\,\bigr)\,(\ub\scal\n)\;\ud\s\ =\ \zb\,,
\end{equation}
where $\ud\s$ is an element of the curve $\partial\,\D\,$. The contour integral $\oint_{\partial\,\D}\,(\cdot)\;\ud\s$ vanishes thanks to the wall boundary Condition~\eqref{eq:bz}. Moreover, due to the assumption that the bottom is steady, we have:
\begin{equation*}
  \od{}{t}\;\iint_{\,\D}\,\rho\,g\;\frac{h^{\,2}}{2}\;\ud\x\ =\ \zb\,.
\end{equation*}
Hence, Equation~\eqref{eq:enint} becomes:
\begin{equation*}
  \od{}{t}\;\iint_{\,\D}\,\Et\;\ud\x\ =\ \zb\,,
\end{equation*}
where
\begin{equation*}
  \Et\ \eqdef\ \rho\,\H\;\frac{\abs{\ub}^{\,2}}{2}\ +\ \frac{\rho}{2}\;\int_{\,-h}^{\,\eta}\,v^{\,2}\,(\scal,\,y,\,\cdot)\,\ud y\ +\ \rho\,g\,\frac{\eta^{\,2}}{2}\,.
\end{equation*}
The quantity $\Et$ is obviously positive definite, it is conserved and it can be used, in principle, to control the solution norm in order to show model well-posedness properties \cite{BCL}.

Consequently, in this Section we demonstrated that it is always possible to redefine the potential energy density $\Pot$ to have a positive-definite norm-like quantity $\Et\,$. The presence of the moving bottom effects ($h_{\,t}\ \neq\ \zb$) breaks the total energy conservation \eqref{eq:enint}. However, the evolution of the quantity $\iint_{\,\D}\,\Et\;\ud\x$ can still be followed by integrating the energy \emph{balance} Equation~\eqref{eq:en} over a bounded domain $\D$ (if the domain is not bounded by walls, we shall have some additional boundary terms in \eqref{eq:enint}).


\subsubsection{Inner domain}

By applying the same derivation technique as highlighted in \cite{Khakimzyanov2016c}, we can obtain the $\SGN$ model equations in the inner domain $\D^{\,\blackdiamond}\,$, \ie between the floating body $\B$ bottom $y\ =\ d\,(\x,\,t)$ and the solid bottom $y\ =\ -\,h\,(\x,\,t)\,$. The averaging procedure across this fluid layer bounded from below and above by two solid surfaces, we obtain in $\D^{\,\blackdiamond}\,$:
\begin{equation}\label{eq:gn1i}
  \Q_{\,t}\ +\ \div\bigl(\Q\,\ut\bigr)\ =\ \zb\,,
\end{equation}
\begin{equation}\label{eq:gn2i}
  \ut_{\,t}\ +\ (\ut\scal\grad)\,\ut\ +\ \frac{\grad\,\Pr}{\rho\,\Q}\ =\ \frac{\pc^{\,\blackdiamond}\,\grad\,h\ +\ \pt^{\,\blackdiamond}\,\grad\,d}{\rho\,\Q}\,,
\end{equation}
where $\Q\ \eqdef\ d\ +\ h$ is the function $\D^{\,\blackdiamond}\,\times\,\R^{\,+}_{\,0}\ \overset{\Q}{\longrightarrow}\ \R^{\,+}\,$, which returns the local fluid layer height confined between two solid surfaces. In this study we shall assume that the floating body $\B\,(t)$ does not touch the solid bottom, \ie
\begin{equation*}
  \Q\,(\x,\,t)\ =\ d\,(\x,\,t)\ +\ h\,(\x,\,t)\ \geq\ q_{\,0}\ >\ 0\,, \qquad
  \forall\,\x\ \in\ \D^{\,\blackdiamond}\,, \qquad \forall\, t\ \geq\ 0\,.
\end{equation*}
The functions $\pc^{\,\blackdiamond},\ \pt^{\,\blackdiamond}\,:\ \D^{\,\blackdiamond}\,\times\,\R^{\,+}_{\,0}\ \longrightarrow\ \R$ describe the distribution of fluid pressure along lower and upper solid boundaries respectively. The fluid pressure inside the fluid layer can be reconstructed by the following asymptotic formula (accurate to order $\O\,(\mu^{\,4})$):
\begin{multline}\label{eq:pr}
  p\,(\x,\,y,\,t)\ \eqass\ \pc^{\,\blackdiamond}\,(\x,\,t)\ -\ \rho\,\bigl(y\ +\ h\,(\x,\,t)\bigr)\,\bigl(g\ -\ \Rrt_{\,2}\,(\x,\,t)\bigr)\ +\\ \rho\;\frac{\bigl(y\ +\ h\,(\x,\,t)\bigr)^{\,2}}{2}\;\Rrt_{\,1}\,(\x,\,t)\,, \qquad \x\ \in\ \D^{\,\blackdiamond}\,.
\end{multline}
The accelerations $\Rrt_{\,1,\,2}$ are defined precisely in the same way as specified in Equation~\eqref{eq:rr}: one only has to make a substitution $\ub\ \leftsquigarrow\ \ut\,$. The \emph{column-integrated} pressure function $\Pr\,:\ \D^{\,\blackdiamond}\,\times\,\R^{\,+}_{\,0}\ \longrightarrow\ \R$ can be readily found from \eqref{eq:pr}:
\begin{equation}\label{eq:pr1}
  \Pr\ \eqdef\ \int_{\,-\,h}^{\,d}\,p\,\vert_{\,\D^{\,\blackdiamond}}\,(\cdot,\,y,\,\cdot)\;\ud y\ \equiv\ \pc^{\,\blackdiamond}\,\Q\ -\ \frac{\rho\,g}{2}\;\Q^{\,2}\ +\ \rho\;\biggl(\frac{\Q^{\,3}}{6}\;\Rrt_{\,1}\ +\ \frac{\Q^{\,2}}{2}\;\Rrt_{\,2}\biggr)\,.
\end{equation}
We proceed similarly to find the fluid pressure at the object $\B\,(t)$ bottom:
\begin{equation}\label{eq:pr2}
  \pt^{\,\blackdiamond}\ \eqdef\ p\,\vert_{\,\D^{\,\blackdiamond}}\,\bigl(\cdot,\,d,\,\cdot\bigr)\ \equiv\ \pc^{\,\blackdiamond}\ -\ \rho\,g\,\Q\ +\ \rho\;\biggl(\frac{\Q^{\,2}}{2}\;\Rrt_{\,1}\ +\ \Q\,\Rrt_{\,2}\biggr)\,.
\end{equation}
Under the $\Po$ model underlying assumptions, the horizontal components of the velocity field can be reconstructed in domain $\D^{\,\blackdiamond}$ to order $\O\,(\mu^{\,4})$ as
\begin{multline}\label{eq:40}
  \u\,\vert_{\,\D^{\,\blackdiamond}}\ \eqass\ \ut\ +\ \biggl(\frac{\Q}{2}\ -\ (y\ +\ h)\biggr)\cdot\bigl(\grad\,(\Ddt\,h)\ +\ \grad\,h\,(\div\ut)\bigr)\\ 
  +\ \biggl(\frac{\Q^{\,2}}{6}\ -\ \frac{(y\ +\ h)^{\,2}}{2}\biggr)\,\grad\,(\div\ut)\,,
\end{multline}
where $\Ddt$ is the total (material) derivative operator based on the advection velocity $\ut\,$, \ie
\begin{equation*}
  \Ddt\,(\cdot)\ \eqdef\ (\cdot)_{\,t}\ +\ \ut\scal\grad\,
  (\cdot)\,.
\end{equation*}
The vertical component $v$ is reconstructed using the same formula \eqref{eq:vert} by substituting $\ub\ \leftsquigarrow\ \ut\,$. It is not difficult to see that $\ut$ is the \emph{column-averaged} velocity, since
\begin{equation}\label{eq:41}
  \ut\ \equiv\ \frac{1}{\Q}\;\int_{\,-\,h}^{\,d}\,\u\,\vert_{\,\D^{\,\blackdiamond}}\,(\cdot,\,y,\,\cdot)\;\ud y\,.
\end{equation}
Equation~\eqref{eq:gn2i} can be recast in the conservative form:
\begin{equation}\label{eq:momi}
  (\Q\,\ut)_{\,t}\ +\ \div\Bigl(\,\Q\,\ut\,\otimes\,\ut\ +\ \frac{\Pr}{\rho}\;\Id\,\Bigr)\ =\ \frac{\pc^{\,\blackdiamond}\,\grad\,h\ +\ \pt^{\,\blackdiamond}\,\grad\,d}{\rho}\,.
\end{equation}
The last balance law can be easily derived by following this scheme:
\begin{equation*}
  \eqref{eq:gn1i}\,\cdot\,\ut\ +\ \eqref{eq:gn2i}\,\cdot\,\Q\ \Longrightarrow\ \eqref{eq:momi}\,.
\end{equation*}
Between two flat solid surfaces ($\grad\,h\ \equiv\ \zbv\,$, $\grad\,d\ \equiv\ \zbv$) Equation~\eqref{eq:momi} becomes a conservation law.

The structure of the $\SGN$ Model~\eqref{eq:gn1i}, \eqref{eq:gn2i} in the inner domain $\D^{\,\blackdiamond}$ is very similar to Equations~\eqref{eq:gn1}, \eqref{eq:gn2} in the outer domain $\D^{\,\diamond}\,$. However, there are some important differences as well. For instance, in the outer domain $\D^{\,\diamond}$ the unknown functions are $\H$ and $\ub\,$. In the inner domain $\D^{\,\blackdiamond}$ the counterpart of $\H$ is $\Q$ and $\Q\ =\ d\ +\ h$ is supposed to be known in our problem formulation. Hence, one can conclude that System~\eqref{eq:gn1i}, \eqref{eq:gn2i} is overdetermined. This would be a false impression since we have the pressure $\Pr$ to be determined. The last function will be known if we find $\pc^{\,\blackdiamond}$ \emph{or} $\pt^{\,\blackdiamond}$ in accordance with Equations~\eqref{eq:pr1} and \eqref{eq:pr2}. Thus, the number of unknowns coincides with the number of equations.

We can describe the balance of total energy for inner $\SGN$ model as follows:
\begin{equation}\label{eq:eni}
  (\Q\,\E)_{\,t}\ +\ \div\bigl((\,\Q\,\E\ +\ \Pr\,)\,\ut\,\bigr)\ =\ -\,\pc^{\,\blackdiamond}\,h_{\,t}\ -\ \pt^{\,\blackdiamond}\,d_{\,t}\,.
\end{equation}
The derivation of the last equation is provided in Section~\ref{app:AE}. The total energy $\E$ is defined as in Definition~\eqref{eq:defen} with only two differences: the total water depth $\H$ has to be replaced with the column height $\Q\ \rightsquigarrow\ \H$ and $\ut\ \rightsquigarrow\ \ub\,$. The structure of energy balance is very similar to Equation~\eqref{eq:en} in the outer domain. There is only one important difference: in the inner domain the energy can be also created by the motion of the floating object as well. This explains the presence of \textbf{two} source terms in the right-hand side of Equation~\eqref{eq:eni} comparing to \textbf{one} in \eqref{eq:en}. The boundary conditions for System~\eqref{eq:gn1i}, \eqref{eq:gn2i} are replaced by compatibility conditions along the boundary $\Gamma^{\,\blackdiamond}\,$. They will be discussed below.

\begin{remark}
From Equation~\eqref{eq:exprpc} it follows that for a quiescent fluid layer ($\ub\ =\ \zbv\,$, $h_{\,t}\ =\ \zb$ and $\eta_{\,t}\ =\ \zb$) the pressure at the bottom is equal to
\begin{equation}\label{eq:rep}
  \pc^{\,\diamond}\,(\x)\ =\ \rho\,g\,h\,(\x)\,, \qquad \x\ \in\ \D^{\,\diamond}
\end{equation}
in perfect agreement with hydrostatics. However, in the inner domain $\D^{\,\blackdiamond}$ the quantity $\pc^{\,\blackdiamond}$ is not completely determined, which prevents us from reconstructing completely the pressure $p$ of the $3$D flow. However, from the Equation of motion \eqref{eq:gn2i} and from formulas~\eqref{eq:pr1}, \eqref{eq:pr2}, it follows that in a quiescent fluid state in the inner domain $\D^{\,\blackdiamond}$ we necessarily have:
\begin{equation*}
  \grad\,\pc^{\,\blackdiamond}\ =\ \rho\,g\,\grad\,h\,.
\end{equation*}
Consequently, in the inner domain $\D^{\,\blackdiamond}$ we can seemingly use the same representation \eqref{eq:rep}:
\begin{equation*}
  \pc^{\,\blackdiamond}\,(\x)\ =\ \rho\,g\,h\,(\x)\,, \qquad \x\ \in\ \D^{\,\blackdiamond}\,.
\end{equation*}
Moreover, the last formula guarantees also the continuity of the bottom pressure $\pc^{\,\blackdiamond}$ on the common boundary $\Gamma^{\,\blackdiamond}\ =\ \cl\,\bigl(\,\D^{\,\diamond}\,\bigr)\ \bigcap\ \cl\,\bigl(\,\D^{\,\blackdiamond}\,\bigr)\,$.
\end{remark}


\subsubsection{Total energy balance in the inner domain}
\label{app:AE}

One of the central properties of physically sound models is that they admit several important \emph{differential consequences}, which are often referred to in applications as \emph{conservation laws}. In the case of an ideal homogeneous fluid, the \textsc{Euler} equations $\Eu$ possess the conservation laws for the mass, momenta ($\times\,3$) and energy. While deriving approximations, in particular in the shallow water regime relevant to this study, some of these properties might be destroyed. For instance, the energy conservation property is quite easy to loose (see an example of such derivation in \cite[Section~\textsection2.2]{Clamond2015c} and how to recover this property using variational methods in \cite[Section~\textsection3.2]{Clamond2015c}). However, we believe that a physically sound approximate model should possess the total energy conservation property. There are good physical reasons to require this property, but also from the mathematical point of view, the (coercive) energy conservation might be used to show well-posedness of approximate models \cite{BCL}. Finally, the energy conservation might be used to control the accuracy of numerical computations.

For the $\SGN$ model of freely propagating long waves (the outer region $\D^{\,\diamond}$ in terms of the present study), the balance of total energy was obtained in \cite{Fedotova2014} in the presence of moving bottom and without requiring the flow irrotationality. Another important requirement we impose on conservation laws is that they be consistent with the base model ($\Eu$ in our case). The \emph{consistency property} for conservation laws can be stated as follows: the conserved density of the approximate model can be obtained from the corresponding density of the base model by applying the same approximation, which was used to derive the simplified model. The consistency of Equation~\eqref{eq:en} was demonstrated in \cite{Fedotova2014}. Namely, the energy $\E$ of the $\SGN$ model can be obtained from the total energy of the full \textsc{Euler} equations by applying the depth-averaging and truncation operations to the full energy of the $3$D flow. To give a negative example, we can mention the classical \textsc{Boussinesq} equations \cite{Peregrine1967}. During long time it was believed that they do not admit any energy conservation law. However, recently a total energy conservation law for the classical \textsc{Boussinesq} system was found in \cite[Section~\textsection2.1]{Duran2011}. Unfortunately, the conserved energy is not consistent with the base model $\Eu$.

The goal of this Section is to derive the total energy balance Equation~\eqref{eq:eni} in the inner domain $\D^{\,\blackdiamond}\,$, which corresponds to the fluid layer contained between two solid moving surfaces corresponding to the wave tank and floating object bottoms. The motion of these surfaces is prescribed. A consistent energy balance law can be obtained by depth-averaging and simplifying the corresponding balance of the total energy to the full \textsc{Euler} equations, written in the domain \emph{under} the immersed body $\B\,(t)\,$. It is also possible to derive the same energy balance equation by computing a judicious differential consequence of the $\SGN$ model \eqref{eq:gn1i}, \eqref{eq:gn2i}. In this case two methods give the same\footnote{Normally, this claim has to be demonstrated. However, for the sake of manuscript compactness, we provide only one derivation and we beg the reader to trust us.} result. In this Section we detail the latter derivation only.

First of all, we multiply\footnote{The multiplication here is understood in the sense of the standard scalar product in $\Ee^{\,2}\,$.} Equation~\eqref{eq:gn2i} by the function $\ut$ and we use the following obvious identities:
\begin{equation*}
  \ut\scal(\,\ut\scal\grad)\,\ut\ \equiv\ \half\;\ut\scal\grad\,(\,\ut\scal\ut\,)\,, \qquad \ut\scal\grad\,h\ \equiv\ \Ddt\,h\ -\ h_{\,t}\,, \qquad \ut\scal\grad\,d\ \equiv\ \Ddt\,d\ -\ d_{\,t}\,.
\end{equation*}
As a result, after some simple computations we obtain the following differential consequence:
\begin{equation}\label{eq:92}
  \Ddt\,\Bigl(\rho\;\frac{\abs{\ut}^{\,2}}{2}\Bigr)\ +\ \frac{1}{\Q}\;\div(\Pr\,\ut)\ -\ \underbrace{\Bigl[\,\frac{\Pr}{\Q}\;\div\ut\ +\ \frac{\pc^{\,\blackdiamond}}{\Q}\;\Ddt\,h\ +\ \frac{\pt^{\,\blackdiamond}}{\Q}\;\Ddt\,d\,\Bigr]}_{\displaystyle{(\bigstar)}}\ =\ -\,\frac{\pc^{\,\blackdiamond}\,h_{\,t}\ +\ \pt^{\,\blackdiamond}\,d_{\,t}}{\Q}\,.
\end{equation}
Then, we transform the expression $(\bigstar)$ using Formulae~\eqref{eq:pr1}, \eqref{eq:pr2} along with identity $\Ddt\,d\ \equiv\ \Ddt\,\Q\ -\ \Ddt\,h\,$:
\begin{multline*}
  (\bigstar)\ =\ \biggl[\,\pc^{\,\blackdiamond}\ -\ \frac{1}{2}\;\rho\,g\,\Q\ +\ \rho\;\biggl(\frac{\Q^{\,2}}{6}\;\Rrt_{\,1}\ +\ \frac{\Q}{2}\;\Rrt_{\,2}\biggr)\,\biggr]\,\div\ut\ +\ \frac{\pc^{\,\blackdiamond}}{\Q}\;\,\Ddt\,h\ +\\
  \biggl[\,\frac{\pc^{\,\blackdiamond}}{\Q}\ -\ \rho\,g\ +\ \rho\;\biggl(\frac{\Q}{2}\;\Rrt_{\,1}\ +\ \Rrt_{\,2}\biggr)\,\biggr]\;(\Ddt\,\Q\ -\ \Ddt\,h)\,.
\end{multline*}
By using an equivalent form of the mass conservation Equation~\eqref{eq:gn1i}
\begin{equation}\label{eq:mass1}
  \div\ut\ =\ -\,\frac{\Ddt\,\Q}{\Q}\,,
\end{equation}
we obtain the following relations:
\begin{equation*}
  \pc^{\,\blackdiamond}\,\div\ut\ +\ \frac{\pc^{\,\blackdiamond}}{\Q}\;\Ddt\,h\ +\ \frac{\pc^{\,\blackdiamond}}{\Q}\;(\Ddt\,\Q\ -\ \Ddt\,h)\ =\ 0\,,
\end{equation*}
\begin{equation*}
  -\,\frac{1}{2}\;\rho\,g\,\Q\;\div\ut\ -\ \rho\,g\,(\Ddt\,\Q\ -\ \Ddt\,h)\ =\ -\,\rho\,g\,\Ddt\,\Bigl(\frac{\Q\ -\ 2\,h}{2}\Bigr)\,.
\end{equation*}
Consequently, we obtain:
\begin{multline*}
  (\bigstar)\ =\ -\,\rho\,g\,\Ddt\,\Bigl(\frac{\Q\ -\ 2\,h}{2}\Bigr)\ +\ \rho\;\biggl(\frac{\Q^{\,2}}{6}\;\Rrt_{\,1}\ +\ \frac{\Q}{2}\;\Rrt_{\,2}\biggr)\;\div\ut\\
  +\ \rho\;\biggl(\frac{\Q}{2}\;\Rrt_{\,1}\ +\ \Rrt_{\,2}\biggr)\;(\Ddt\,\Q\ -\ \Ddt\,h)\,.
\end{multline*}
By using now the definitions of $\Rrt_{\,1,\,2}$ along with the equivalent form of the mass conservation Equation~\eqref{eq:mass1}, we obtain three additional relations:
\begin{align*}
  \frac{\Q^{\,2}}{6}\;\Rrt_{\,1}\,\div\ut\ +\ \frac{\Q}{2}\;\Rrt_{\,1}\,\Ddt\,\Q\ &=\ -\,\Ddt\,\Bigl(\frac{\Q^{\,2}}{6}\,(\div\ut)^{\,2}\Bigr)\,, \\
  \frac{\Q}{2}\;\Rrt_{\,2}\,\div\ut\ +\ \Rrt_{\,2}\,\Ddt\,\Q\ -\ \frac{\Q}{2}\;\Rrt_{\,1}\,\Ddt\,h\ &=\ -\,\Ddt\,\Bigl(\frac{\Q}{2}\,(\div\ut)\,\Ddt\,h\Bigr)\,, \\
  \Rrt_{\,2}\,\Ddt\,h\ &=\ \Ddt\,\Bigl(\frac{(\Ddt\,h)^{\,2}}{2}\Bigr)\,.
\end{align*}
Using these relations, Equation~\eqref{eq:92} becomes:
\begin{multline}\label{eq:92bis}
  \rho\,\Ddt\;\biggl(\frac{\abs{\ut}^{\,2}}{2}\ +\ \frac{\Q^{\,2}}{6}\;(\div\ut)^{\,2}\ +\ \frac{\Q}{2}\;(\div\ut)\,\Ddt\,h\ +\ \frac{(\Ddt\,h)^{\,2}}{2}\ +\ g\;\frac{\Q\ -\ 2\,h}{2}\biggr)\\
  +\ \frac{\div(\Pr\,\ut)}{\Q}\ =\ -\,\frac{\pc^{\,\blackdiamond}\,h_{\,t}\ +\ \pt^{\,\blackdiamond}\,d_{\,t}}{\Q}\,.
\end{multline}
After introducing the total energy as
\begin{equation*}
  \frac{\E}{\rho}\ \eqdef\ \frac{\abs{\ut}^{\,2}}{2}\ +\ \frac{\Q^{\,2}}{6}\;\bigl(\div\ut\bigr)^{\,2}\ +\ \frac{\Q}{2}\;(\div\ut)\,\Ddt\,h\ +\ \frac{(\Ddt\,h)^{\,2}}{2}\ +\ g\;\frac{\Q\ -\ 2\,h}{2}\,,
\end{equation*}
Equation~\eqref{eq:92bis} can be rewritten in a more compact form:
\begin{equation*}
  \E_{\,t}\ +\ \ut\scal\grad\,\E\ +\ \frac{\div(\Pr\,\ut)}{\Q}\ =\ -\,\frac{\pc^{\,\blackdiamond}\,h_{\,t}\ +\ \pt^{\,\blackdiamond}\,d_{\,t}}{\Q}\,.
\end{equation*}
Finally, if we multiply the last equation by $\Q\,$, the mass conservation Equation~\eqref{eq:gn1i} by $\E$ and sum up the results, we obtain the total energy balance Equation~\eqref{eq:eni}. This equation describes the evolution of the total energy in the inner domain $\D^{\,\blackdiamond}\,$. The change in the total energy $\E$ is due to the motion of the immersed body $\B\,(t)$ and, eventually, of the wave tank bottom. If these two surfaces are fixed, then Equation~\eqref{eq:eni} becomes a conservation law.


\subsection{Potential vorticity in the inner domain}
\label{app:vort}

First of all, we would like to mention that even if the base model ($\Po$) is exactly irrotational, the shallow water equations $\SW$ provide a rotational approximation to an irrotational flow. Here we remind the Definition~\eqref{eq:vort} of the horizontal vorticity $\omega\,$:
\begin{equation*}
  \omega\ \bydef\ \pd{\under{u}_{\,2}}{x_{\,1}}\ -\ \pd{\under{u}_{\,1}}{x_{\,2}}\,,
\end{equation*}
where $\ut\ \cong\ (\under{u}_{\,1},\,\under{u}_{\,2})\,$. The notion of the \emph{potential vorticity} $\dfrac{\omega}{\H}$ is very important for the description of long wave motions in the ocean and in the atmosphere \cite{Rossby1940}. It is well-known that in dispersionless equations $\SW^{\,\diamond}$ the potential vorticity is preserved along the fluid particle trajectories \cite{Dolzhansky2013}:
\begin{equation*}
  \Ddb\,\Bigl(\,\frac{\omega}{\H}\,\Bigr)\ =\ \zb\,.
\end{equation*}
These results are known for the free propagation of long waves, corresponding to the outer domain $\D^{\,\diamond}$ in our study. That is why we used the total water depth variable $\H$ and, \emph{stricto sensu}, one has to use the corresponding horizontal depth-averaged velocity field $\ub$ in the Definition~\eqref{eq:vort} of $\omega\,$. The goal of this Section is to show the derivation of Equation~\eqref{eq:57} for $\SGN^{\,\blackdiamond}$ model in the inner domain $\D^{\,\blackdiamond}\,$, since it is used as one of the main building blocks of our numerical solver \cite{Khakimzyanov2018b}.


\subsubsection{Vector analysis on the plane}

To simplify our task of derivation, we shall use a restricted version of the vector analysis on a plane $\R^{\,2}\,$. Here we provide the main operators definitions and identities whose proofs are left to the reader.

Let $\a,\,\b\,:\ \R^{\,2}\ \longrightarrow\ \R^{\,2}$ be two smooth vector fields with components $\a\ \cong\ (a_{\,1},\,a_{\,2})\,$, $\b\ \cong\ (b_{\,1},\,b_{\,2})\,$. We introduce also two operators, which return \textbf{scalars} in contrast to the $3$D case:
\begin{align*}
  \rot\a\ &\eqdef\ \pd{a_{\,2}}{x_{\,1}}\ -\ \pd{a_{\,1}}{x_{\,2}}\,, \\
  \a\bwedge\b\ &\eqdef\ a_{\,1}\cdot b_{\,2}\ -\ a_{\,2}\cdot b_{\,1}\,.
\end{align*}
Using these operators, we can write the following simple identities:
\begin{equation*}
  \omega\ =\ \rot\ut\,, \qquad \omega_{\,t}\ =\ \rot\ut_{\,t}\,,
\end{equation*}
\begin{equation}\label{eq:96a}
  \rot\bigl((\ut\scal\grad)\,\ut\bigr)\ =\ \omega\,\div\ut\ +\ \ut\scal\grad\,\omega\ \stackrel{\text{\eqref{eq:mass1}}}{=}\ -\,\omega\;\frac{\Ddt\,\Q}{\Q}\ +\ \ut\scal\omega\,.
\end{equation}

Let us take also two smooth scalar fields defined on the plane $f,\,g\,:\ \R^{\,2}\ \longrightarrow\ \R\,$. Then, it is not difficult to check that the following identities hold:
\begin{equation}\label{eq:97}
  \rot\grad\,f\ =\ \zb\,, \qquad \grad\,f\bwedge\grad\,f\ =\ \zb\,, \qquad \rot(f\,\grad\,g)\ =\ \grad\,f\bwedge\grad\,g\,,
\end{equation}
\begin{equation}\label{eq:98}
  \grad\,f\bwedge\grad\,(\,\Ddt\,g\,)\ =\ \Ddt\,\bigl(\,\grad\,f\bwedge\grad\,g\,\bigr)\ -\ \grad\,(\,\Ddt\,f\,)\bwedge\grad\,g\ +\ (\grad\,f\bwedge\grad\,g)\,\div\ut\,,
\end{equation}
\begin{equation}\label{eq:99}
  \grad\,f\bwedge\grad\,(\,\Ddt\,g\,)\ =\ \Q\:\Ddt\,\Bigl(\;\frac{\grad\,f\bwedge\grad\,g}{\Q}\;\Bigr)\ -\ \grad\,(\,\Ddt\,f\,)\bwedge\grad\,g\,.
\end{equation}
Identity~\eqref{eq:99} follows directly from \eqref{eq:98} if one substitutes $\div\ut$ in accordance with Equation~\eqref{eq:mass1}.


\subsubsection{Derivation}

In the derivation of Equation~\eqref{eq:105} we start from the momentum balance Equation~\eqref{eq:gn2i} in the inner domain $\D^{\,\blackdiamond}\,$. First, we rewrite Equation~\eqref{eq:gn2i} by expressing pressure traces $\pc^{\,\blackdiamond}$ and $\pt^{\,\blackdiamond}$ through the depth-integrated pressure $\Pr$ according to Formulae~\eqref{eq:pr1} and \eqref{eq:pr2}:
\begin{multline*}
  \text{\eqref{eq:gn2i}}\ \rightsquigarrow\ \ut_{\,t}\ +\ (\ut\scal\grad)\,\ut\ +\ \frac{1}{\rho\,\Q}\;\biggl(\Q\,\grad\,\Bigl(\,\frac{\Pr}{\Q}\,\Bigr)\ +\ \frac{\Pr}{\Q}\;\grad\,\Q\biggr)\ = \\
  \frac{1}{\rho\,\Q}\;\biggl[\,\frac{\Pr}{\Q}\ +\ \rho\,g\,\frac{\Q}{2}\ -\ \rho\,\biggl(\frac{\Q^{\,2}}{6}\;\Rrt_{\,1}\ +\ \frac{\Q}{2}\;\Rrt_{\,2}\biggr)\,\biggr]\,\grad\,h\ +\ \frac{1}{\rho\,\Q}\;\biggl[\,\frac{\Pr}{\Q}\ -\ \rho\,g\,\frac{\Q}{2}\ +\\
  \rho\,\biggl(\frac{\Q^{\,2}}{3}\;\Rrt_{\,1}\ +\ \frac{\Q}{2}\;\Rrt_{\,2}\biggr)\,\biggr]\,\grad\,d\,.
\end{multline*}
By noticing that $\grad\,d\ =\ \grad\,(\Q\ -\ h)\,$, the last equation can be simplified as follows:
\begin{multline*}
  \ut_{\,t}\ +\ (\ut\scal\grad)\,\ut\ +\ \grad\,\biggl(\frac{\Pr}{\rho\,\Q}\biggr)\ =\\ 
  \frac{g}{2}\;\grad\,(h\ -\ d)\ -\ \biggl(\frac{\Q}{6}\;\Rrt_{\,1}\ +\ \frac{\Rrt_{\,2}}{2}\biggr)\,\grad\,h\ +\ \biggl(\frac{\Q}{3}\;\Rrt_{\,1}\ +\ \frac{\Rrt_{\,2}}{2}\biggr)\,\grad\,(\Q\ -\ h)\,.
\end{multline*}
If we apply the operator $\rot$ to the last equation and if we use identities~\eqref{eq:96a} and \eqref{eq:97}, we obtain:
\begin{multline*}
  \omega_{\,t}\ -\ \omega\;\frac{\Ddt\,\Q}{\Q}\ +\ \ut\scal\grad\,\omega\ =\\ 
  -\,\rot\biggl(\frac{\Q}{6}\;\Rrt_{\,1}\,\grad\,h\biggr)\ +\ \rot\biggl(\frac{\Q}{3}\;\Rrt_{\,1}\,\grad\,(\Q\ -\ h)\biggr)\ +\ \rot\biggl(\frac{\Q}{2}\;\grad\,(\Q\ -\ 2\,h)\biggr)\,,
\end{multline*}
By using again Equation~\eqref{eq:97}, we transform it into a more compact equation:
\begin{equation}\label{eq:100}
  \Q\,\Ddt\,\biggl(\frac{\omega}{\Q}\biggr)\ +\ \grad\,\biggl(\frac{\Q}{6}\;\Rrt_{\,1}\biggr)\bwedge\grad\,h\ -\ \grad\,\biggl(\frac{\Q}{3}\;\Rrt_{\,1}\biggr)\bwedge\grad\,(\Q\ -\ h)\ -\ \grad\,\biggl(\frac{\Rrt_{\,2}}{2}\biggr)\bwedge\grad\,(\Q\ -\ 2\,h)\ =\ \zb\,.
\end{equation}
To go further, we shall rewrite the acceleration $\Rrt_{\,1}\,$:
\begin{multline*}
  \Rrt_{\,1}\ \bydef\ \Ddt\,(\div\ut)\ -\ (\div\ut)^{\,2}\ \stackrel{\text{\eqref{eq:mass1}}}{=}\ \Ddt\,(\div\ut)\ +\ (\div\ut)\;\frac{\Ddt\,\Q}{\Q}\ =\\ 
  \frac{\Q\,\Ddt\,(\div\ut)\ +\ (\div\ut)\,\Ddt\,\Q}{\Q}\ =\ \frac{\Ddt\,(\Q\,\div\ut)}{\Q}\,.
\end{multline*}
Then, by using Identities~\eqref{eq:97} -- \eqref{eq:99}, we obtain new ones:
\begin{align*}
  \grad\,\biggl(\frac{\Q}{6}\;\Rrt_{\,1}\biggr)\bwedge\grad\,h\ &=\ \frac{\Q}{6}\;\Ddt\,\biggl(\frac{\grad\,(\Q\,\div\ut)\bwedge\grad\,h}{\Q}\biggr)\ +\ \frac{1}{6}\;\grad\,(\Ddt\,\Q)\bwedge\grad\,(\Ddt\,h)\,, \\
  \grad\,\biggl(\frac{\Q}{3}\;\Rrt_{\,1}\biggr)\bwedge\grad\,(\Q\ -\ h)\ &=\ \frac{\Q}{3}\;\Ddt\,\biggl(\frac{\grad\,(\Q\,\div\ut)\bwedge\grad\,(\Q\ -\ h)}{\Q}\biggr)\ +\\ 
  & \frac{1}{3}\;\grad\,(\Ddt\,\Q)\bwedge\grad\,(\Ddt\,\Q\ -\ \Ddt\,h)\,, \\
  \grad\,\biggl(\frac{\Rrt_{\,2}}{2}\biggr)\bwedge\grad\,(\Q\ -\ 2\,h)\ &=\ -\,\frac{\Q}{2}\;\Ddt\,\biggl(\frac{\grad\,(\Q\ -\ 2\,h)\bwedge\grad\,(\Ddt\,h)}{\Q}\biggr)\ +\\
  & \frac{1}{2}\;\grad\,(\Ddt\,\Q)\bwedge\grad\,(\Ddt\,h)\,.
\end{align*}
Consequently, Equation~\eqref{eq:100} can be recast in the following form:
\begin{equation*}
  \Ddt\,\biggl[\,\frac{\omega}{\Q}\ +\ \frac{\grad\,(\Q\,\div\ut)\bwedge\grad\,h}{6\,\Q}\ -\ \frac{\grad\,(\Q\,\div\ut)\bwedge\grad\,(\Q\ -\ h)}{3\,\Q}\ +\ \frac{\grad\,(\Q\ -\ 2\,h)\bwedge\grad\,(\Ddt\,h)}{2\,\Q}\,\biggr]\ =\ \zb\,,
\end{equation*}
or, after a simplification:
\begin{equation}\label{eq:102}
  \Ddt\,\biggl[\,\underbrace{\frac{\omega}{\Q}\ +\ \frac{\Q\,\grad\,\Q\bwedge\grad\,(\div\ut)}{3\,\Q}\ +\ \frac{\grad\,(\Q\,\div\ut)\bwedge\grad\,h}{2\,\Q}\ +\ \frac{\grad\,(\Q\ -\ 2\,h)\bwedge\grad\,(\Ddt\,h)}{2\,\Q}}_{\displaystyle{(\text{\ding{96}})}}\,\biggr]\ =\ \zb\,.
\end{equation}
From the last equation one can see that the $\SGN^{\,\blackdiamond}$ model does not transport $\dfrac{\omega}{\Q}\,$, but something more complex $(\text{\ding{96}})\,$, which is contained in square brackets in Equation~\eqref{eq:102}. By following the same steps and neglecting extra terms, one can easily derive the following result for the $\SW^{\,\blackdiamond}$ model:
\begin{equation}\label{eq:103}
  \Ddt\,\biggl(\frac{\omega}{\Q}\biggr)\ =\ \zb\,.
\end{equation}

Let us decrypt the physical sense of the quantity $(\text{\ding{96}})\,$. We shall do it in the case of a $3$D potential flow $\Po$, since in this case the horizontal components of the velocity $\u\,\vert_{\,\D^{\,\blackdiamond}}$ can be reconstructed to the \emph{optimal} asymptotic order $\O\,(\mu^{\,4})$ using Formula~\eqref{eq:40}. Let us compute the asymptotic representation for the third component $\omega_{\,3}$ of the vorticity vector:
\begin{multline*}
  \omega_{\,3}\ \bydef\ \rot\u\,\vert_{\,\D^{\,\blackdiamond}}\ \stackrel{\text{\eqref{eq:40}}}{\eqass}\ \omega\ +\ \grad\,\biggl(\frac{\Q}{2}\ -\ h\biggr)\bwedge\grad\,\Ddt\,h\ +\\ 
  \grad\,\biggl(\Bigl(\frac{\Q}{2}\ -\ y\ -\ h\Bigr)\,\div\ut\biggr)\bwedge\grad\,h\ +\ \biggl(\frac{\Q}{3}\;\grad\,\Q\ -\ (y\ +\ h)\,\grad\,h\biggr)\bwedge\grad\,(\div\ut)\,.
\end{multline*}
For two last terms on the right hand side, the following identities can be verified:
\begin{gather*}
  \grad\,\biggl(\Bigl(\frac{\Q}{2}\ -\ y\ -\ h\Bigr)\,\div\ut\biggr)\bwedge\grad\,h\ \equiv\ \grad\,\biggl(\frac{\Q\,\div\ut}{2}\biggr)\bwedge\grad\,h\\ 
  -\ (y\ +\ h)\,\grad\,(\div\ut)\bwedge\grad\,h\,, \\
  \biggl(\frac{\Q}{3}\;\grad\,\Q\ -\ (y\ +\ h)\,\grad\,h\biggr)\bwedge\grad\,(\div\ut)\ \equiv\ \frac{\Q}{3}\;\grad\,\Q\bwedge\grad\,(\div\ut)\ +\\ 
  (y\ +\ h)\,\grad\,(\div\ut)\bwedge\grad\,h\,.
\end{gather*}
Consequently, we have:
\begin{equation}\label{eq:104}
  \omega_{\,3}\ \eqass\ \omega\ +\ \underbrace{\frac{\Q\,\grad\,\Q\bwedge\grad\,(\div\ut)}{3}\ +\ \frac{\grad\,(\Q\,\div\ut)\bwedge\grad\,h}{2}\ +\ \frac{\grad\,(\Q\ -\ 2\,h)\bwedge\grad\,(\Ddt\,h)}{2}}_{\displaystyle{\ \defeq\ \wideparen{\delta\omega}}}\,.
\end{equation}
The last equation establishes an asymptotic connection between $\omega$ and $\omega_{\,3}\,$. Thus, the expression $(\text{\ding{96}})$ coincides with the quantity $\dfrac{\omega_{\,3}}{\Q}\,$. Thus, Equation~\eqref{eq:102} can be recast in the following compact form:
\begin{equation}\label{eq:105}
  \Ddt\,\biggl(\frac{\omega_{\,3}}{\Q}\biggr)\ =\ \zb\,,
\end{equation}
which elucidates the potential vorticity transport in $\SGN^{\,\blackdiamond}$ equations. The form of Equation~\eqref{eq:103} coincides identically with Equation~\eqref{eq:105}. However, the \emph{content} is different. Namely, the quantity $\omega_{\,3}$ approximates much better the horizontal vorticity of the $3$D flow, which vanishes in the limit when we approach by successive approximations the base model $\Po$.

Equation~\eqref{eq:105} can be re-written in a more \textsc{Eulerian} fashion:
\begin{equation*}
  \biggl(\frac{\omega_{\,3}}{\Q}\biggr)_{\,t}\ +\ \ut\scal\grad\,\biggl(\frac{\omega_{\,3}}{\Q}\biggr)\ =\ \zb\,.
\end{equation*}
By using the mass conservation Equation~\eqref{eq:gn1i}, we can recast it in the conservative form:
\begin{equation}\label{eq:106}
  (\omega_{\,3})_{\,t}\ +\ \div(\omega_{\,3}\,\ut)\ =\ \zb\,.
\end{equation}
The last form is more suitable for the discretization using conservative methods such as finite volumes \cite{Dutykh2011e, Dutykh2011a, Khakimzyanov2016}.

Equation~\eqref{eq:106} can be rewritten in terms of the variable $\omega$ using Relation~\eqref{eq:104}:
\begin{equation}\label{eq:107}
  \omega_{\,t}\ +\ \div(\omega\,\ut)\ =\ \digamma\,,
\end{equation}
with $\digamma\ \eqdef\ -\,\bigl\{\wideparen{\delta\omega}_{\,t}\ +\ \div(\wideparen{\delta\omega}\,\ut)\bigr\}\,$, where $\wideparen{\delta\omega}$ was defined above, but we repeat its definition here for the sake of reader convenience:
\begin{equation*}
  \wideparen{\delta\omega}\ \bydef\ \frac{\Q\,\grad\,\Q\bwedge\grad\,(\div\ut)}{3}\ +\ \frac{\grad\,(\Q\,\div\ut)\bwedge\grad\,h}{2}\ +\ \frac{\grad\,(\Q\ -\ 2\,h)\bwedge\grad\,(\Ddt\,h)}{2}\,.
\end{equation*}
In contrast with the appearance, the last expression contains only first order derivatives of the velocity $\ut\,$. Indeed, it can be simplified using Equation~\eqref{eq:mass1} as follows:
\begin{equation*}
  \wideparen{\delta\omega}\ \equiv\ \frac{1}{6}\;\Bigl\{\grad\,h\bwedge\grad\,\bigl(\Ddt\,(d\ -\ 2\,h)\bigr)\ +\ \grad\,d\bwedge\grad\bigl(\Ddt\,(h\ -\ 2\,d)\bigr)\Bigr\}\,.
\end{equation*}
From the last equation it is clear that in the case of even, but possibly moving, bottoms, \ie $h\,(\x,\,t)\ \equiv\ h\,(t)$ and $d\,(\x,\,t)\ \equiv\ d\,(t)\,$, the correction $\wideparen{\delta\omega}\ \equiv\ \zb$ vanishes. Thus, in this case, Equation~\eqref{eq:107} becomes particularly simple:
\begin{equation*}
  \omega_{\,t}\ +\ \div(\omega\,\ut)\ =\ \zb\,.
\end{equation*}
This completes our derivation and discussion of Equation~\eqref{eq:105}.


\subsection{Dispersionless equations}
\label{sec:sv}

In this Section we detail the $\SW$ model equations. They can be obtained in a straightforward manner from the $\SGN$ model by erasing acceleration terms $\Rrt_{\,1,\,2}\ \leftsquigarrow\ \zb$ in the inner domain $\D^{\,\blackdiamond}\,$, and $\Rrb_{\,1,\,2}\ \leftsquigarrow\ \zb$ in the outer domain $\D^{\,\diamond}\,$.


\subsubsection{Outer domain}

In the outer domain $\D^{\,\diamond}$ the Nonlinear Shallow Water (or \textsc{Saint}-\textsc{Venant}) $\SW$ equations have the standard form \eqref{eq:gn1}, \eqref{eq:gn2} with the following \emph{hydrostatic} expressions for the depth-integrated $\Pp$ and bottom $\pc^{\,\diamond}$ pressures:
\begin{equation}\label{eq:45}
  \Pp\ \eqdef\ \rho\,g\;\frac{\H^{\,2}}{2}\,, \qquad
  \pc^{\,\diamond}\ \eqdef\ \rho\,g\,\H\,.
\end{equation}
The last definitions are readily obtained from Equations~\eqref{eq:exprpp}, \eqref{eq:exprpc} by applying the transformation $\Rrb_{\,1,\,2}\ \leftsquigarrow\ \zb\,$. The $\SW$ model is usually written in the conservative form directly, since it is known to develop weak solutions in finite time \cite{Lax1973}:
\begin{align}
  \H_{\,t}\ +\ \div(\H\,\ub)\ &=\ \zb\,,\label{eq:sv1} \\
  (\H\,\ub)_{\,t}\ +\ \div\Bigl(\,\H\,\ub\,\otimes\,\ub\ +\ \frac{\Pp}{\rho}\;\Id\,\Bigr)\ &=\ \frac{\pc^{\,\diamond}\,\grad\,h}{\rho}\,.\label{eq:sv2}
\end{align}
This form turns out to be very beneficial for the computation of weak solutions \cite{Dutykh2007a, Dutykh2009a}. It is interesting to note that the energy balance Equation~\eqref{eq:en} is true for $\SW$ model \eqref{eq:sv1}, \eqref{eq:sv2} as well. However, the definition of energy $\E$ has to be simplified accordingly:
\begin{equation*}
  \frac{\E}{\rho}\ \eqdef\ \frac{\abs{\ub}^{\,2}}{2}\ +\ g\;\frac{\H\ -\ 2\,h}{2}\,.
\end{equation*}
For an alternative definition of the wave energy in view of applications to tsunami wave energy estimation we refer to \cite{Dutykh2009b}.

The initial conditions for $\SW$ model \eqref{eq:sv1}, \eqref{eq:sv2} are posed in the same way as for $\SGN$. The impermeability boundary Condition~\eqref{eq:bz} for $\SGN$ on $\Gamma^{\:\boxdot}$ along with Condition~\eqref{eq:poly} are transposed to $\SW$ as well. On the interior boundary $\Gamma^{\,\blackdiamond}$ the conditions on the normal derivative $\pd{\eta}{\n}$ remain precisely the same as for the $\SGN$ model regardless whether the boundary is curvilinear or polygonal. Moreover, on the boundary $\Gamma^{\,\blackdiamond}$ we have to impose also the compatibility (or gluing) conditions.


\subsubsection{Inner domain}

Under the immersed body $\B\,(t)$ the $\SW$ equations are not different in the form from $\SGN$ Equations~\eqref{eq:gn1i}, \eqref{eq:gn2i} written in $\D^{\,\blackdiamond}\,$. However, the pressure-related variables are greatly simplified due to the hydrostaticity assumption:
\begin{equation}\label{eq:45i}
  \pc^{\,\blackdiamond}\ =\ \frac{\Pr}{\Q}\ +\ \rho\,g\;\frac{\Q}{2}\,, \qquad
  \pt^{\,\blackdiamond}\ =\ \frac{\Pr}{\Q}\ -\ \rho\,g\;\frac{\Q}{2}\,.
\end{equation}
Thus, the $\SW$ model in the inner domain reads as:
\begin{equation}\label{eq:sv1i}
  \Q_{\,t}\ +\ \div\bigl(\Q\,\ut\bigr)\ =\ \zb\,,
\end{equation}
\begin{equation}\label{eq:sv2i}
  (\Q\,\ut)_{\,t}\ +\ \div\Bigl(\,\Q\,\ut\,\otimes\,\ut\ +\ \frac{\Pr}{\rho}\;\Id\,\Bigr)\ =\ \frac{\pc^{\,\blackdiamond}\,\grad\,h\ +\ \pt^{\,\blackdiamond}\,\grad\,d}{\rho}\,.
\end{equation}
The energy balance Equation~\eqref{eq:eni} holds as well in $\D^{\,\blackdiamond}\,$, provided that we redefine the energy $\E$ as
\begin{equation*}
  \frac{\E}{\rho}\ \eqdef\ \frac{\abs{\ut}^{\,2}}{2}\ +\ g\;\frac{\Q\ -\ 2\,h}{2}\,.
\end{equation*}
In the next Section, we discuss some mathematical properties of $\SW$ Equations~\eqref{eq:sv1}, \eqref{eq:sv2}.


\subsubsection{Analogy with gas dynamics}

Let us consider in this Section the \emph{isentropic} compressible \textsc{Euler} system of gas dynamics in $2$D. As its name suggests, it is obtained as a simplification of the full compressible \textsc{Euler} equations by assuming the entropy to be constant \cite{Batchelor1967}. The state of the gas is described by two functions $\rho\,:\ \R^{\,2}\,\times\,\R^{\,+}_{\,0}\ \longrightarrow\ \R^{\,+}$ and $\v\,:\ \R^{\,2}\,\times\,\R^{\,+}_{\,0}\ \longrightarrow\ \R^{\,2}\,$, which represent the local fluid density and the velocity field correspondingly. The system of isentropic \textsc{Euler} equations then reads:
\begin{equation}\label{eq:eul1}
  \rho_{\,t}\ +\ \div(\rho\,\v)\ =\ 0\,,
\end{equation}
\begin{equation}\label{eq:eul2}
  (\rho\,\v)_{\,t}\ +\ \div(\rho\,\v\,\otimes\,\v\ +\ p\,(\rho)\,\Id)\ =\ 0\,.
\end{equation}
The last two equations have to be supplemented by initial conditions in order to obtain a well-defined \textsc{Cauchy} problem:
\begin{equation*}
  \rho\,(\x,\,0)\ =\ \rho_{\,0}\,(\x)\,, \qquad \v\,(\x,\,0)\ =\ \v_{\,0}\,(\x)\,.
\end{equation*}
The pressure $p$ is a function of the density $\rho\,$, which is called the \emph{barotropic equation of state}, determined from constitutive thermodynamic relations. We assume that the function $p\,:\ \R^{\,+}\ \longrightarrow\ \R^{\,+}$ satisfies the condition
\begin{equation*}
  p^{\,\prime}\,(\rho)\ \eqdef\ \od{p}{\rho}\ >\ 0\,, \qquad \forall\,\rho\ >\ 0\,,
\end{equation*}
under which the isentropic \textsc{Euler} System \eqref{eq:eul1}, \eqref{eq:eul2} is hyperbolic \cite{Lax1973}. Usually, it is assumed that $p\ \propto\ \rho^{\,\gamma}$ with constant $\gamma\ \geq\ 1\,$. Obviously, this choice is not unique.

\begin{remark}
It is not difficult to see that the $\SW$ model belongs to the class of isentropic \textsc{Euler} Systems \eqref{eq:eul1}, \eqref{eq:eul2} if we replace $\H\ \leftsquigarrow\ \rho\,$, $\ub\ \leftsquigarrow\ \v$ and we take the quadratic pressure law with $\gamma\ =\ 2\,$ defined in the first half of Equation~\eqref{eq:45}. In order to remove the source term in the right-hand side of Equation~\eqref{eq:sv2}, it is enough to assume the bottom to be even, \ie $h\ =\ \const$.
\end{remark}

It is also well-known that hyperbolic systems of conservation laws develop discontinuities in finite time even if we start from smooth initial data \cite{Lax1973}. In water wave theory they are known as \emph{hydraulic jumps} (or undular bores\footnote{The presence of undulations behind the front requires the inclusion of slight dispersion effects.}) \cite{Stoker1957}. This phenomenon is known as the \emph{gradient catastrophe} or \emph{breakdown of classical solutions}. This obstacle was overcome by introducing the so-called \emph{weak} solutions. Unfortunately, weak solutions fail to be unique. The help comes from Physics, namely from the second law of Thermodynamics. One can stipulate that admissible solutions satisfy some additional \emph{entropy inequalities}. The quest for well-posedness theory of the \textsc{Cauchy} problem for hyperbolic conservation laws is more than one century old. Unfortunately, it was shown recently in \cite{Chiodaroli2015} that entropy conditions do not single out unique weak solutions in $2$D even under very strong assumptions on the initial data $(\rho_{\,0},\,\v_{\,0})\ \in\ W^{\,1,\,\infty}\,(\,\R^{\,2})$ (here $\abs{\infty}\ =\ \aleph_{\,0}$) \cite{Chiodaroli2016}:

\begin{theorem}
There are \textsc{Lipschitz} continuous initial data $(\rho_{\,0},\,\v_{\,0})$ for which there are infinitely many bounded admissible solutions $(\rho,\,\v)$ to System \eqref{eq:eul1}, \eqref{eq:eul2} on $\R^{\,2}\,\times\,\R^{\,+}_{\,0}$ with $\inf\,\rho\ >\ 0\,$. These solutions are locally \textsc{Lipschitz} on a finite interval on which they all coincide with the unique classical solution.
\end{theorem}

The solutions described in the last Theorem were called \emph{non-standard solutions} in \cite{Chiodaroli2017}. To the best of our knowledge, the question of more efficient selection criteria for weak solutions is still an open problem. In the light of the last result we can conclude that the $\SW$ model in $2$D is not well-posed in the sens of \textsc{Hadamard}, since the requirement of solution uniqueness is not fulfilled. However, the full water wave problem is well-posed (see \eg \cite{Wu2011}) Our recommendation is to use in practice the $\SGN$ model, which can be seen as a dispersive regularization of $\SW$ equations \cite{Bhat2007, Bhat2009, Clamond2018}. For the mathematical justification of the $\SGN$ model we refer to \cite{Makarenko1986, Li2006}.


\subsection{Modified weakly nonlinear weakly dispersive equations}

Above we presented two mathematical models for the fluid flow in the outer domain:
\begin{equation*}
  \SW\ \Longrightarrow\ \SGN\ \Longrightarrow\ \ldots\,,
\end{equation*}
where we neglected completely the dispersive effects by following the direction opposite to arrows. However, a more accurate description of the hierarchy of mathematical models, at least in the outer domain\footnote{We shall explain below why this hierarchy does not make sense in the inner domain $\D^{\,\blackdiamond}\,$.} $\D^{\,\diamond}\,$, would consist of the following scheme:
\begin{equation}\label{eq:scheme}
  \SW^{\,\diamond}\ \Longrightarrow\ \Bouss^{\,\diamond}\ \Longrightarrow\ \SGN^{\,\diamond}\ \Longrightarrow\ \ldots\,,
\end{equation}
where the arrows indicate the increasing level of complexity, but also of model completeness. Without considering \textsc{Boussinesq}-type equations, our work would not be complete since these equations are being used in practice for studying long wave propagation across the oceans \cite{Tkalich2007, Lovholt2008, Lovholt2010}. They occupy the intermediate place among $\SW$ and $\SGN$ models as indicated in \eqref{eq:scheme}. It is believed that $\Bouss$ models, such as well-known classical \textsc{Peregrine} \cite{Peregrine1967} or \textsc{Nwogu} \cite{Nwogu1993} systems, might be used for tsunami propagation far from the coasts, where the flow cannot be considered to be dominated by nonlinear effects solely as in the coastal areas \cite{Tkalich2007}. However, some simplifications is achieved with respect to $\SGN$ model since the nonlinear effects in the dispersive terms are neglected \cite{Shokin2015}. Currently \textsc{Boussinesq}-type models are also used to simulate the wave run-up on the coasts \cite{Roeber2010, Dutykh2011e, Duran2011}.

Weakly nonlinear models can be obtained using a variational principle \cite{Clamond2009, Fedotova1996}. However, \emph{currently} a more popular approach seems to be based on double-asymptotic expansions in two small parameters $\eps$ and $\mu^{\,2}\,$, where \cite{Khakimzyanov2016c, DMII}
\begin{description}
  \item[$\eps\ $] nonlinearity parameter defined as the ratio between the characteristic wave amplitude to the typical water depth;
  \item[$\mu^{\,2}\ $] the shallowness parameter equal to the squared ratio between a typical water depth to the characteristic wavelength.
\end{description}
The so-called \textsc{Boussinesq} regime\footnote{Sometimes the \textsc{Boussinesq} regime is equivalently characterized through the \textsc{Stokes}--\textsc{Ursell} number $\St$ \cite{Ursell1953}, which has to be of order one, \ie
\begin{equation*}
  \St\ \eqdef\ \frac{\eps}{\mu^{\,2}}\ =\ \O\,(1)\,.
\end{equation*}} assumes that
\begin{equation}\label{eq:bouss}
  \eps\ =\ \O\,(\mu^{\,2})
\end{equation}
and the simplification is achieved by neglecting all terms of the order $o\,(\mu^{\,2})\,$. By applying this simplification to dispersive terms $\dfrac{\grad\,\Pp}{\rho\,\H}\,$, $\dfrac{\pc^{\,\diamond}\,\grad\,h}{\rho\,\H}$ in Equation~\eqref{eq:gn2}, we can obtain the corresponding \textsc{Boussinesq}-type system. However, if we proceed in this naive and direct way, we shall loose some important physical properties of the $\SGN$ model. In particular, the \textsc{Galilean} invariance\footnote{We, together with many other authors, consider this property of capital importance for any physically sound model in classical (\ie non-relativistic, non-quantum) physics \cite{Souriau1997}.} can be lost as long as some conservation laws such as the total energy conservation. To give an example, the classical \textsc{Peregrine} system \cite{Peregrine1967} is not \textsc{Galilean} invariant. However, this situation can be corrected \emph{a posteriori} by adding some peculiar higher order terms (in the asymptotic sense) to recover this lost property. The examples of such invariantization process can be found in \cite{Duran2013}, where an invariant \textsc{Peregrine} system can be found as well.

In this Section we demonstrate how a weakly nonlinear model can be derived in multiple spatial dimensions and over uneven bottoms so that the \textsc{Galilean} invariance and energy conservation properties be preserved. To achieve these goals, the standard simplification technique has to be modified accordingly.

Let us rewrite the pressure-related terms \eqref{eq:exprpp}, \eqref{eq:exprpc} in the following completely equivalent way, where we just separate the hydrostatic and non-hydrostatic parts and we introduce new variables:
\begin{align*}
  \Pp\ &=\ \rho\,g\,\frac{\H^{\,2}}{2}\ -\ \rho\,\H\,Q\,, \\
  \pc^{\,\diamond}\ &=\ \rho\,g\,\H\ -\ \rho\,\H\,\qc\,,
\end{align*}
where $Q$ and $\qc$ read
\begin{align*}
  Q\ &=\ \frac{\H}{3}\;\Ddb\,(\H\,\div\ub)\ +\ \frac{\H}{2}\;\Ddb^{\,2}\,h\,, \\
  \qc\ &=\ \frac{1}{2}\;\Ddb\,(\H\,\div\ub)\ +\ \Ddb^{\,2}\,h\,.
\end{align*}
We substituted also the expressions\footnote{It is more convenient to use an alternative form of the expression \eqref{eq:rr} for $\Rrb_{\,1}\,$. As a second ingredient, it is based on an equivalent form of the mass conservation Equation~\eqref{eq:gn1}:
\begin{equation*}
  \div\ub\ =\ -\,\frac{\Ddb\,\H}{\H}\,.
\end{equation*}
Hence, $\Rrb_{\,1}$ can be rewritten as
\begin{equation*}
  \Rrb_{\,1}\ =\ \frac{\Ddb\,(\H\,\div\ub)}{\H}\,.
\end{equation*}} \eqref{eq:rr} for $\Rrb_{\,1,\,2}$ into the definitions of $Q$ and $\qc\,$. In contrast to the standard simplification approach, we apply the \textsc{Boussinesq} approximation \eqref{eq:bouss} only in $Q$ and $\qc\,$:
\begin{align*}
  Q\ &\leftsquigarrow\ \frac{h}{3}\;\Ddb\,(h\,\div\ub)\ +\ \frac{h}{2}\;\Ddb^{\,2}\,h\,, \\
  \qc\ &\leftsquigarrow\ \frac{1}{2}\;\Ddb\,(h\,\div\ub)\ +\ \Ddb^{\,2}\,h\,.
\end{align*}
Thus, to summarize, the pressure terms in the $\Bouss$ model should be taken as:
\begin{align}\label{eq:prppB}
  \Pp\ &=\ \rho\,g\,\frac{\H^{\,2}}{2}\ -\ \rho\,\H\,\biggl[\,\frac{h}{3}\;\Ddb\,(h\,\div\ub)\ +\ \frac{h}{2}\;\Ddb^{\,2}\,h\,\biggr]\,, \\
  \pc^{\,\diamond}\ &=\ \rho\,g\,\H\ -\ \rho\,\H\,\biggl[\,\frac{1}{2}\;\Ddb\,(h\,\div\ub)\ +\ \Ddb^{\,2}\,h\,\biggr]\,.\label{eq:prpcB}
\end{align}
It is remarkable that the general shape of governing Equations~\eqref{eq:gn1}, \eqref{eq:gn2} in $\Bouss$ and $\SGN$ models coincides. The difference consists in the content of the pressure terms $\Pp$ and $\pc\,$, which are defined in \eqref{eq:prppB}, \eqref{eq:prpcB}. We would like to underline the fact that $\Bouss$ model proposed in this study (and also earlier in \cite{Fedotova2014}) is different from the standard weakly nonlinear \textsc{Boussinesq}-type equations. Namely, the dispersive terms of $\Bouss$ model are linear in $\H$ (as the standard ones do as well), but nonlinear in $\ub$ (in contrast to standard equations). This is needed to keep important structural properties of the $\SGN$ model such as the \textsc{Galilean} invariance\footnote{The \textsc{Galilean} invariance of the $\Bouss$ models follows from the observation that expressions \eqref{eq:prppB} and \eqref{eq:prpcB} contain only complete (or material) derivatives of the velocity variable $\ub\,$, which is an invariant quantity in contrast to the standard partial derivative $\ub_{\,t}\,$.} and energy conservation. These advantages are strong enough, in our humble opinion, to keep some degree of nonlinearity in dispersive terms. For this reason, we cannot call \emph{stricto sensu} the $\Bouss$ model to be \emph{weakly nonlinear}. However, we shall continue to do it to distinguish from fully nonlinear and weakly dispersive model $\SGN$. Another advantage of the $\Bouss$ model is that it inherits also the same conservative formulation given in Equation~\eqref{eq:mom} (with the only difference that the pressure terms have to be taken from Definitions~\eqref{eq:prppB} and \eqref{eq:prpcB}). This allows to transpose all standard numerical technology based on the finite volume method already developed for $\SW$ equations (see \eg \cite{Barakhnin1999, Dutykh2009a, Dutykh2011e, Dutykh2011}).


\subsubsection{Energy conservation in weakly nonlinear weakly dispersive equations}

The reduced $\Bouss$ model has another major advantage --- it conserves the total energy, in contrast to most other standard \textsc{Boussinesq}-type equations. The energy conservation Equation~\eqref{eq:en} shape is preserved as well. However, the expression of the total energy $\E$ has to be modified accordingly:
\begin{equation*}
  \frac{\E}{\rho}\ \eqdef\ \frac{\abs{\ub}^{\,2}}{2}\ +\ \frac{h^{\,2}}{6}\;\bigl(\div\ub\bigr)^{\,2}\ +\ \frac{h}{2}\;(\div\ub)\,\Ddb\,h\ +\ \frac{(\Ddb\,h)^{\,2}}{2}\ +\ g\;\frac{\H\ -\ 2\,h}{2}\,.
\end{equation*}
The idea behind the derivation of the energy conservation equation in the $\Bouss$ model is slightly different from the $\SGN$ derivation sketched in Section~\ref{app:AE}. Consequently, we sketch briefly here the derivation.

First of all, we multiply\footnote{The multiplication is understood here in the sense of the standard scalar product in the \textsc{Euclidean} space $\Ee^{\,2}\,$.} Equation~\eqref{eq:gn2} by $\ub$ and, by taking into account the identity $\ub\scal(\ub\scal\grad)\,\ub\ \equiv\ \ub\scal\grad\,\Bigl(\,\dfrac{\abs{\ub}^{\,2}}{2}\,\Bigr)\,$, we obtain:
\begin{equation*}
  \Ddb\,\Bigl(\,\dfrac{\abs{\ub}^{\,2}}{2}\,\Bigr)\ +\ \frac{1}{\rho\,\H}\;\div(\Pp\,\ub)\ \underbrace{-\ \frac{\Pp}{\rho\,\H}\;\div\ub\ -\ \frac{\pc^{\,\diamond}}{\rho\,\H}\;\Ddb\,h}_{\displaystyle{(\,\text{\textleaf}\,)}}\ =\ -\,\frac{\pc^{\,\diamond}}{\rho\,\H}\;h_{\,t}\,.
\end{equation*}
The last two terms $(\,\text{\textleaf}\,)$ in the left hand side can be greatly simplified by taking into account Definitions~\eqref{eq:prppB}, \eqref{eq:prpcB}:
\begin{multline*}
  (\,\text{\textleaf}\,)\ =\ g\,\Ddb\,\biggl(\,\frac{\H\ -\ 2\,h}{2}\,\biggr)\ +\ \Ddb\,\biggl(\,\frac{h^{\,2}}{6}\;\bigl(\div\ub\bigr)^{\,2}\,\biggr)\ +\\ 
  \Ddb\,\biggl(\,\frac{h}{2}\;(\div\ub)\,\Ddb\,h\,\biggr)\ +\ \Ddb\,\biggl(\,\frac{(\Ddb\,h)^{\,2}}{2}\,\biggr)\ =\ \Ddb\,\biggl(\,\frac{\E}{\rho}\ -\ \frac{\abs{\ub}^{\,2}}{2}\,\biggr)\,.
\end{multline*}
Thus, for the $\Bouss$ model the energy balance can be written as:
\begin{equation*}
  \E_{\,t}\ +\ \ub\scal\grad\E\ +\ \frac{\div(\Pp\,\ub)}{\H}\ =\ -\,\frac{\pc^{\,\diamond}}{\H}\;h_{\,t}\,.
\end{equation*}
As the final step of the derivation, it is not difficult to transform the last equation into the conservative form given in \eqref{eq:en}.

\begin{remark}
In this Section we proposed a weakly nonlinear and weakly dispersive model $\Bouss$ in the outer domain $\D^{\,\diamond}\,$. All along the simplifications we exploited the \textsc{Boussinesq} assumption that the wave amplitude is a small parameter of the same order of smallness as the frequency dispersion parameter (see Assumption~\eqref{eq:bouss}). In the inner domain the analogue of $\H$ is the function $\Q\,$, which measures the local fluid layer height between the solid (ocean) bottom and the solid body $\B\,(t)\,$. This function $\Q$ is prescribed (\ie makes part of problem data). Thus, we cannot really make any assumptions on its magnitude. Henceforth, in our study the $\Bouss$ model is not considered in the inner domain $\D^{\,\blackdiamond}\,$.
\end{remark}


\subsection{On the structure of $\SGN$ equations in the inner domain}
\label{sec:struct}

In order to construct efficient solvers (and even to analyze mathematically) the modular structure of the governing equations should be understood. This point of view is exploited extensively in various operator-splitting-type approaches, see \eg \cite{Kovenya1981, Barakhnin1999}. We claim that this regard on mathematical models turns out to be useful even outside the broad framework of splitting methods \cite{Holden2010}. This philosophy was successfully applied recently to the numerical simulation of nonlinear dispersive wave propagation on globally flat \cite{Khakimzyanov2016c, Khakimzyanov2016} and globally spherical \cite{Khakimzyanov2016b, Khakimzyanov2016a} geometries. Namely, instead of tackling the dispersive equations directly, at every time step we solve two simpler sub-problems: the hyperbolic system of balance laws and one scalar elliptic equation. In the present work we follow the same philosophy in the outer domain $\D^{\,\diamond}\,$. In this Section we make a similar discussion of the $\SGN$ system properties in the inner domain $\D^{\,\blackdiamond}\,$. This information will be used later in the construction of the numerical algorithm \cite{Khakimzyanov2018b}.

As we already mentioned, the $\SGN$ model \eqref{eq:gn1i}, \eqref{eq:gn2i} in the inner domain $\D^{\,\blackdiamond}$ is different from the $\SGN$ model \eqref{eq:gn1}, \eqref{eq:gn2} in the outer domain $\D^{\,\diamond}$ in the following respect:
\smallskip
\begin{quote}
  \textit{In the outer domain $\D^{\,\diamond}$ the unknown functions are $\H$ and $\ub\,$, while in the inner domain $\D^{\,\blackdiamond}$ we seek to determine $\Pr$ and $\ut\,$.}
\end{quote}
\smallskip
This apparently little change implies important differences in properties of two systems that we can denote by $\SGN^{\,\diamond}$ and $\SGN^{\,\blackdiamond}$ correspondingly to underline this difference. To give an example, Equation~\eqref{eq:gn1i} is not of evolutionary type, since it does not contain the derivatives of unknown functions $\Pr$ and $\ut$ with respect to time $t\,$. Moreover, Equation~\eqref{eq:gn1i} is \emph{linear} in contrast to Equation~\eqref{eq:gn1}, which is its counterpart in the $\SGN^{\,\diamond}$ realm. However, the most important difference is that the momentum balance Equation~\eqref{eq:gn2i} does not contain mixed derivatives of the velocity $\ut$ of the \textbf{third order}, as it is the case in Equation~\eqref{eq:gn2}. To see it, we can rewrite Equation~\eqref{eq:gn2i} in the following way:
\begin{equation*}
  \ut_{\,t}\ +\ (\ut\scal\grad)\,\ut\ +\ \frac{1}{\rho\,\Q}\;\grad\,\Pr\ =\ \frac{\Pr}{\rho\,\Q^{\,2}}\;\grad\,\Q\ -\ \frac{\Q\,\Rrt_{\,1}}{6}\;\grad\,(h\ -\ 2\,d)\ +\ \frac{g\ -\ \Rrt_{\,2}}{2}\;\grad\,(h\ -\ d)\,,
\end{equation*}
where we expressed the pressure traces $\pc^{\,\blackdiamond}$ and $\pt^{\,\blackdiamond}$ through $\Pr\,$. Now, by Definitions~\eqref{eq:pr1} and of quantities $\Rrt_{\,1,\,2}$ it can be easily seen that the last equation contains second order mixed spatial derivative of $\ut$ at most.

However, there are even more differences. Let us consider a very simple situation where the solid bottom $h\ =\ \breve{h}_{\,0}$ and the object bottom $d\ =\ \breve{d}_{\,0}$ are even and fixed. In this limiting situation the $\SGN^{\,\blackdiamond}$ system becomes:
\begin{align*}
  \div\ut\ &=\ \zb\,, \\
  \ut_{\,t}\ +\ (\ut\scal\grad)\,\ut\ +\ \frac{1}{\rho\,(h_{\,0}\ +\ d_{\,0})}\;\grad\,\Pr\ &=\ \zb\,.
\end{align*}
One can easily recognize the system of $2$D incompressible \textsc{Euler} equations governing the flow of an ideal fluid \cite{Majda2001}, which has a mixed elliptic-hyperbolic type.

All these facts indicate that en efficient modular view on $\SGN^{\,\blackdiamond}$ should be different from $\SGN^{\,\diamond}$ \cite{Khakimzyanov2018b}. Our approach consists in introducing a new variable $\omega\,:\ \D^{\,\blackdiamond}\,\times\,\R^{\,+}_{\,0}\ \longrightarrow\ \R\,$, which plays the r\^ole of the \emph{horizontal vorticity} in incompressible flows \cite{Majda2001}:
\begin{equation}\label{eq:vort}
  \omega\ \eqdef\ \pd{\under{u}_{\,2}}{x_{\,1}}\ -\ \pd{\under{u}_{\,1}}{x_{\,2}}\,,
\end{equation}
where $\ut\ \cong\ (\under{u}_{\,1},\,\under{u}_{\,2})\,$. The main advantage of introducing the vorticity function $\omega$ is that it allows to eliminate pressure\footnote{If one needs to reconstruct the depth-integrated pressure $\Pr\,$, it can be achieving using a marching pressure scheme \cite{Richards1980}.} $\Pr$ from the governing Equation~\eqref{eq:gn2i}. Thus, we change the set of unknown functions $(\Pr,\,\ut)\ \leftsquigarrow\ (\omega,\,\ut)\,$. As a result, we have a natural splitting of our problem. To determine the velocity function $\ut\,$, we have a system of non-homogeneous \textsc{Cauchy}--\textsc{Riemann}-type System~\eqref{eq:gn1i}, \eqref{eq:vort}. The evolution of vorticity $\omega$ is given by the following hyperbolic equation derived in Section~\ref{app:vort}:
\begin{equation}\label{eq:57}
  \omega_{\,t}\ +\ \div(\omega\,\ut)\ =\ \digamma\,.
\end{equation}
These two simple problems are solved at each time step of our numerical algorithm. The advantage of the proposed splitting becomes particularly clear in a special case when the fluid layer height under the body $\B\,(t)$ is steady, \ie $\Q_{\,t}\ =\ \zb\,$. Then, the mass conservation Equation~\eqref{eq:gn1i}$\ \rightsquigarrow\ \div(\Q\,\ut)\ =\ \zb\,$. Taking into account Equation~\eqref{eq:vort}, one can derive the following elliptic \textsc{Poisson} equation to determine the \emph{stream function} $\psi\,:\ \D^{\,\blackdiamond}\,\times\,\R^{\,2}_{\,0}\ \longrightarrow\ \R$ \cite{Serrin1959}:
\begin{equation*}
  \biggl(\frac{\psi_{\,x_{\,1}}}{\Q}\biggr)_{\,x_{\,1}}\ +\ \biggl(\frac{\psi_{\,x_{\,2}}}{\Q}\biggr)_{\,x_{\,2}}\ =\ -\,\omega\,.
\end{equation*}
The velocity field $\ut$ can be easily reconstructed from the stream function $\psi\,$:
\begin{equation*}
  \Q\,\under{u}_{\,1}\ =\ \psi_{\,x_{\,2}}\,, \qquad \Q\,\under{u}_{\,2}\ =\ -\,\psi_{\,x_{\,1}}\,.
\end{equation*}
Consequently, in the case $\Q_{\,t}\ =\ \zb$ the splitting approach yields a scalar elliptic equation instead of a non-homogeneous \textsc{Cauchy}--\textsc{Riemann}-type system to determine the stream function $\psi\,$. The computational technology to solve these \emph{standard} PDEs is well understood nowadays \cite{Fedorenko1994}. On the other hand, the introduction of new variables requires the re-formulation of boundary conditions. We shall explain this peculiar point in the second Part of our study \cite{Khakimzyanov2018b}.


\section{Compatibility conditions}
\label{sec:compa}

On the common boundary $\Gamma^{\,\blackdiamond}\ \bydef\ \cl\bigl(\,\D^{\,\blackdiamond}\bigr)\ \bigcap\ \cl\bigl(\,\D^{\,\diamond}\bigr)$ of inner $\D^{\,\blackdiamond}$ and outer $\D^{\,\diamond}$ domains the fluid layer heights $\H$ and $\Q\,$, velocities $\ub$ and $\ut\,$, and depth-integrated pressures $\Pp$ and $\Pr$ are, in general, discontinuous. However, the jumps of these quantities across $\Gamma^{\,\blackdiamond}$ cannot assume arbitrary values. They are subject to some additional physical constraints, which are called \emph{compatibility} (or transmission) conditions. They play in a certain sense the r\^ole of celebrated \textsc{Rankine}--\textsc{Hugoniot} conditions in gas dynamics \cite{Rankine1870, Hugoniot1887}. Below, we shall obtain these conditions in $\SGN$ and $\SW$ models. We would like to underline the fact that $\Eu$ and $\Po$ models do not require any compatibility conditions since they treat the fluid domain $\Om\,(t)$ in its entirety.

As a general guiding principle, we quote here a recent study devoted to similar problems in gas dynamics \cite{Coquel2016}:
\smallskip
\begin{quote}
  [\;\dots\,] \textit{The method of interface coupling allows us to represent the evolution of such flows, where different models are separated by fixed interfaces. First, coupling conditions are specified at the interface to exchange information between the systems. The definition of transmission conditions generally results from physical consideration, \eg, the conservation or the continuity of given variables. Then, the transmission conditions are represented at the discrete level. The study of interface coupling for nonlinear hyperbolic systems has received attention for several years.} [\;\dots\,]

  \textit{From previous studies, coupling conditions [\;\dots\,] can be classified in three categories: flux coupling, state coupling and coupling with measure source term. The flux coupling method is a conservative approach which ensures the continuity of the physical flux through the coupling interface. Conversely, the state coupling method is a nonconservative approach which imposes (at least weakly) the continuity property of either the (conservative) variables or a nonlinear transformation of them (say, primitive variables). Finally, the coupling condition can be modelled thanks to a bounded vector-valued Dirac measure concentrated at the coupling interface. The coupling condition is then prescribed from the definition of the mass of the measure.} [\;\dots\,]
\end{quote}
\smallskip
In this study we consider both the state and flux couplings.


\subsection{Fully nonlinear weakly dispersive $\SGN$ case}

As we already mentioned, the unknowns in the $\SGN$ model are the depth-averaged velocities $\ub\,$, $\ut\,$; the total water depth $\H$ in $\D^{\,\diamond}$ and the depth-integrated pressure $\Pr$ in $\D^{\,\blackdiamond}\,$. Consequently, we need to derive three relations in order to be able to reconstruct these three quantities on one side of $\Gamma^{\,\blackdiamond}$ by having their values on the other side. In this Section we shall obtain only two such relations, which will be sufficient for the numerical implementation of our algorithm. In order to obtain the $3$\up{rd} relation, one has to make additional assumptions on the $3$D flow behaviour in the vicinity of $\partial_{\,\parallel}\,\B\,(t)\ \eqdef\ \Gamma^{\,\blackdiamond}\,\times\,\bigl[\,d\,\vert_{\,\Gamma^{\,\blackdiamond}}\,(\x,\,t),\,+\,\infty\,\bigr]\ \subseteq\ \R^{\,3}$ (here $\abs{\infty}\ =\ \aleph_{\,1}$). For a more general treatment in $3$D space we refer to Appendix~\ref{app:comp}.


\subsubsection{Pressure condition}
\label{sec:press}

First we derive the compatibility condition for the pressure variables. We can assume that the pressure in the $3$D flow is at least continuous everywhere in the fluid domain, \ie $p\ \in\ C^{\,0}\,\bigl(\Om\,(t)\bigr)\,$. In particular, it is continuous on the vertical surface $\partial_{\,\parallel}\,\Om\,(t)\ \eqdef\ \Gamma^{\,\blackdiamond}\,\times\,\bigl[\,-\,h\,\vert_{\,\Gamma^{\,\blackdiamond}}\,(\x,\,t),\,d\,\vert_{\,\Gamma^{\,\blackdiamond}}\,(\x,\,t)\,\bigr]\ \subset\ \cl\,\bigl(\,\Om\,(t)\,\bigr)\,$. Thus, we can write it mathematically:
\begin{equation}\label{eq:lim}
  \lim_{\substack{\x\ \to\ \x_{\,\parallel} \\ \x\ \in\ \D^{\,\diamond}}} p\,(\x,\,y,\,\cdot)\ =\ \lim_{\substack{\x\ \to\ \x_{\,\parallel} \\ \x\ \in\ \D^{\,\blackdiamond}}} p\,(\x,\,y,\,\cdot)\,, \qquad (\x_{\,\parallel},\,y)\ \in\ \partial_{\,\parallel}\,\Om\,(t)\,.
\end{equation}
We underline the fact that the above equality has to hold $\forall\,y\ \in\ \bigl[\,-\,h\,\vert_{\,\Gamma^{\,\blackdiamond}}\,(\x,\,t),\,d\,\vert_{\,\Gamma^{\,\blackdiamond}}\,(\x,\,t)\,\bigr]\,$. By substituting in continuity Condition~\eqref{eq:lim} the asymptotic representations \eqref{eq:press}, \eqref{eq:pr} of the $3$D pressure field $p$ on both sides of the boundary $\Gamma^{\,\blackdiamond}\,$, we obtain that \eqref{eq:lim} is equivalent to the following three conditions $\forall\,\x\ \in\ \Gamma^{\,\blackdiamond}$ and $\forall\,t\ \in\ \R^{\,+}_{\,0}\,$:
\begin{gather*}
  \pc^{\,\blackdiamond}\ =\ \rho\,g\,\H\ -\ \rho\,\biggl(\frac{\H^{\,2}}{2}\;\Rrb_{\,1}\ +\ \H\,\Rrb_{\,2}\biggr)\,, \\
  \Rrb_{\,1}\ =\ \Rrt_{\,1}\,, \\
  \Rrb_{\,2}\ =\ \Rrt_{\,2}\,.
\end{gather*}
Then, from Equation~\eqref{eq:pr1} and from the last three conditions it follows that
\begin{multline*}
  \Pr\ =\ \rho\,g\,\H\,\Q\ -\ \rho\,\Q\;\biggl(\frac{\H^{\,2}}{2}\;\Rrb_{\,1}\ +\ \H\,\Rrb_{\,2}\biggr)\ -\ \rho\,g\,\frac{\Q^{\,2}}{2}\ +\ \rho\;\biggl(\frac{\Q^{\,3}}{6}\;\Rrt_{\,1}\ +\ \frac{\Q^{\,2}}{2}\;\Rrt_{\,2}\biggr)\ =\\
  \rho\,\Q\;\biggl(\H\ -\ \frac{\Q}{2}\biggr)\cdot(g\ -\ \Rrb_{\,2})\ -\ \rho\,\Q\;\biggl(\frac{\H^{\,2}}{2}\ -\ \frac{\Q^{\,2}}{6}\biggr)\;\Rrb_{\,1}\,.
\end{multline*}
This gives us the first compatibility condition for the depth-integrated pressure, which relates in a highly nonlinear way quantities from inner and outer domains:
\begin{empheq}[box={\mymath}]{equation}
  \Pr\, =\, \rho\,\Q\;\biggl(\H\, -\, \frac{\Q}{2}\biggr)\cdot(g\, -\, \Rrb_{\,2})\, -\, \rho\,\Q\;\biggl(\frac{\H^{\,2}}{2}\, -\, \frac{\Q^{\,2}}{6}\biggr)\;\Rrb_{\,1}\,, \ \x\, \in\, \Gamma^{\,\blackdiamond}\,, \ \forall\,t\, \in\, \R^{\,+}_{\,0}\,.\label{eq:cond1}
\end{empheq}


\subsubsection{Flux condition}

The second compatibility condition can be obtained by making a natural assumption that the $3$D (horizontal) velocity field $\u$ across the surface $\partial_{\,\parallel}\,\Om\,(t)$ is continuous. Let $\n$ denote the projection of a unit normal $\n_{\,3}$ to $\partial_{\,\parallel}\,\Om\,(t)$ on horizontal directions, \ie $\n\ =\ \pi_{\,\x}\,(\n_{\,3})\,$. Then, from Definitions~\eqref{eq:ave} and \eqref{eq:41} of the depth-averaged horizontal velocities $\ub$ and $\ut$ correspondingly, we obtain the following sequence of equalities, which hold on $\Gamma^{\,\blackdiamond}\,$:
\begin{equation*}
  \H\,(\ub\scal\n)\ \stackrel{\text{\eqref{eq:ave}}}{=}\ \int_{\,-\,h}^{\,\eta}\,(\u\scal\n)\,(\scal,\,y,\,\cdot)\;\ud y\ \stackrel{\text{\eqref{eq:sides}}}{=}\ \int_{\,-\,h}^{\,d}\,(\u\scal\n)\,(\scal,\,y,\,\cdot)\;\ud y\ \stackrel{\text{\eqref{eq:41}}}{=}\ \Q\,(\ut\scal\n)\,.
\end{equation*}
By identifying two ends of the last sequence of equalities, we obtain the second compatibility condition:
\begin{empheq}[box={\mymath}]{equation}
  \rho\,\H\,(\ub\scal\n)\ =\ \rho\,\Q\,(\ut\scal\n)\,, \qquad \x\ \in\ \Gamma^{\,\blackdiamond}\,, \qquad \forall\,t\ \in\ \R^{\,+}_{\,0}\,,\label{eq:cond2}
\end{empheq}
where we multiplied both sides by fluid density $\rho$ to have the equality of fluid fluxes (in terms of physical units) through the surface $\partial_{\,\parallel}\,\Om\,(t)\,$. We would like to underline that compatibility Condition~\eqref{eq:cond2} ensures the mass conservation in domain $\D^{\:\boxdot}\,$.


\subsection{Nonlinear shallow water $\SW$ case}

The compatibility conditions derived in Section~\ref{sec:compa} can be directly transposed to the case of nonlinear shallow water equations $\SW$. The pressure-related Condition~\eqref{eq:cond1} simplifies greatly in the $\SW$ realm:
\begin{empheq}[box={\mymath}]{equation*}
  \Pr\ =\ \rho\,g\,\Q\;\biggl(\H\ -\ \frac{\Q}{2}\biggr)\,, \qquad \x\ \in\ \Gamma^{\,\blackdiamond}\,, \qquad \forall\,t\ \in\ \R^{\,+}_{\,0}\,.
\end{empheq}
The flux Condition~\eqref{eq:cond2} is transposed directly to $\SW$ case without any modifications.


\section{Discussion}\label{sec:concl}

In the present work we considered the problem of wave interaction with a floating, but fixed partially immersed solid body. The main conclusions and perspectives of this study are outlined below.


\subsection{Conclusions}

The present manuscript is the first part of a series of studies devoted to the problem of wave/partially immersed body interaction. In this part we considered the hierarchy of mathematical models, which can be used to model this situation:
\begin{equation*}
  \SW\ \Longrightarrow\ \SGN\ \Longrightarrow\ \Po\ \Longrightarrow\ \Eu\ \Longrightarrow\ \NS
\end{equation*}
The models considered in this study are conservative (inviscid), since the dissipative effects can be added later if needed. A special emphasis was made on long wave models $\SGN$ and $\SW$, since their reduced computational complexity is very attractive for various practical applications. The derivation of these models is well understood, especially for free wave propagation. The case of the moving rigid lid is covered in our study. Moreover, we showed that there are important differences between these equations written in the free space (projects on the outer domain $\D^{\,\diamond}$) and under the floating body $\B\,(t)$ (projects on the inner domain $\D^{\,\blackdiamond}$). While the free wave propagation in relatively well understood nowadays \cite{Khakimzyanov2016, Khakimzyanov2016c}, the main peculiarity in this study is the surgery of solutions at the boundary $\Gamma^{\,\blackdiamond}\ =\ \cl\,(\,\D^{\,\diamond}\,)\ \bigcap\ \cl\,(\,\D^{\,\blackdiamond}\,)$ between two domains in order to obtain a globally valid physical solution in $\D^{\:\boxdot}\,$. This is achieved using specifically derived compatibility conditions on $\Gamma^{\,\blackdiamond}\,$. They allow us to ensure, in particular, the global mass and/or energy conservation, depending on the choice of imposed conditions. Moreover, we derive also for $\SW^{\,\blackdiamond}$ and $\SGN^{\,\blackdiamond}$ models in the inner domain the \emph{local} balance equations for the momentum and total energy, which become conservation laws if the solid boundaries are fixed (\ie do not move in time). The structure of these equations in the inner domain is discussed and we showed that a successful splitting strategy for the numerical and theoretical analyses is fundamentally different in the inner and outer domains. This understanding will be used in the following Part~II \cite{Khakimzyanov2018b} to construct efficient numerical solvers for fully nonlinear weakly dispersive wave propagation in the presence of a partially immersed floating body.


\subsection{Perspectives}

The modelling approaches described in the present study can be further extended in two main directions. First of all, the continuous models $\Eu$, $\Po$, $\SGN$ and $\SW$ we described above have to be properly discretized (in space and in time) and implemented in numerical codes. In the following part of our study \cite{Khakimzyanov2018b} we shall take care of $\SW$ and $\SGN$ models. First numerical results will be presented in $2$D. However, one may want to solve $3$D problems as well. The peculiarity in $3$D wave/body interaction is that a wave can overcome the obstacle on the lateral sides as well, while in $2$D there are only two possibilities: \emph{reflection} and \emph{underflow}. Thus, the $3$D case will represent a much richer hydrodynamic behaviour. In the future, we would like to study and shed some light on these phenomena.

Another possible generalization will consist in allowing the obstacle to move under the wave loading. We remind that the obstacle was assumed to be fixed in the present study. Of course, it was a simplifying assumption, which has to be released in future investigations. This generalization will propose new simulation tools for ship and naval hydrodynamics.


\subsection*{Acknowledgments}
\addcontentsline{toc}{subsection}{Acknowledgments}

This work was supported by the Program of Presidium of RAS \No~27 ``\textit{Fundamental problems of solving sophisticated practical problems using supercomputers}''. The Authors would like to acknowledge the backing from University \textsc{Savoie Mont Blanc} under the project \textrm{MNBT} retained during AAP/2018 call. Moreover, we would like to thank Dr.~R\'emi \textsc{Carmigniani} together with his PhD thesis advisor Dr.~Damien \textsc{Violeau} (\textsc{Saint-Venant} Laboratory for Hydraulics, EDF R\&D Department) for urging us to work on this exciting topic. We would like to thank the anonymous Referee for helping us to shape this manuscript. Finally, the Authors thank also Dr.~Laurent~\textsc{Gosse} (IAC -- CNR, Italy) and Dr.~Francesco~\textsc{Fedele} (\textsc{Georgia} Institute of Technology, USA) for discussing with us various theoretical and numerical matters.


\appendix
\section*{Nomenclature}
\addcontentsline{toc}{section}{Nomenclature}
\label{sec:N}

\begin{description}
  \item[$\leftsquigarrow\ $] the left hand side is substituted by the right hand side
  \item[$\rightsquigarrow\ $] the right hand side is substituted by the left hand side
  \item[$\equiv\ $] equal identically
  \item[$\cong\ $] isomorphism relation
  \item[$\eqdef\ $] the left hand side is defined by the right hand side
  \item[$\defeq\ $] the right hand side is defined by the left hand side
  \item[$\bydef\ $] equal by definition
  \item[$\eqass\ $] left hand side is asymptotically equal to \dots (after truncation of higher order terms)
  \item[$\asseq\ $] right hand side is asymptotically equal to \dots (after truncation of higher order terms)
  \item[$\propto\ $] the left hand side is proportional to the right hand side
  \item[$\otimes\ $] tensor product
  \item[$\Id\ $] identity square matrix in $\Mat_{\,n}\,(\R)$
  \item[$\N\ $] the set of natural numbers, \ie $\bigl\{1,\,2,\,\ldots\bigr\}$ (a monoid with respect to multiplication and a semi-group with respect to addition operations)
  \item[$\N_{\,0}\ $] the set of natural numbers with zero, \ie $\bigl\{0,\,1,\,2,\,\ldots\bigr\}$ (a semi-group with respect to multiplication and a monoid with respect to addition operations)
  \item[$\boldsymbol{n}^{\,\leq}\ $] for any $n\ \in\ \N$ denotes a finite set of natural numbers $\boldsymbol{n}^{\,\leq}\ \eqdef\ \bigl\{1,\,2,\,\ldots,\,n\ -\ 1,\,n\bigr\}$
  \item[$\boldsymbol{n}^{\,<}\ $] for any $n\ \in\ \N$ denotes a finite set of natural numbers $\boldsymbol{n}^{\,<}\ \eqdef\ \bigl\{1,\,2,\,\ldots,n\ -\ 2,\,\,n\ -\ 1\bigr\}$
  \item[$\boldsymbol{n}^{\,\leq}_{\,0}\ $] for any $n\ \in\ \N$ denotes a finite set of natural numbers $\boldsymbol{n}^{\,\leq}_{\,0}\ \eqdef\ \bigl\{0,\,1,\,2,\,\ldots,\,n\ -\ 1,\,n\bigr\}$
  \item[$\boldsymbol{n}^{\,<}_{\,0}\ $] for any $n\ \in\ \N$ denotes a finite set of natural numbers $\boldsymbol{n}^{\,<}_{\,0}\ \eqdef\ \bigl\{0,\,1,\,2,\,\ldots,n\ -\ 2,\,\,n\ -\ 1\bigr\}$
  \item[$\R\ $] the set of all real numbers
  \item[$\R^{\,d}\ $] the standard vector space of $d-$tuples of real numbers
  \item[$\Ee^{\,d}\ $] the standard (real) \textsc{Euclidean} space, \ie $\R^{\,d}$ supplied with the standard scalar product
  \item[$\R^{\,+}\ $] the set of all positive real numbers (a group with respect to the multiplication and a semi-group with respect to the addition operation)
  \item[$\R^{\,-}\ $] the set of all negative real numbers (a semi-group with respect to addition operation)
  \item[$\R^{\,+}_{\,0}\ $] the set of all non-negative real numbers (a monoid with respect to the addition operation)
  \item[$\infty\ $] $\aleph_{\,0}$ or $\aleph_{\,1}$ depending on the context
  \item[$\rb\ $] for any fixed $r\ \in\ \R$ denotes a function, which takes a constant value $r\,$, \ie $\rb\,:\ \R^{\,d}\ \longrightarrow\ \R$ such that $\forall\, \x\ \in\ \R^{\,d}\,$: $\rb\,(\x)\ \equiv\ r\,$, $d\ \geq\ 1$
  \item[$\rbv\ $] for any fixed $r\ \in\ \R$ denotes a vector-function, which takes a constant vector value $\underbrace{(r,\,r,\,\ldots,r)}_{n\ \text{times}}\,$, \ie $\rbv\,:\ \R^{\,d}\ \longrightarrow\ \R^{\,n}$ such that $\forall\, \x\ \in\ \R^{\,d}\,$: $\rbv\,(\x)\ \equiv\ \underbrace{(r,\,r,\,\ldots,r)}_{n\ \text{times}}\,$, $d\ \geq\ 1$ and $n\ >\ 1$
  \item[$\x\ $] horizontal components of \textsc{Cartesian} coordinates in $\R^{\,3}$
  \item[$\abs{\cdot}\ $] depending on the argument $(\cdot)\,$: it is the absolute value of a real number or \textsc{Euclidean} norm of a vector
  \item[$y\ $] vertical component of \textsc{Cartesian} coordinates in $\R^{\,3}$
  \item[$\partial_{\,\alpha}\ $] is the partial derivative with respect to $x_{\,\alpha}\,$, \ie $\partial_{\,\alpha}\,(\cdot)\ \eqdef\ \pdd{(\cdot)}{x_{\,\alpha}}\,$, $\alpha\ \in\ \Fin{2}$
  \item[$\Ddb\ $] total (material) derivative operator in $2$D based on velocity $\ub$
  \item[$\Ddt\ $] total (material) derivative operator in $2$D based on velocity $\ut$
  \item[$\Om\ $] fluid domain, $\Om\ \subseteq\ \R^{\,3}$
  \item[$\text{\mancube}\ $] a parallelepiped, $\text{\mancube}\ \subseteq\ \R^{\,3}$
  \item[$\curvearrowbotright(\alpha\,\beta)\ $] a piecewise smooth oriented curve (or curvilinear segment) connecting points $\alpha$ and $\beta$ with the natural orientation from $\alpha$ to $\beta$
  \item[$\curvearrowleft(\alpha\,\beta)\ $] a piecewise smooth oriented curve (or curvilinear segment) connecting points $\alpha$ and $\beta$ with the inverse orientation from $\beta$ to $\alpha$
  \item[$\uparrow\,(\alpha,\,\beta)\ $] a vector in $\R^{\,3}$ directed vertically upwards going from point $\alpha\ \in\ \R^{\,3}$ to point $\beta\ \in\ \R^{\,3}\,$, necessarily we have $\pi_{\,\x}\,(\alpha)\ =\ \pi_{\,\x}\,(\beta)$
  \item[$\circlearrowleft(\alpha\,\beta\,\ldots\,\delta)\ $] a simple, piecewise smooth, closed oriented curve connecting points $\alpha\,$, $\beta\,$, $\ldots\,$, $\delta$ and returning to $\alpha$ with this natural orientation
  \item[$\D\,(\alpha\,\beta\,\ldots\,\delta)\ $] a plain bounded domain surrounded by the closed contour $\circlearrowleft(\alpha\,\beta\,\ldots\,\delta)$
  \item[$C^{\,p}\,(\Om)\ $] space of continuously differentiable functions up to order $p\ \in\ \N_{\,0}$ defined on domain $\Om\,$. If $p\ \equiv\ 0$ then we have the space of just continuous functions on $\Om$
  \item[$\under{\Om}\ $] canonical projection of a $3$D domain $\Om\ \subseteq\ \R^{\,3}$ on horizontal directions $O\,x\,y\,$. Thus, $\under{\Om}\ \subseteq\ \R^{\,2}$
  \item[$\D\ $] a domain in $\R^{\,2}$
  \item[$\cl(\D)\ $] closure of a domain $\D$ in $\R^{\,2}$
  \item[$f\,\vert_{\,\D}\ $] restriction of a map $f$ to the sub-domain $\D$
  \item[$\Gamma\ $] boundary of a domain in $\R^{\,2}\,$, \ie $\Gamma\ \eqdef\ \partial\,\D$
  \item[$\partial_{\,\parallel}\,\Om\ $] lateral\footnote{By lateral boundary we understand the part of the surface of a cylindrical domain, which is parallel to the central axis. In our study this axis coincides with $O\,y\,$.} boundary of a domain $\Om\ \subseteq\ \R^{\,3}$
  \item[$\n\ $] horizontal components of a unit exterior normal to a boundary in a point
  \item[$\kappa\ $] curvature of a $2$D boundary in a point
  \item[$\kappar\ $] signed curvature of a $2$D boundary in a point
  \item[$\rho\ $] constant fluid density
  \item[$g\ $] constant gravity acceleration
  \item[$\eta\ $] free surface excursion
  \item[$\H\ $] total water depth, \ie $\H\ \eqdef\ h\ +\ \eta$ ($\SGN$, $\Bouss$ and $\SW$ models)
  \item[$h\ $] a function, which describes the bottom of the fluid domain $\Om$
  \item[$\B\ $] partially immersed floating body, $\B\ \subseteq\ \R^{\,3}$
  \item[$d\ $] a function, which describes the bottom of the immersed body $\B$
  \item[$\Q\ $] fluid layer height under the body $\B\,$, \ie $\Q\ \eqdef\ d\ +\ h$
  \item[$\u\ $] horizontal velocity field in the $3$D flow ($\Eu$ model)
  \item[$v\ $] vertical velocity in the $3$D flow ($\Eu$ model)
  \item[$p\ $] pressure field in the $3$D flow ($\Eu$ and $\Po$ models)
  \item[$\phi\ $] velocity potential function ($\Po$ model)
  \item[$\ub\ $] depth-averaged horizontal velocity in the outer domain $\D^{\,\diamond}\,$, $\ub\ =\ \ub\,(\x,\,t)$ ($\SGN$, $\Bouss$ and $\SW$ models)
  \item[$\ut\ $] depth-averaged horizontal velocity in the inner domain $\D^{\,\blackdiamond}\,$, $\ud\ =\ \ud\,(\x,\,t)$ ($\SGN$ and $\SW$ models)
  \item[$\Pp\ $] depth-integrated fluid pressure in the outer domain $\D^{\,\diamond}$ ($\SGN$ and $\Bouss$ models)
  \item[$\Pr\ $] depth-integrated fluid pressure in the inner domain $\D^{\,\blackdiamond}$ ($\SGN$ model)
  \item[$\pc^{\,\blackdiamond}\ $] fluid pressure at the bottom in the inner domain $\D^{\,\blackdiamond}$ ($\SGN^{\,\blackdiamond}$ and $\SW^{\,\blackdiamond}$ models)
  \item[$\pc^{\,\diamond}\ $] fluid pressure at the bottom in the outer domain $\D^{\,\diamond}$ ($\SGN^{\,\diamond}$, $\Bouss^{\,\diamond}$ and $\SW^{\,\diamond}$ models)
  \item[$\pt^{\,\blackdiamond}\ $] fluid pressure at the bottom of the object $\B\,(t)$ ($\SGN^{\,\blackdiamond}$ and $\SW^{\,\blackdiamond}$ models)
  \item[$\Rrb_{\,1,\,2}\ $] acceleration due to uneven bottom effects in the outer domain $\D^{\,\diamond}$ ($\SGN^{\,\diamond}$ model)
  \item[$\Rrt_{\,1,\,2}\ $] acceleration due to uneven bottom effects in the inner domain $\D^{\,\blackdiamond}$ ($\SGN^{\,\blackdiamond}$ model)
  \item[$\Qq_{\,\n}\ $] the mass flux along a certain direction $\n\,$, which defined in the outer domain as $\Qq_{\,\n}\ \eqdef\ \rho\,\H\,(\ub\scal\n)\,$. In the inner domain $\H\ \leftsquigarrow\ \Q$ and $\ub\ \leftsquigarrow\ \ut$
  \item[$\E\ $] the total energy in various models ($\SW$, $\Bouss$, $\SGN$, $\Po$, $\Eu$). In some situations we employ the superscript $\E^{\,\blackdiamond}$ and $\E^{\,\diamond}$ to differentiate the energy contained in the inner and outer domains respectively ($\SW$ and $\SGN$ models)
  \item[$\eps\ $] the nonlinearity parameter equal to the ratio between a typical wave amplitude to the characteristic water depth ($\SGN$, $\Bouss$ and $\SW$ models)
  \item[$\mu\ $] the shallowness parameter equal to the ratio between a typical water depth to the characteristic wavelength ($\SGN$, $\Bouss$ and $\SW$ models)
\end{description}


\section{An alternative derivation of compatibility conditions}
\label{app:comp}

In this Appendix we provide a more detailed, but also somehow alternative, derivation of the compatibility conditions given in Section~\ref{sec:compa}. Let us have a critical view on the derivation presented above. First of all, in the derivation of the flux Condition~\eqref{eq:cond2} we did not use the reconstruction Formulae \eqref{eq:hor}, \eqref{eq:40} for the horizontal velocity field, only the Definitions \eqref{eq:ave} and \eqref{eq:41} of the depth-averaged velocities $\ub$ and $\ut$ correspondingly and the impermeability Condition \eqref{eq:sides} on the body lateral surface $\partial_{\,\parallel}\,\B\,(t)\,$. This is rather a strong point since reconstruction Formulae \eqref{eq:hor}, \eqref{eq:40} are valid for irrotational flows only, while the compatibility conditions are being derived in the general case. On the other hand, in the derivation of the pressure continuity Condition~\eqref{eq:cond1} we used the reconstruction Formulae~\eqref{eq:press}, \eqref{eq:pr} for the fluid pressure $p\,$. The validity of these reconstructions near the boundary $\Gamma^{\,\blackdiamond}$ may be questioned\footnote{We provide here one reason for this: the long wave character of the flow might be perturbed in the vicinity of such abrupt changes in the flow geometry. Thus, the reconstructed flow might be well different from the corresponding $3$D solution to $\Eu$ model, for example.}. Thus, in this Appendix we re-consider our previous derivation. Our goal is to avoid using any kind of reconstruction formulae, which are questionable in the region of interest. Moreover, we shall include in our consideration the energy balance equations on both sides of the boundary $\Gamma^{\,\blackdiamond}\,$, which are formal differential consequences of the mass and momentum balances for sufficiently smooth solutions. However, the situation might be different for abrupt changes. Thus, some authors\footnote{These authors considered the \textsc{Riemann} problem \cite{Kroner1997} for nonlinear shallow water equations $\SW$ in the presence of a discontinuity in the bathymetry. Our situation is similar in the sense that the discontinuity is located at the free surface (in contrast to the bottom) and caused by the presence of a floating body $\B\,(t)\,$.} favour the usage of the energy conservation law in such situations in lieu of the momentum conservation \cite{Alcrudo2001, Bernetti2008}. We follow this philosophy in great lines as well.


\subsection{Nonlinear dispersive case}

We already mentioned in Section~\ref{sec:struct} that the unknown functions in $\SGN$ models are:
\begin{description}
  \item[Outer domain $\D^{\,\diamond}\,$] two components of the velocity vector $\ub$ and the water height $\H\,$,
  \item[Inner domain $\D^{\,\blackdiamond}\,$] two components of the velocity vector $\ut$ and the depth-integrated pressure $\Pr\,$.
\end{description}

Thus, we need to have three compatibility conditions in $3$D and two in $2$D correspondingly. These conditions will allow us to deduce the flow parameters on the other side of $\Gamma^{\,\blackdiamond}$ from knowing them on one side.

We shall work with conservative forms of equations only. Let us remind them here. We take advantage of this occasion to rewrite the governing equations in the tensorial form:
\begin{description}
  \item[Mass conservation]
  \begin{align*}
    \H_{\,t}\ +\ \partial_{\,\jmath}\,(\H\,\bar{u}_{\,\jmath})\ &=\ \zb\,, \qquad \x\ \in\ \D^{\,\diamond}\,, \\
    \Q_{\,t}\ +\ \partial_{\,\jmath}\,\bigl(\Q\,\under{u}_{\,\jmath}\bigr)\ &=\ \zb\,, \qquad \x\ \in\ \D^{\,\blackdiamond}\,,
  \end{align*}
  \item[Momentum balance]
  \begin{align*}
    (\rho\,\H\,\bar{u}_{\,\imath})_{\,t}\ +\ \partial_{\,\jmath}\,\Bigl(\,\H\,\bar{u}_{\,\imath}\,\bar{u}_{\,\jmath}\ +\ \delta_{\,\jmath}^{\,\imath}\,\Pp\,\Bigr)\ &=\ \pc^{\,\diamond}\,h_{\,x_{\,\imath}}\,, \qquad \x\ \in\ \D^{\,\diamond}\,, \\
    (\rho\,\Q\,\under{u}_{\,\imath})_{\,t}\ +\ \partial_{\,\jmath}\,\Bigl(\,\Q\,\under{u}_{\,\imath}\,\under{u}_{\,\jmath}\ +\ \delta_{\,\jmath}^{\,\imath}\,\Pr\,\Bigr)\ &=\ \pc^{\,\blackdiamond}\,h_{\,x_{\,\imath}}\ +\ \pt^{\,\blackdiamond}\,d_{\,x_{\,\imath}}\,, \qquad \x\ \in\ \D^{\,\blackdiamond}\,,
  \end{align*}
  \item[Energy balance]
  \begin{align*}
    (\H\,\E)_{\,t}\ +\ \partial_{\,\jmath}\,\Bigl(\bigl(\,\H\,\E\ +\ \Pp\,\bigr)\,\bar{u}_{\,\jmath}\Bigr)\ &=\ -\,\pc^{\,\diamond}\,h_{\,t}\,, \qquad \x\ \in\ \D^{\,\diamond}\,, \\
    (\Q\,\E)_{\,t}\ +\ \partial_{\,\jmath}\,\bigl((\,\Q\,\E\ +\ \Pr\,)\,\under{u}_{\,\jmath}\,\bigr)\ &=\ -\,\pc^{\,\blackdiamond}\,h_{\,t}\ -\ \pt^{\,\blackdiamond}\,d_{\,t}\,, \qquad \x\ \in\ \D^{\,\blackdiamond}\,,
  \end{align*}
\end{description}
where $\imath\ \in\ \Fin{2}$ and $\delta_{\,\jmath}^{\,\imath}$ is the \textsc{Kronecker} $\delta-$function. In equations above we used the \textsc{Einstein} summation convention over the repeated index (here $\jmath\ \in\ \Fin{2}$) and $\partial_{\,\jmath}\,(\cdot)\ \eqdef\ \pdd{(\cdot)}{x_{\,\jmath}}$ is the usual partial derivative with respect to $x_{\,\jmath}\,$. One can see that all equations listed above can be written \emph{symbolically} as
\begin{equation}\label{eq:cons}
  \qs_{\,t}\ +\ \partial_{\,\jmath}\,\fs_{\,\jmath}\ =\ \rs\,, \qquad \x\ \in\ \D^{\:\boxdot}\,, \qquad \jmath\ \in\ \Fin{2}\,,
\end{equation}
for some accordingly chosen functions $\qs\,$, $\fs_{\,\jmath}$ and $\rs$ defined on the corresponding domain $\D^{\:\boxdot}\ \bydef\ \D^{\,\diamond}\ \bigcup\ \D^{\,\blackdiamond}\,$. However, all these functions are discontinuous across $\Gamma^{\,\blackdiamond}\,$. Thus, Equation~\eqref{eq:cons} should be understood in the integral form only. 

\begin{figure}
  \centering
  \includegraphics[width=1.02\textwidth]{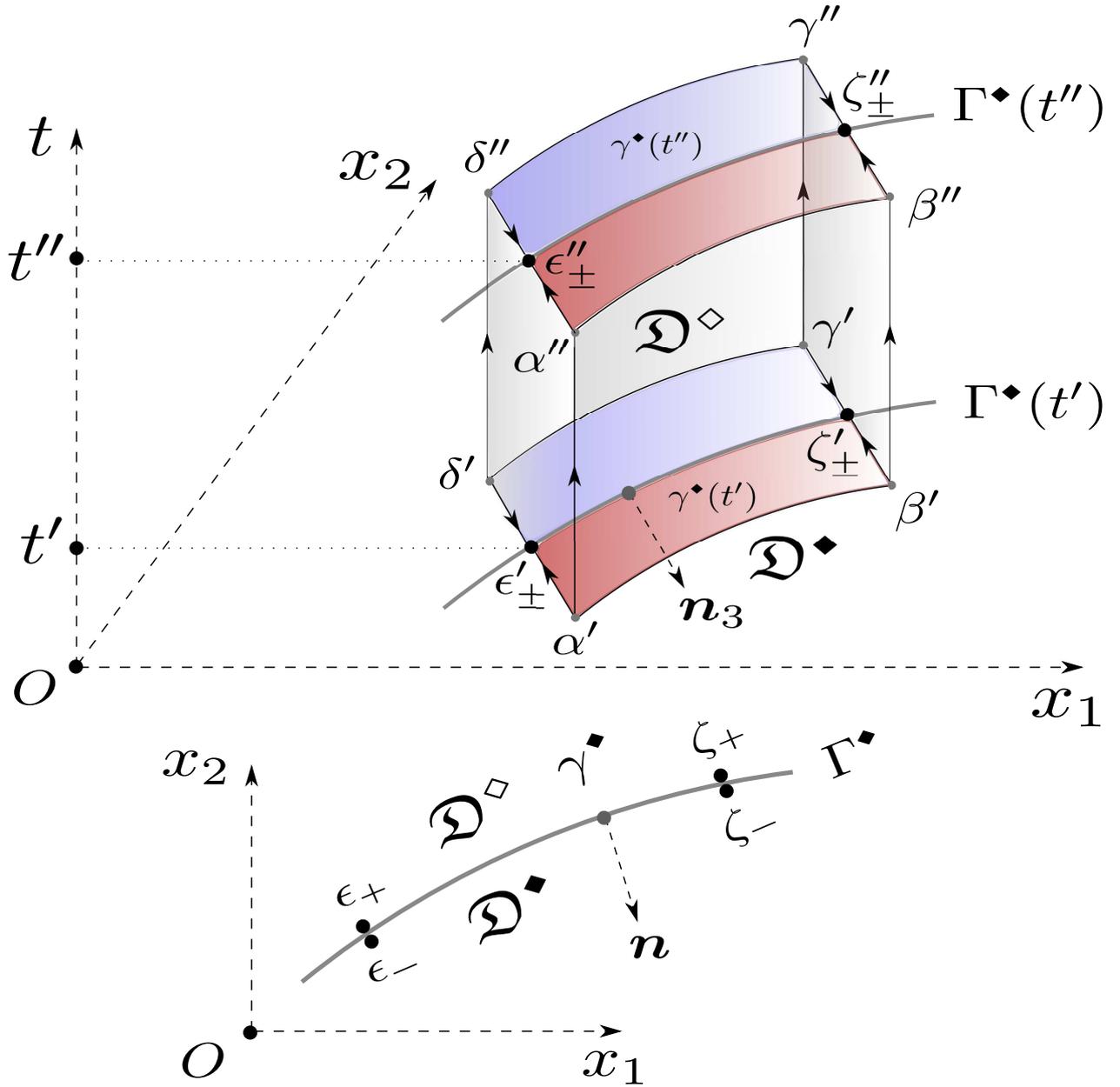}
  \caption{\small\em A curvilinear parallelepiped constructed around an element $\gamma^{\,\blackdiamond}\ \subseteq\ \Gamma^{\,\blackdiamond}$ to illustrate the derivation of compatibility conditions.}
  \label{fig:spacetime}
\end{figure}

Consider a \textsc{Cartesian} reference frame $O\,\x\,t\,$. To make explicit this integral form, let us choose first two arbitrary instances of time $0\ \leq\ t^{\,\prime}\ \leq\ t^{\,\dprime}$ and a small portion $\gamma^{\,\blackdiamond}\ \subseteq\ \Gamma^{\,\blackdiamond}$ together with a neighbourhood $\D\,(\alpha\,\beta\,\gamma\,\delta)$ delimited by the contour $\circlearrowleft(\alpha\,\beta\,\gamma\,\delta)$ such that the lines $\curvearrowbotright(\alpha\,\beta)\ \subseteq\ \D^{\,\blackdiamond}$ and $\curvearrowleft(\gamma\,\delta)\ \subseteq\ \D^{\,\diamond}$ lie on two sides of $\gamma^{\,\blackdiamond}\,$. The curve $\gamma^{\,\blackdiamond}$ along with its contour $\circlearrowleft(\alpha\,\beta\,\gamma\,\delta)$ do not change in time. However, to make difference among these geometric objects at different time layers, we shall denote them with the corresponding number of primes at superscripts, \ie in space-time \emph{stricto sensu}
\begin{equation*}
  \circlearrowleft(\alpha^{\,\prime}\,\beta^{\,\prime}\,\gamma^{\,\prime}\,\delta^{\,\prime})\ \neq\ \circlearrowleft(\alpha^{\,\dprime}\,\beta^{\,\dprime}\,\gamma^{\,\dprime}\,\delta^{\,\dprime})\,,
\end{equation*}
even if they encode the same geometrical object in \emph{space}. We refer to Figure~\ref{fig:spacetime} for an illustration. As a result, in the space $O\,\x\,t$ we obtain a curvilinear parallelepiped $\text{\mancube}\,(\alpha^{\,\prime}\,\beta^{\,\prime}\,\gamma^{\,\prime}\,\delta^{\,\prime}\,\alpha^{\,\dprime}\,\beta^{\,\dprime}\,\gamma^{\,\dprime}\,\delta^{\,\dprime})\,$. Its upper and lower faces are parallel to the coordinate plane $O\,\x\,$, while the lateral boundary, that we denote by $\partial_{\,\parallel}\,\text{\mancube}\,$, consists of a cylindrical surface formed by $\circlearrowleft(\alpha^{\,\prime}\,\beta^{\,\prime}\,\gamma^{\,\prime}\,\delta^{\,\prime})$ as a generatrix and $\uparrow\,(\alpha^{\,\prime},\,\alpha^{\,\dprime})$ as a directrix:
\begin{equation*}
  \partial_{\,\parallel}\,\text{\mancube}\ \eqdef\ \D\,(\alpha^{\,\prime}\,\alpha^{\,\dprime}\,\delta^{\,\dprime}\,\delta^{\,\prime})\ \bigcup\ \D\,(\alpha^{\,\prime}\,\beta^{\,\prime}\,\beta^{\,\dprime}\,\alpha^{\,\dprime})\ \bigcup\ \D\,(\beta^{\,\prime}\,\beta^{\,\dprime}\,\gamma^{\,\dprime}\,\gamma^{\,\prime})\ \bigcup\ \D\,(\gamma^{\,\prime}\,\gamma^{\,\dprime}\,\delta^{\,\dprime}\,\delta^{\,\prime})\,.
\end{equation*}
Since the directrix of this cylinder is parallel to the axis $O\,t\,$, the exterior unitary normal $\n_{\,3}$ to $\partial_{\,\parallel}\,\text{\mancube}$ will have the form $\n_{\,3}\ \cong\ (n_{\,1},\,n_{\,2},\,0)\ \in\ \R^{\,3}$ and the vector $\n\ \cong\ (n_{\,1},\,n_{\,2})\ \in\ \R^{\,2}$ is independent of time. However, in general $\n\ =\ \n\,(\x)\,$.

We integrate Equation~\eqref{eq:cons} over the domain $\text{\mancube}\,(\alpha^{\,\prime}\,\beta^{\,\prime}\,\gamma^{\,\prime}\,\delta^{\,\prime}\,\alpha^{\,\dprime}\,\beta^{\,\dprime}\,\gamma^{\,\dprime}\,\delta^{\,\dprime})$ to obtain the following integral relation:
\begin{multline}\label{eq:int}
  \sqiint\limits_{\D\,(\alpha^{\,\dprime}\,\beta^{\,\dprime}\,\gamma^{\,\dprime}\,\delta^{\,\dprime})}\,\qs\,(\scal,\,t^{\,\dprime})\;\ud\x\ -\ \sqiint\limits_{\D\,(\alpha^{\,\prime}\,\beta^{\,\prime}\,\gamma^{\,\prime}\,\delta^{\,\prime})}\,\qs\,(\scal,\,t^{\,\prime})\;\ud\x\ +\\
  \oiint\limits_{\partial_{\,\parallel}\,\text{\mancube}}\,\fs_{\,\jmath}\,(\scal,\,\tau)\,n_{\,\jmath}\,(\scal)\;\ud\bigma \ =\ \iiint_{\text{\mancube}}\,\rs\;\ud\x\,\ud t\,,
\end{multline}
where $\jmath\ \in\ \Fin{2}\,$, $t^{\,\prime}\ \leq\ \tau\ \leq\ t^{\,\dprime}\,$, $\ud\bigma$ is the surface element and we omitted the lists of points ($\alpha\,$, $\beta\,$, $\gamma\,$, $\delta$) for the sake of notation compactness. The integral relation obtained above makes sense for bounded and continuous almost everywhere functions. Thus, they might admit a discontinuity along $\gamma^{\,\blackdiamond}\,$. Now, we can tend from each side of $\gamma^{\,\blackdiamond}$ the points $\alpha\ \to\ \epsilon_{\,-}\,$, $\delta\ \to\ \epsilon_{\,+}\,$, $\beta\ \to\ \zeta_{\,-}$ and $\gamma\ \to\ \zeta_{\,+}\,$. In the limit only two surface integrals over the faces $\D\,(\alpha^{\,\prime}\,\beta^{\,\prime}\,\beta^{\,\dprime}\,\alpha^{\,\dprime})$ and $\D\,(\delta^{\,\prime}\,\gamma^{\,\prime}\,\gamma^{\,\dprime}\,\delta^{\,\dprime})$ will remain in integral Relation~\eqref{eq:int}. These two faces will becomes two opposite sides (with opposite exterior normals $\n_{\,3}$) of the surface $\D\,(\epsilon^{\,\prime}\,\zeta^{\,\prime}\,\epsilon^{\,\dprime}\,\zeta^{\,\dprime})\,$. Let us assume the unitary vector $\n_{\,3}$ points into the inner domain $\D^{\,\blackdiamond}\,$. Then, in the limit the integral Relation~\eqref{eq:int} will become:
\begin{multline}\label{eq:59}
  \sqiint\limits_{\D\,(\epsilon^{\,\prime}\,\zeta^{\,\prime}\,\zeta^{\,\dprime}\,\epsilon^{\,\dprime})}\,\fs_{\,\jmath}^{\,\blackdiamond}\,(\scal,\,\tau)\,n_{\,\jmath}\,(\scal)\;\ud\bigma\ -\ \sqiint\limits_{\D\,(\epsilon^{\,\prime}\,\zeta^{\,\prime}\,\zeta^{\,\dprime}\,\epsilon^{\,\dprime})}\,\fs_{\,\jmath}^{\,\diamond}\,(\scal,\,\tau)\,n_{\,\jmath}\,(\scal)\;\ud\bigma\ =\ \zb\,, \quad \Longrightarrow \\
  \sqiint\limits_{\D\,(\epsilon^{\,\prime}\,\zeta^{\,\prime}\,\zeta^{\,\dprime}\,\epsilon^{\,\dprime})}\underbrace{\Bigl\{\fs_{\,\jmath}^{\,\blackdiamond}\,(\scal,\,\tau)\,n_{\,\jmath}\,(\scal)\ -\ \fs_{\,\jmath}^{\,\diamond}\,(\scal,\,\tau)\,n_{\,\jmath}\,(\scal)\Bigr\}}_{\displaystyle{(\text{\manstar})}}\;\ud\bigma\ =\ \zb\,, \qquad \jmath\ \in\ \Fin{2}\,,
\end{multline}
where
\begin{equation*}
  \lim_{\substack{\x\ \to\ \bepsilon \\ \x\ \in\ \D^{\,\blackdiamond}}} \fs_{\,\jmath}\,(\x,\,\cdot)\ \defeq\ \fs_{\,\jmath}^{\,\blackdiamond}\,(\bepsilon,\,\cdot)\,, \qquad \lim_{\substack{\x\ \to\ \bepsilon \\ \x\ \in\ \D^{\,\diamond}}} \fs_{\,\jmath}\,(\x,\,\cdot)\ \defeq\ \fs_{\,\jmath}^{\,\diamond}\,(\bepsilon,\,\cdot)\,, \qquad \jmath\ \in\ \Fin{2}\,.
\end{equation*}
Since the portion $\gamma^{\,\blackdiamond}\ \subseteq\ \Gamma^{\,\blackdiamond}$ is arbitrary along with the time instances $t^{\,\prime}\ \leq\ t^{\,\dprime}\,$, the integral Identity~\eqref{eq:59} can be fulfilled only if the expression $(\text{\manstar})$ vanishes identically\footnote{The proper justification of this step is given by the Localization theorem, which requires only the continuity of the expression under the integral sign.} in all points $\x\ \in\ \Gamma^{\,\blackdiamond}$ and for all times, \ie
\begin{equation*}
  \fs_{\,\jmath}^{\,\blackdiamond}\,(\x,\,t)\cdot n_{\,\jmath}\,(\x)\ =\ \fs_{\,\jmath}^{\,\diamond}\,(\x,\,t)\cdot n_{\,\jmath}\,(\x)\,, \qquad \forall\,\x\ \in\ \Gamma^{\,\blackdiamond}\,, \qquad \forall\,t\ \in\ \R^{\,+}_{\,0}\,.
\end{equation*}
The just obtained general result can be applied to the mass, momentum and energy balance laws mentioned hereinabove to obtain the following set of compatibility conditions, which hold $\forall\,\x\ \in\ \Gamma^{\,\blackdiamond}$ and $\forall\,t\ \in\ \R^{\,+}_{\,0}\,$:
\begin{empheq}[box={\mymath}]{align}
  \rho\,\Q\,(\ut\scal\n)\ &=\ \rho\,\H\,(\ub\scal\n)\ \defeq\ \Qq_{\,\n}\,, \label{eq:comp1} \\
  \rho\,\Q\,(\ut\scal\n)\,\ut\ +\ \Pr\,\n\ &=\ \rho\,\H\,(\ub\scal\n)\ +\ \Pp\,\n\,, \label{eq:comp2} \\
  \rho\,\Q\,(\ut\scal\n)\,\Bigl(\,\E^{\,\blackdiamond}\ +\ \frac{\Pr}{\Q}\,\Bigr)\ &=\ \rho\,\H\,(\ub\scal\n)\,\Bigl(\,\E^{\,\diamond}\ +\ \frac{\Pp}{\H}\,\Bigr)\,, \label{eq:comp3}
\end{empheq}
where the quantity $\Qq_{\,\n}$ is the \emph{mass flux} through the boundary $\Gamma^{\,\blackdiamond}$ in the normal direction $\n\,$. Using the notion of the mass flux $\Qq_{\,\n}\,$, we can greatly simplify two other families of compatibility conditions:
\begin{empheq}[box={\mymath}]{align}
  \Qq_{\,\n}\,\ut\ +\ \Pr\,\n\ &=\ \Qq_{\,\n}\,\ub\ +\ \Pp\,\n\,, \label{eq:64} \\
  \E^{\,\blackdiamond}\ +\ \frac{\Pr}{\Q}\ &=\ \E^{\,\diamond}\ +\ \frac{\Pp}{\H}\,. \label{eq:65}
\end{empheq}
We can derive a useful consequence of Condition~\eqref{eq:64} if we multiply\footnote{The multiplication here is understood in the sense of the standard scalar product in $\Ee^{\,2}\,$.} it from both sides by a vector $\taub \perp \n$ tangential to $\Gamma^{\,\blackdiamond}\,$:
\begin{empheq}[box={\mymath}]{equation}\label{eq:66}
  \ut\scal\taub\ =\ \ub\scal\taub\,, \qquad \forall\,\x\ \in\ \Gamma^{\,\blackdiamond}\,, \qquad \forall\,t\ \in\ \R^{\,+}_{\,0}\,.
\end{empheq}
Any subset\footnote{The number of compatibility conditions to be employed depends on the dimension of the problem under consideration ($2$D or $3$D) and, in likelihood, on the employed numerical algorithm. The choice of employed conditions depends mostly on modelling convictions and taste of the investigator.} of relations obtained in this Section can be used as compatibility/transmission conditions to solve the wave/floating body interaction problem.

We would like to underline that Condition~\eqref{eq:comp1} ensures the global mass conservation and we highly recommend its usage. Then, Conditions~\eqref{eq:comp1}, \eqref{eq:comp2} together with boundary Condition~\eqref{eq:sides} ensures the global conservation of the momentum in $\D^{\:\boxdot}\ \bydef\ \D^{\,\blackdiamond}\ \bigcup\ \D^{\,\diamond}\,$, while the combination \eqref{eq:comp1}, \eqref{eq:comp3} $\bigcup$ \eqref{eq:sides} ensures the global conservation of the total energy in $\D^{\:\boxdot}\,$. As we already mentioned, some authors privilege the energy conservation over momentum \cite{Alcrudo2001, Bernetti2008}. There exists also another school of thinking\footnote{First of all, the presence of energy dissipation in wave/floating obstacle interaction process due to various mechanisms was clearly demonstrated recently in \cite{Nelli2017}. However, the problem of wave interaction with submerged obstacles is much better studied. We can quote here another study devoted to this problem \cite{Lin2004}:
\smallskip
\begin{quote}
  \textit{[\;\dots\,] Energy dissipation has a profound role during the solitary wave interaction with the obstacle. As mentioned earlier, the main energy losses come from the vortex generation around the corners of the obstacle and wave breaking above the obstacle, which are much more significant than the bottom friction in this problem. While the vortex shedding takes place for all cases, wave breaking only occurs for the obstacle with large height. [\;\dots\,]}
\end{quote}
\smallskip
The influence of various dissipation mechanisms was studied in \cite{Stamos2001}:
\smallskip
\begin{quote}
  \textit{[\;\dots\,] The wave reflection and transmission in the presence of a rigid or flexible fluid-filled breakwater structure would occur with the loss of some portion of the incident wave energy by the following processes: losses due to wave breaking over the structure, loses due to turbulence induced by flow separation over and near the structure, and internal losses due to turbulence of the fluid contained inside the structure as well as nonelastic deformations of the structure itself. [\;\dots\,]}
\end{quote}
\smallskip
Henceforth, the presence of turbulent effects is widely acknowledged. In the framework presented in our study, we can take into account at the level of compatibility conditions, since not all choices conserve the energy.}, which consists in including the hydrodynamic turbulence into consideration. This phenomenon destroys the energy conservation principle, due to the inherent dissipation at \textsc{Kolmogorov} scales \cite{Frisch1995}. Thus, the pressure continuity principle described in Section~\ref{sec:press} and the resulting Condition~\eqref{eq:cond1} is preferred, since it does not yield the energy conservation. We cannot deny no one's preference. Our goal in this work is to describe various existing possibilities, which can be used in modeling practice.


\subsection{Nonlinear shallow water $\SW$ case}

In the case of hydrostatic equations $\SW$, the compatibility Conditions~\eqref{eq:comp1} and \eqref{eq:66} are preserved without any modifications. The Conditions~\eqref{eq:comp2}, \eqref{eq:comp3} and \eqref{eq:65} can be further simplified by neglecting all non-hydrostatic terms in the pressure terms using Formulae~\eqref{eq:45} and \eqref{eq:45i}. To give an example of a dynamic compatibility condition for the $\SW$ model, we can combine \eqref{eq:65} with \eqref{eq:66} to obtain the following scalar condition:
\begin{empheq}[box={\mymath}]{equation*}
  \frac{\Qq_{\,\n}^{\,2}}{2\,\rho\,\Q^{\,2}}\ +\ \rho\,g\;\frac{\Q}{2}\ +\ \frac{\Pr}{\Q}\ =\ \frac{\Qq_{\,\n}^{\,2}}{2\,\rho\,\H^{\,2}}\ +\ \rho\,g\,\H\,, \qquad \forall\,\x\ \in\ \Gamma^{\,\blackdiamond}\,, \qquad \forall\,t\ \in\ \R^{\,+}_{\,0}\,.
\end{empheq}
This completes our discussion on the so-called compatibility / transmission or even gluing conditions. We shall return to this point for the practical implementation in the following Part~II \cite{Khakimzyanov2018b}.


\bigskip
\addcontentsline{toc}{section}{References}
\bibliographystyle{abbrv}
\bibliography{biblio}
\bigskip\bigskip

\end{document}